\documentclass[twocolumn,tighten,twocolappendix]{aastex63}

\mathchardef\mhyphen="2D

\received{May 5, 2020}
\revised{June 15, 2020}
\accepted{June 24, 2020}
\submitjournal{ApJ}

\usepackage{url, soul, bm}
\usepackage{amsmath,comment}      

\shorttitle{Star and Gas Kinematics in NGC~7000/IC~5070}
\shortauthors{Kuhn et al.}

\begin{document} 

\title{The Formation of a Stellar Association in the NGC~7000/IC~5070 Complex:\\ Results from Kinematic Analysis of Stars and Gas}

\correspondingauthor{Michael A. Kuhn}
\email{mkuhn@astro.caltech.edu}

\author[0000-0002-0631-7514]{Michael A. Kuhn}
\affil{Department of Astronomy, California Institute of Technology, Pasadena, CA 91125, USA}

\author{Lynne A. Hillenbrand}
\affil{Department of Astronomy, California Institute of Technology, Pasadena, CA 91125, USA}

\author{John M. Carpenter}
\affil{Joint ALMA Observatory, Alonso de C\'ordova 3107 Vitacura, Santiago, Chile}

\author{Angel Rodrigo Avelar Menendez}
\affil{Department of Astronomy, California Institute of Technology, Pasadena, CA 91125, USA}

\begin{abstract} 
We examine the clustering and kinematics of young stellar objects (YSOs) in the North America/Pelican Nebulae, as revealed by {\it Gaia} astrometry, in relation to the structure and motions of the molecular gas, as indicated in molecular line maps. The {\it Gaia} parallaxes and proper motions allow us to significantly refine previously published lists of YSOs, demonstrating that many of the objects previously thought to form a distributed population turn out to be non-members. The members are subdivided into at least 6 spatio-kinematic groups, each of which is associated with its own molecular cloud component or components. Three of the groups are expanding, with velocity gradients of 0.3--0.5~km~s$^{-1}$~pc$^{-1}$, up to maximum velocities of $\sim$8~km~s$^{-1}$ away from the groups' centers. The two known O-type stars associated with the region, 2MASS~J20555125+4352246 and HD~199579, are rapidly escaping one of these groups, following the same position--velocity relation as the low-mass stars. We calculate that a combination of gas expulsion and tidal forces from the clumpy distribution of molecular gas could impart the observed velocity gradients within the groups. However, on a global scale, the relative motions of the groups do not appear either divergent or convergent. The velocity dispersion of the whole system is consistent with the kinetic energy gained due to gravitational collapse of the complex. Most of the stellar population has ages similar to the free-fall timescales for the natal clouds. Thus, we suggest the nearly free-fall collapse of a turbulent molecular cloud as the most likely scenario for star formation in this complex.
\end{abstract}


\section{Introduction}\label{intro.sec}

The way that stars and gas are distributed in a star-forming region can provide useful constraints on the conditions in which the stars were formed \citep[e.g.,][]{2002ApJ...577..206E,2014MNRAS.438..620P,2018PASP..130g2001G,2018ASSL..424..119F}. 
For the nearest star-forming regions, astrometric measurements by ESA's {\it Gaia} spacecraft \citep{2016A&A...595A...1G} can provide a multi-dimensional picture of how the young stars are clustered. Spatial and kinematic clustering has already been examined in several of the major star-forming regions within 1~kpc, including Orion \citep{2018AJ....156...84K,2018A&A...619A.106G,2019MNRAS.487.2977G}, Taurus \citep{2018AJ....156..271L,2019arXiv190406980F,2019A&A...630A.137G}, $\rho$~Oph \citep{2019A&A...626A..80C}, and Serpens \citep{2019ApJ...878..111H}. We aim to do a similar analysis for North America and Pelican Nebulae (hereafter NAP). 
Given the correlation between stars and gas within many star-forming regions \citep[e.g.,][]{2011ApJ...739...84G,2009ApJ...697.1103T,2013ApJ...778..133L}, better understanding of the clustering of the stars can help decipher the complex velocity structures seen in radio molecular-line maps of star-forming clouds \citep[e.g.,][]{1981MNRAS.194..809L,2015ARA&A..53..583H}.

The NAP region contains a molecular cloud complex \citep{1980ApJ...239..121B}, an H\,{\sc ii} region \citep[W80;][]{1958BAN....14..215W}, and a population of YSOs, many of which have been extensively studied \citep{2008hsf1.book...36R}. This complex is fairly extended, with a diameter of $\sim$3$^\circ$ ($\approx$40~pc) encompassing several sites of active star formation. The North America (NGC 7000) and Pelican (IC 5070) nebulae make up the east/west components of the H\,{\sc ii} region, which is bisected in projection on the sky by a dark lane known as L933/935 \citep{1962ApJS....7....1L}, giving rise to the characteristic shape of the nebulae as seen in optical light. The YSOs are predominantly located in the dark lane \citep{2014AJ....148..120B}, with the southeastern portion known as the ``Gulf of Mexico'' 
and the northern part of the cloud containing the ``Atlantic Ocean'' and the ``Pelican'' regions.\footnote{The boundaries of the named subregions are given by \citet{2011ApJS..193...25R} and \citet{2014AJ....147...46Z}.} 
Many of the first-identified NAP YSOs were optically visible emission-line stars \citep[e.g.,][]{1958ApJ...128..259H}. More recently, {\it Spitzer} has uncovered thousands of embedded stars and protostars in the ``Gulf of Mexico,'' ``Pelican,'' and ``Pelican's Hat''  \citep{2009ApJ...697..787G,2011ApJS..193...25R}. \citet{2002AJ....123.2559C} identified several dense clusters of stars in the NAP region based on near-infrared star counts. Dozens of outflows from young stellar objects (YSOs) attest to ongoing star formation throughout the complex \citep{2003AJ....126..893B,2014AJ....148..120B}. 

Another intriguing aspect of this star-forming region is its primary ionizing source, 2MASS~J20555125 +4352246 \citep[dubbed the Bajamar Star;][]{2016ApJS..224....4M}. This source was discovered by \citet{2005AaA...430..541C} lying to the northwest of the ``Gulf of Mexico'' behind $\sim$9.6~mag of optical extinction. The star was recently classified as an O3.5((f*))+O8:\ spectroscopic binary  \citep{2016ApJS..224....4M}, which would make the primary the nearest known massive star with spectral type earlier than O4.\footnote{As a point of comparison, the nearest O4 star, $\zeta$~Pup, is less than half as far as the Bajamar Star \citep{2019MNRAS.484.5350H}.}
Another O star, HD~199579, is projected on the North America Nebula. This star, with spectral type O6.5 V ((f))z \citep{2011ApJS..193...24S}, has generally been regarded as being too far from the center of the H\,{\sc ii} region to be a main ionizing source \citep[e.g.,][]{1958ApJ...128..259H}. Whether HD~199579 is a member of the NAP association has been uncertain \citep{1983A&A...124..116W}.

The NAP complex is in the Orion-Cygnus Arm of the Galaxy, positioned ahead of our Sun in the direction of Galactic rotation at a distance of about 795$\pm$25~pc (Section~\ref{distance.sec}). In projection on the sky, NAP is located adjacent to the massive star-forming regions of Cygnus X, but it is generally thought that NAP is nearer than Cygnus X. A study of molecular clouds in the Solar neighborhood by \citet{2020arXiv200100591Z} and \citet{2020Natur.578..237A} has confirmed that this complex is part of a string of star-forming regions stretching from the Orion Molecular Clouds to Cygnus X. 

The NAP region provides a useful laboratory for studying both star formation in individual clouds as well as the evolution of the whole complex. 
Section~\ref{data.sec} presents the YSO catalogs, {\it Gaia} data, radio data, and spectroscopy used for this study. Section~\ref{ysoc_identified.sec} describes the methods used to identify stellar members and reject contaminants.  Section~\ref{groups.sec} describes the properties of multiple stellar groups. Section~\ref{distance.sec} computes distance estimates for the components of the association. Section~\ref{cloud.sec} describes the structure of the molecular clouds and their relation to the stars. Section~\ref{kinematics.sec} provides an analysis of the stellar kinematics. 
Section~\ref{gaia_identified.sec} identifies new candidate members identified from the {\it Gaia} catalogs.
Section~\ref{discussion.sec} discusses the formation and dynamical evolution of the system. And, Section~\ref{conclusion.sec} summarizes the main conclusions.

\section{Data}\label{data.sec}

\subsection{Initial Stellar Catalog}\label{stars.sec}

We searched for lists of candidate YSOs from published studies of the NAP region using the ``YSO Corral'' \citep{2015IAUGA..2253943H}, a curated database for YSOs.

The previous studies have used a variety of techniques to identify candidate members, as listed in Table~\ref{pub.tab}. These included selection based on H$\alpha$ emission, infrared excess, X-ray emission, placement on color-magnitude diagrams, and spatial clustering. H$\alpha$ emission and infrared excess  are mostly sensitive to disks or accretion, while X-ray emission can be used to detect pre--main-sequence stars both with and without disks. Each of these methods yields different types of contaminants and rates of contamination, so we aim to re-assess memberships of their candidate members using astrometry from {\it Gaia}, as described below. Contaminants to YSO searches in the Galactic plane include both field stars and extragalactic sources, in particular dusty (post-) asymptotic giant branch (AGB) stars and star-forming galaxies may be major sources of contaminants for selection based on infrared excess \citep{2008AJ....136.2413R}, while active galactic nuclei (AGN) and foreground active M dwarfs may dominate X-ray contaminants \citep[][]{2013ApJS..209...32B}. 

\subsection{Astrometric Data}\label{gaia_data.sec}

We obtained stellar astrometry from {\it Gaia}'s second data release \citep[DR2;][]{GaiaBrown}, which provides 5-parameter astrometric solutions for 1.3~billion objects to as faint as $G\approx21$~mag \citep{2018arXiv180409366L}. These solutions provide source positions in right ascension ($\alpha$) and declination ($\delta$), parallax ($\varpi$), and proper motion\footnote{We follow the convention of \citet{1997A&A...323L..49P}, and define $\mu_{\alpha^\star}\equiv\mu_\alpha\cos{\delta}$. This quantity is called $\mathtt{PMRA}$ in the {\it Gaia} DR2 tables.}
($\mu_{\alpha^\star}$, $\mu_\delta$). For DR2, most stars in the direction of NAP were observed during $\sim$15 visibility periods; 
typical formal uncertainties for a $G=15$~mag star are $\sim$0.03~mas in parallax and $\sim$0.05~mas~yr$^{-1}$ in proper motion. In addition to the formal uncertainty, $\sim$0.04~mas and $\sim$0.07~mas~yr$^{-1}$ correlated systematic uncertainties affect these measurments \citep{2018arXiv180409366L}. There have been a variety of efforts to estimate the systematic zero-point offsets \citep[e.g.,][and references therein]{2019MNRAS.489.2079L}; in most cases we try to work with the observed quantities (i.e.\ $\varpi$), but when distances are required we apply a 0.0523~mas parallax correction  estimated in the aforementioned paper. 

We use the {\it Gaia} catalog both for cross matching to the previously identified YSO candidates as well as to search for new member candidates. For cross matching, we use a match radius of 1.2$^{\prime\prime}$ and select the nearest source as a match, yielding 1939 matches out of $\sim$3,500 candidates. This match radius is large enough that changes in position from proper motion are unlikely to affect whether a counterpart is found. The relatively low match rate is because, due to extinction, the NAP population extends to faint optical magnitudes, far below the {\it Gaia} detection limits.
To estimate the rate of false matches, we artificially shifted the DR2 coordinates 2$^\prime$ north and ran a cross match using these coordinates. We found 52 matches to the shifted coordinates, suggesting an incorrect match rate of up to 3\%. This rate should be regarded as an upper limit on match errors, because a true counterpart, if it exists, would likely take precedence over a field star if both lie within the match radius. 

We apply cuts to the {\it Gaia} data to ensure that the measurements are sufficiently precise to be useful for distinguishing between members and non-members. We follow the recommendations\footnote{\url{https://www.cosmos.esa.int/web/gaia/dr2-known-issues}} of the {\it Gaia} Data Processing and Analysis Consortium and use only sources with renormalized unit weight error (RUWE) less than $1.4$. The {\it Gaia} catalog also tabulates excess noise in the astrometric fit. Higher values of excess noise \citep[e.g., $>$1~mas;][]{2019ApJ...870...32K} could indicate some acceleration of a source (e.g., due to a binary) that might affect the astrometric solution, so we only include sources with excess noise $\leq$1.0~mas. And, finally, we require $\mathtt{astrometric\_sigma5d\_max} \leq 0.5$~mas to limit the maximum statistical measurement uncertainties on parallaxes and proper motions. 
These quality cuts only slightly affect the peak of the parallax distribution (shifting it by the equivalent of 4~pc), but the disadvantage of including all sources would be that the modes of the distribution become broader and source classifications become less certain. We also removed objects with $\varpi>4$~mas because they are clearly in the foreground and interfere with the mixture model analysis in Section~\ref{ysoc_identified.sec}.

\subsection{Millimeter Observations}\label{millimeter.sec}

We obtained $^{13}$CO $J=1$--0 (110.201354 GHz) observations of the molecular clouds associated with the NAP region using the Five College Radio Astronomy Observatory (FCRAO) 14-m telescope in 1998 June. The CO map covers most of the region of interest in this study.

The observations were obtained with the Second Quabbin Optical Imaging Array \citep[SEQUOIA;][]{Erickson99}. The array contains 16 pixel elements arranged in a $4\times4$ grid with separation of 88$^{\prime\prime}$ on the sky. At the time of the observations, 12 of the pixels were functional. The spectrometer contained 512 channels with a bandwidth of 40~MHz, which provided a channel spacing of 0.21 km~s$^{-1}$. The FCRAO telescope has a full width at half maximum (FWHM) angular resolution of 47$^{\prime\prime}$ at the $^{13}$CO $J=1$--0 frequency. The forward scattering and spillover efficiency of the telescope at the observed frequency was $\eta_\mathrm{FSS}=0.7$, while the main-beam efficiency was $\eta_\mathrm{mb}=0.45$ \citep{1998ApJ...502..265H}. The data were obtained in position-switching mode with spectra that were sampled every 44$^{\prime\prime}$ on the sky, or approximately the FWHM resolution. The typical RMS noise in the spectra is $\Delta {T_A}^*=0.22$~K per channel. The maps used for our analysis are in units of ${T_R}^*$, calculated from the atmosphere-corrected antenna temperature ${T_A}^*$ by dividing by a correction factor \citep{1981ApJ...250..341K}. Here, we use the correction factor $\eta_\mathrm{FSS}$ since the $^{13}$CO emission is usually more extended than an arcminute in the NAP region. 

Our $^{13}$CO data are similar in sensitivity and spatial resolution to a $^{13}$CO map published by \citet{2014AJ....147...46Z} using the Purple Mountain Observatory Delingha 13.7~m telescope. They also provide $^{12}$CO and C$^{18}$O maps. 

\subsection{Spectroscopic Observation of the Ionizing Source}\label{spec.sec}

Finally, high dispersion optical spectra of the Bajamar Star were taken on 1 December 2019 and 3 January 2020 (UT) 
for the purposes of confirming the existing spectral typing that was based on low resolution spectra, and assessing the source radial velocity.
These data were taken with the Keck I telescope and HIRES \citep{1994SPIE.2198..362V} and cover $\sim4800-9200$ \AA\ at $R\approx 37,000$.

\section{Validation of Membership with Astrometry}\label{ysoc_identified.sec}

\begin{figure*}[t!]
\centering
\includegraphics[width=1\textwidth]{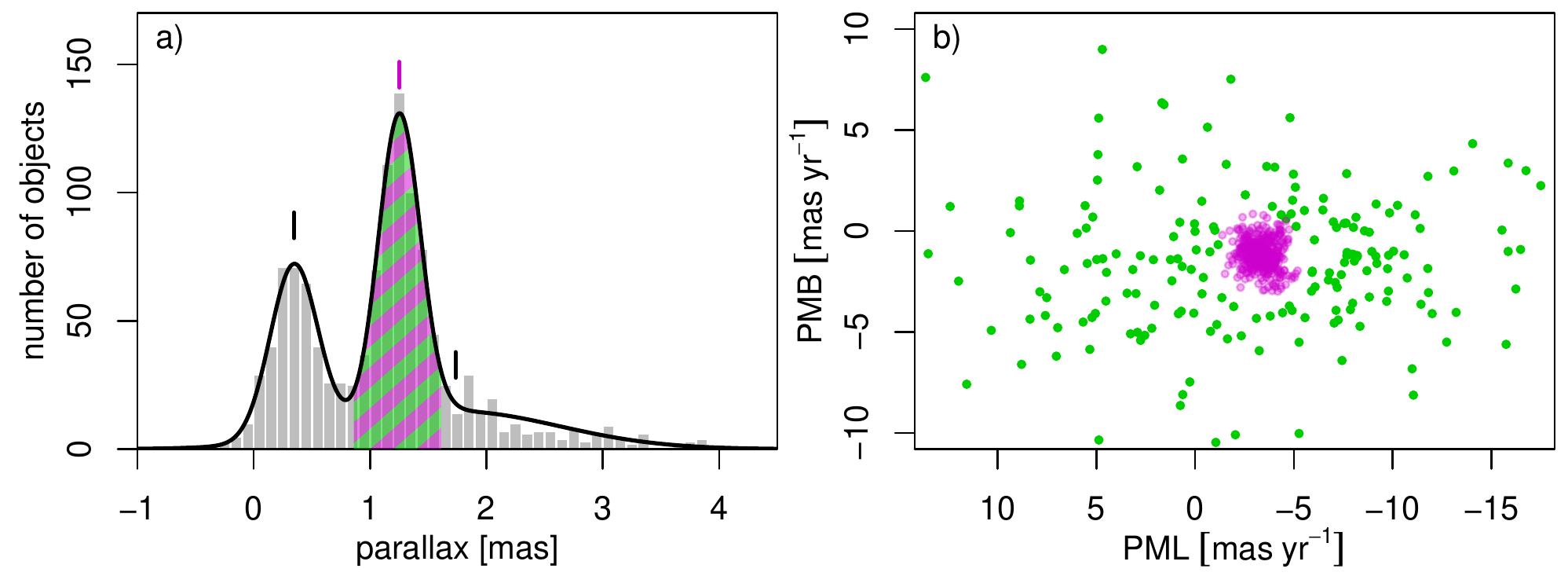} 
\caption{The panels illustrate the two step process for using {\it Gaia} astrometry to validate YSO candidate membership. a) The distribution of parallaxes (histogram) is fit with a Gaussian mixture model (black line). The centers of the three Gaussians are indicated by the vertical marks above the curve. Objects with a probability $>$50\% of being a member of the middle component (associated with NAP) are selected as indicated by the striped green/magenta region. b) Proper motions $\mu_{\ell^\star}$ (PML) and $\mu_b$ (PMB) for the objects selected in step (a). The distribution of these sources is fit with a two-dimensional Gaussian mixture model, and the objects with a probability $>$50\% of being a member of the middle density peak are colored magenta. These comprise our sample of 395 {\it Gaia}-validated NAP members. 
 \label{selection.fig}}
\end{figure*}

\begin{figure*}[b!]
\centering
\includegraphics[width=0.9\textwidth]{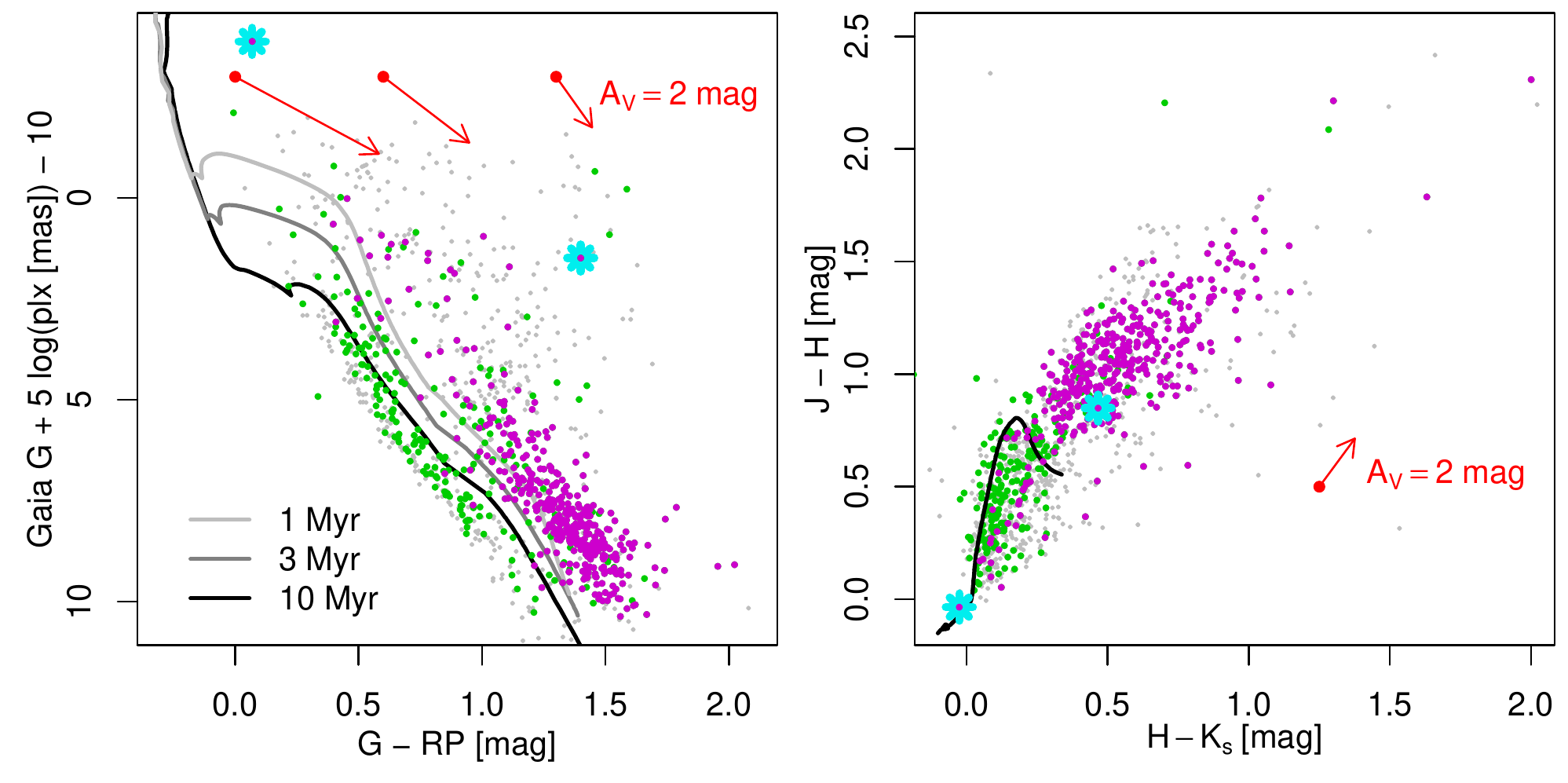} 
\caption{Left: {\it Gaia} color-magnitude diagram for YSO candidates classified in this work. Members are magenta points, sources excluded based on parallax are gray points, and sources excluded based on proper motion are green points. Only sources with $\varpi/\sigma_\varpi>3.0$ are included on the diagram. The curves indicate unreddened PARSEC isochrones for 1, 3, and 10~Myr. The red arrows show approximate reddening vectors for $A_V=2$~mag for pre--main-sequence stars with unreddened colors of $G-RP=0$, 0.6, and 1.3~mag. The two O stars stars are marked with the cyan asterisks; the Bajamar Star is the one with the very red color. Right: 2MASS color-color diagram using the same symbols as the left panel. The black curve indicates the locus of unreddened stellar colors from the PARSEC models, whose shape is similar for young pre--main-sequence stars but extends $\sim$0.15 mag redder in $J-H$ around and beyond the peak \citep{2015ApJ...808...23H}.
 \label{ysoc_cmd.fig}}
\end{figure*}

\begin{figure*}[t]
\centering
\includegraphics[width=1\textwidth]{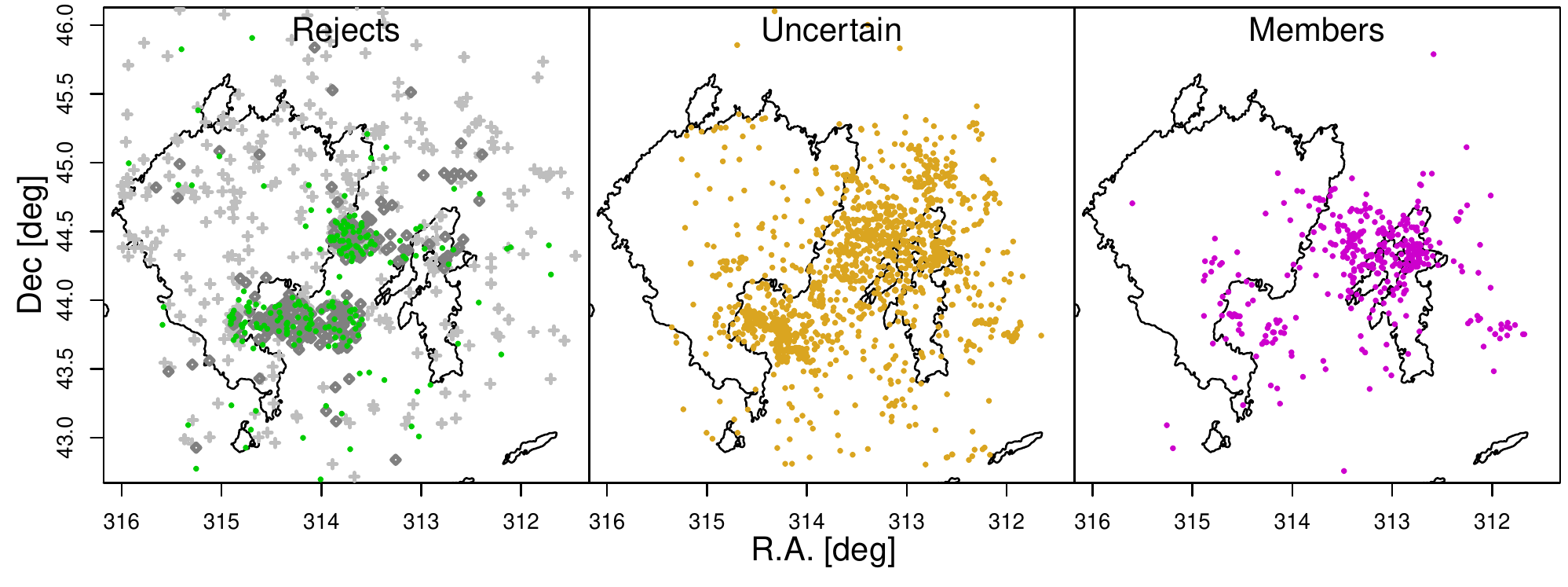} 
\caption{Spatial distributions of YSO candidates stratified by their classification as contaminants or members. The left panel shows the distributions of different classes of contaminant: objects rejected by parallax in gray (background to the NAP as light crosses and foreground as dark diamonds) and objects rejected by proper motion in green. The middle panel shows objects where {\it Gaia}-based classification is uncertain because the source does not exist in the {\it Gaia} catalog or it does not meet our quality criteria. The right panel shows objects classified as {\it bona fide} members. In black we show contours of the optical nebulae from the DSS image. Much of the spatial clustering in the middle panel is reminiscent of structure in the right panel, indicating that our techniques are selecting only a portion of the true young star population, but there is also a broad spatial distribution reminiscent of the left panel, indicating a mix of contaminants in the uncertain population too.
 \label{spatial_contaminants.fig}}
\end{figure*}

We can evaluate membership of candidate YSOs by observing whether they are at the same distance and moving with the same kinematics as the rest of the members in the NAP region. In Figures~\ref{selection.fig}--\ref{spatial_contaminants.fig}, we use the following color scheme: gray -- stars at the wrong distance; green -- stars at a similar distance but with the wrong proper motion; magenta -- probable astrometric members; goldenrod -- stars with uncertain classification.


The distribution of parallaxes of YSO candidates (Figure~\ref{selection.fig}, left panel) has at least two modes as well as a tail of objects with high parallax. The expected distance to NAP based on previous measurements is 500--1000~pc \citep[e.g.,][]{2008hsf1.book...36R}, which coincides with the second peak on the histogram (magenta tic mark). The left-most peak can be attributed to contaminants with small or zero parallax, including giant stars and extragalactic sources. 
The tail of objects to the right includes many bright objects with high-precision parallax measurements, so these are clearly foreground stars. 

To quantitatively divide stars based on parallax, we model the parallax distribution using a Gaussian mixture model, for which we use the Bayesian Information Criterion \citep[BIC;][]{schwarz1978estimating} to determine the number of components. For Gaussian mixture models the BIC can be viewed as an approximation for the Bayes factor which is useful for the statistical problem of model selection \citep{everitt2011cluster}. We find that the distribution can be well modeled with 3 Gaussians, as marked in Figure~\ref{selection.fig}, with $\Delta BIC = 10$ compared to the next best model, implying strong preference for this solution. From low parallax to high parallax, the first component models the peak that we associated with background contaminants, the second component corresponds to the group of stars associated with NAP and has a mean of $\varpi_0=1.25$~mas, and the third component has a slightly higher mean parallax but a much broader distribution that appears to approximate the shape of the high-parallax tail. 
Objects that lie between $0.86 < \varpi < 1.61$~mas (green and magenta striped region) have a $\geq$50\% probability of belonging to the second component; all other sources are classified as non-members. 

In the parallax selected sample, cross-contamination from field stars that happen to have similar parallaxes to the NAP region is inevitable. To reduce this contamination, we perform a second classification step, this time using proper motion.  Figure~\ref{selection.fig} (right panel) reveals a clump of sources with similar motions, and a halo of objects with significantly different motions.  We show these motions in Galactic coordinates ($\mu_{\ell^\star}$, $\mu_b$) to emphasize that the proper-motion dispersion in the halo is largely parallel to the Galactic plane, and thus represents orbital motions of field stars in the Galaxy which are dynamically hotter than newly formed stars and thus have larger dispersions in dynamical phase space. We subdivide the sources in proper-motion space using another Gaussian mixture model, finding two components ($\Delta$BIC=7), corresponding to the clump (smaller dispersion) and the halo (larger dispersion). We classify objects with $\geq$50\% probability of belonging to the first component as members (magenta points), while the others are non members (green points). The members have mean proper motions of $\mu_{\ell^\star} = -3.35$~mas~yr$^{-1}$ and $\mu_b=-1.15$~mas~yr$^{-1}$; other parameters of the classifier are given in Appendix~\ref{classifier.appendix}. The mixture model suggests a 3\% residual contamination rate among the stars classified as members. We note that selection based on proper motion comes at the cost of omitting stars with high proper motions that could have been ejected from the star-forming region.

Figure~\ref{ysoc_cmd.fig} (left) shows the classified sources on a diagram of {\it Gaia} absolute magnitude versus color. When compared to theoretical PARSEC isochrones \citep{2012MNRAS.427..127B}, the members nearly all lie above the 3~Myr isochrone, and many of them lie above the 1~Myr isochrone. In contrast, the objects identified as non-members are scattered throughout the diagram, with many lying below the 10~Myr isochrone, while others are located in the region of post--main-sequence giant stars, and others lie in similar locations to the pre--main-sequence stars. This diagram confirms that nearly all of the astrometrically validated members are pre--main-sequence stars with ages $<$3~Myr, but that some rejected objects that could also be young. The effect of reddening (red arrows) varies with color due to the large width of {\it Gaia}'s $G$ and $G_{RP}$ bands, so we show three approximate vectors for different $G-RP$ colors assuming a typical pre--main-sequence stars spectrum. From the $J-H$ vs.\ $H-K_s$ color-color diagram (Figure~\ref{ysoc_cmd.fig}, right) using photometry from the Two Micron All Sky Survey \citep[2MASS;][]{2006AJ....131.1163S}, we estimate that typical reddening is $E(J-H)\sim0.1$--0.6, corresponding to $A_V\sim1$--6~mag. Both O stars are labeled in the diagrams, but HD~199579 is one of the brightest, bluest sources, while the Bajamar Star is significantly dimmer and redder due to its high extinction.

\subsection{Members and Contaminants}

Out of the sample of YSO candidates, 395 objects are confirmed as members, while 814 objects are reclassified as non-members, and 2264 sources either do not have {\it Gaia} counterpart or do not meet our quality criteria and therefore have uncertain classifications. 
The numbers of candidates in each category are tabulated in Table~\ref{pub.tab} for each of the published lists of YSO candidates. Table~\ref{members.tab} provides a list of the astrometric members. 

It is not surprising that nearly all historical studies have some contamination. For instance, out of the 68 objects from the pioneering study of the NAP region by \citet{1958ApJ...128..259H}, {\it Gaia} provides sufficiently high quality measurements for 27 of them. Of these stars, our analysis has validated the membership of 21 objects, while rejecting membership for 6 others. The rejected members include LkH$\alpha$~132, 133, 147, 183, 192, and 193, which all have parallax values that are too small to be a members of the region, suggesting that they are background Be stars. 

\begin{figure*}
\centering
\includegraphics[width=1.\textwidth]{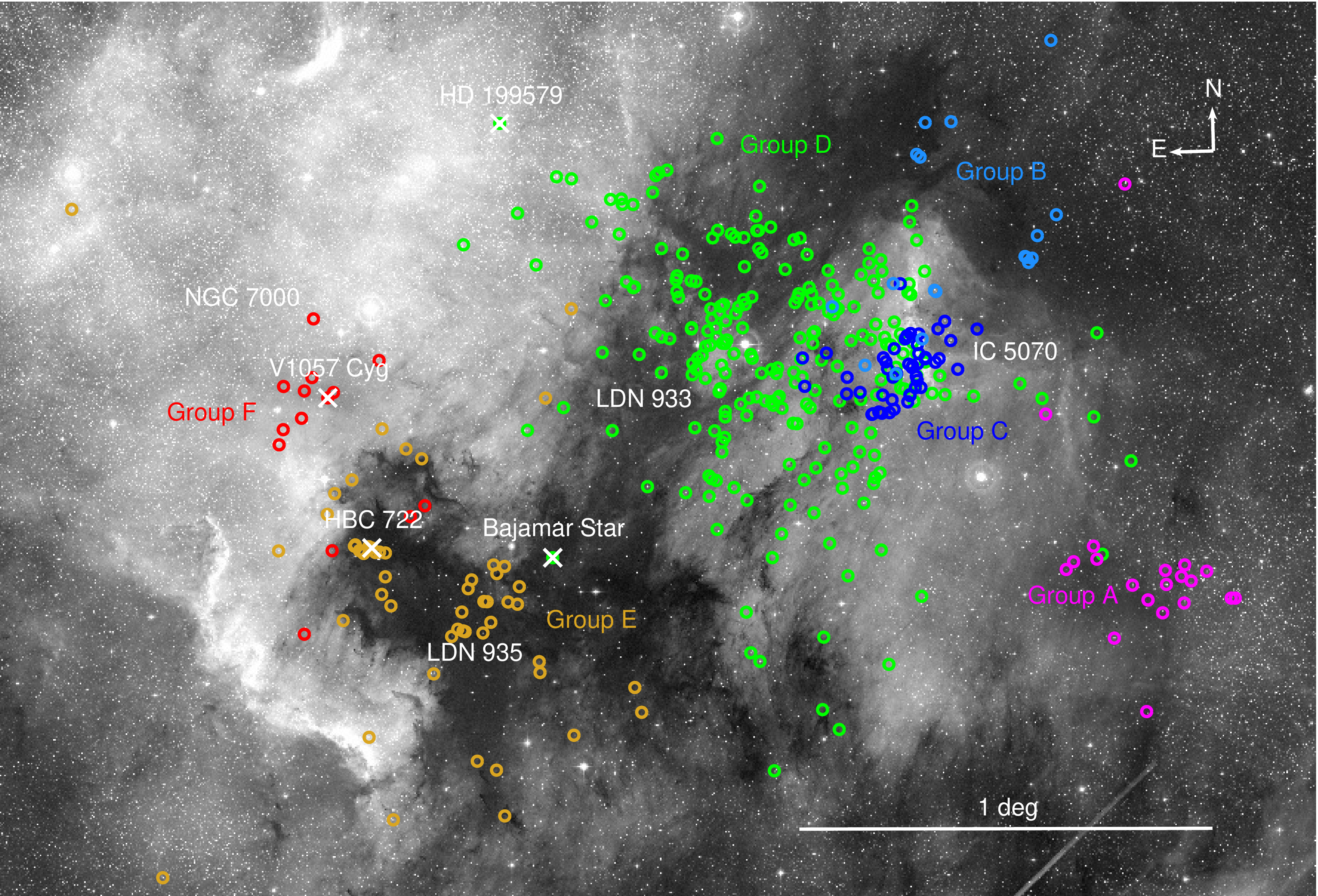} 
\caption{Members, as indicated in the right panel of Figure~\ref{spatial_contaminants.fig}, are over-plotted on a DSS red-band image of the region and color coded based on the kinematic group they belong to. Note that several outlying members are cropped out of the image (see Figure~\ref{spatial_contaminants.fig}) to show enough detail in the central region. Several interesting stars are indicated, including the two O stars and two well-studied outbursting stars. The nebulous regions are those that were outlined in Figure~\ref{spatial_contaminants.fig} -- on the left the nebulosity resembles the shape of North America and on the right it resembles a pelican with its beak facing the Atlantic Coast of North America. The naming of subregions (e.g., Gulf of Mexico, Atlantic, Pelican's Neck) has historically followed these analogies.
 \label{dss.fig}}
\end{figure*}

For some categories of YSO candidate, {\it Gaia}'s requirement that a source must be sufficiently bright in the optical may 
induce a selection bias toward higher contamination rates. For example, out of the 1,286 {\it Spitzer}/MIPS sources classified as YSOs by \citet{2011ApJS..193...25R} on the basis of infrared excess, only 30\% met our {\it Gaia} criteria for inclusion in the analysis. Of those 30\%, 131 were rejected and 251 were validated. However, this ratio is not representative of the whole sample of 1,286 MIPS candidates because the YSOs with the most prominent infrared excesses tend to be very young and deeply embedded, hence not optically visible.    

The spatial distribution of candidates classified members, non-members, or uncertain membership is shown in Figure~\ref{spatial_contaminants.fig}. These points are over-plotted on an outline of the optical nebulosity making up the North America and Pelican components of the nebula.  Non-members are widely dispersed throughout the whole region, while uncertain members are more tightly clustered, and validated members are the most tightly clustered. This suggests that the spatial distribution of stars is dominated by clustered groups, rather than distributed stars. Although a few members are located a degree or more away from the main groups, there is no significant non-clustered population scattered throughout the entire $\sim$3$^\circ$ diameter region.

Our sample of 395 objects is clearly incomplete, missing both objects that were not selected in previous studies (e.g., pre--main-sequence stars without disks) as well as objects that could not be evaluated by {\it Gaia}. There is also evidence that our classifier has rejected a few legitimate YSOs. From a sample of 41 high-confidence YSOs studied by \citet{2013ApJ...768...93F}, 4 were rejected by the classifier, likely spuriously. Altogether, it is likely that the NAP region contains at least several thousand stellar members. This assertion is based on {\it Spitzer} observations that have collectively identified $\sim$2000 YSO candidates \citep{2009ApJ...697..787G,2011ApJS..193...25R}, most of which are not included in our study because they are inaccessible to {\it Gaia}. Furthermore, in Section~\ref{gaia_identified.sec} we identify $>$1000 new candidate members based on {\it Gaia} astrometric data. Strictly speaking, our membership classification is for the {\it Gaia} source, so the chance alignment of a mid-infrared YSO with a {\it Gaia} field star would likely result in rejection of the source.

\subsection{Properties of the Ionizing Sources}\label{sec:bajamar}

Both the Bajamar Star (the primary ionizing source) and HD~199579 pass our {\it Gaia} membership criteria in Section~\ref{ysoc_identified.sec}. However, the parallaxes and proper motions of both of these O stars place them near the edges of parameter space for objects identified as members. For HD~199579 ($\varpi = 1.06\pm0.06$~mas,  $\mu_{\ell^\star}=-1.4\pm0.1$~mas~yr$^{-1}$ and  $\mu_{b}=-1.6\pm0.1$~mas~yr$^{-1}$), its parallax places it behind most other cluster members. It lies on the northeast extreme of the NAP region and is moving in this direction relative to the rest of the NAP members. 

The Bajamar Star ($\varpi = 1.47 \pm 0.08$~mas,  $\mu_{\ell^\star}=-3.4 \pm 0.1$~mas~yr$^{-1}$ and  $\mu_{b}=-2.9 \pm 0.2$~mas~yr$^{-1}$) has a parallax measurement that is 2$\sigma$ greater than the median parallax of the system, and the $\mu_{\ell^\star}$ proper motion is more negative than that of most other members. The {\it Gaia} data do not indicate any obvious problems with the astrometry; the RUWE is 1.04 which is within the recommended $<$1.4 range. The source has a small, but highly statistically significant, excess noise of 0.346~mas, which could be explained if it is a binary system. In Section~\ref{distance.sec} we argue that the star is unlikely to be in front of the complex, so the discrepant parallax measurement is likely to be the result of statistical measurement uncertainty. 

From our HIRES spectrum of the Bajamar Star, we find an O4--O6 spectral type. This estimate is based on the  presence of \ion{He}{2} 5412 \AA\ with $W_\lambda=1.1$ \AA, 
and the barely discernible \ion{He}{1} 4922 \AA\ ($W_\lambda<0.18$ \AA), \ion{He}{1} 5015 \AA\ ($W_\lambda<0.13$ \AA), and \ion{He}{1} 5047 \AA\ ($W_\lambda<0.08$ \AA). 

We can also confirm the statement in \citet{2016ApJS..224....4M} that the different absorption line species have different radial velocities.   However, it is not clear based on our spectra that the source is a spectroscopic binary. While there is evidence for asymmetries in the line profiles, especially \ion{H}{1} and \ion{He}{1}, a high velocity wind, as is expected from such a massive star, would produce similar features.  Furthermore, a cross correlation analysis of our HIRES spectrum does not reveal indications of two peaks, through the lines are extremely broad due to the rapid rotation and possibly other effects, and it certainly would be possible to mask two sets of broad lines within the profiles.  

Using $\tau$~Sco as a radial velocity standard, we derive $v_{helio} = 11.6 \pm 16.5$~km~s$^{-1}$ for our first observation and $v_{helio} = -17.1  \pm 16.8$~km~s$^{-1}$, both based on the single \ion{He}{2} 5412 \AA\ line. Although the errors are large, this is our best estimate of the stellar radial velocity. The velocities derived from the only other measurable and non-interstellar lines in the first spectrum are: $-100$~km~s$^{-1}$ from a single \ion{He}{1} line, and $-90$ and $-35$ from \ion{H}{1} (H$\alpha$ and H$\beta$ respectively), very different from the \ion{He}{2} velocity.
The second spectrum does show a radial velocity shift.  Measured directly, rather than going through a narrow-line star, the shift 
in the strong \ion{He}{2} 5412 \AA\ line is by $28\pm2$~km~s$^{-1}$.  This certainly suggests binarity.  
However, an order containing   \ion{He}{1} 5876 \AA\ and \ion{C}{4} 5801, 5811 \AA\  shows a velocity difference of around $-40$~km~s$^{-1}$
while  \ion{He}{2} 6406, 6527 suggest an 8--10~km~s$^{-1}$ difference and while both
\ion{He}{1} 6678 \AA\ and  \ion{He}{2} 6683 \AA\ are closer to +40~km~s$^{-1}$.  
In H$\alpha$ and H$\beta$ the shift between the spectra is about $15\pm3$~km~s$^{-1}$. It remains unclear whether we should interpret the measurements as a single star having the same velocity as the molecular gas and strong wind
that manifests in confusing absorption velocities, or as a binary with the more massive star producing much of the \ion{He}{2} absorption 
and both stars contributing to the relatively blue-shifted metal, \ion{He}{1} and \ion{H}{1} profiles.

 \section{Kinematic Groups}\label{groups.sec}

\begin{figure*}[t]
\centering
\includegraphics[width=1\textwidth]{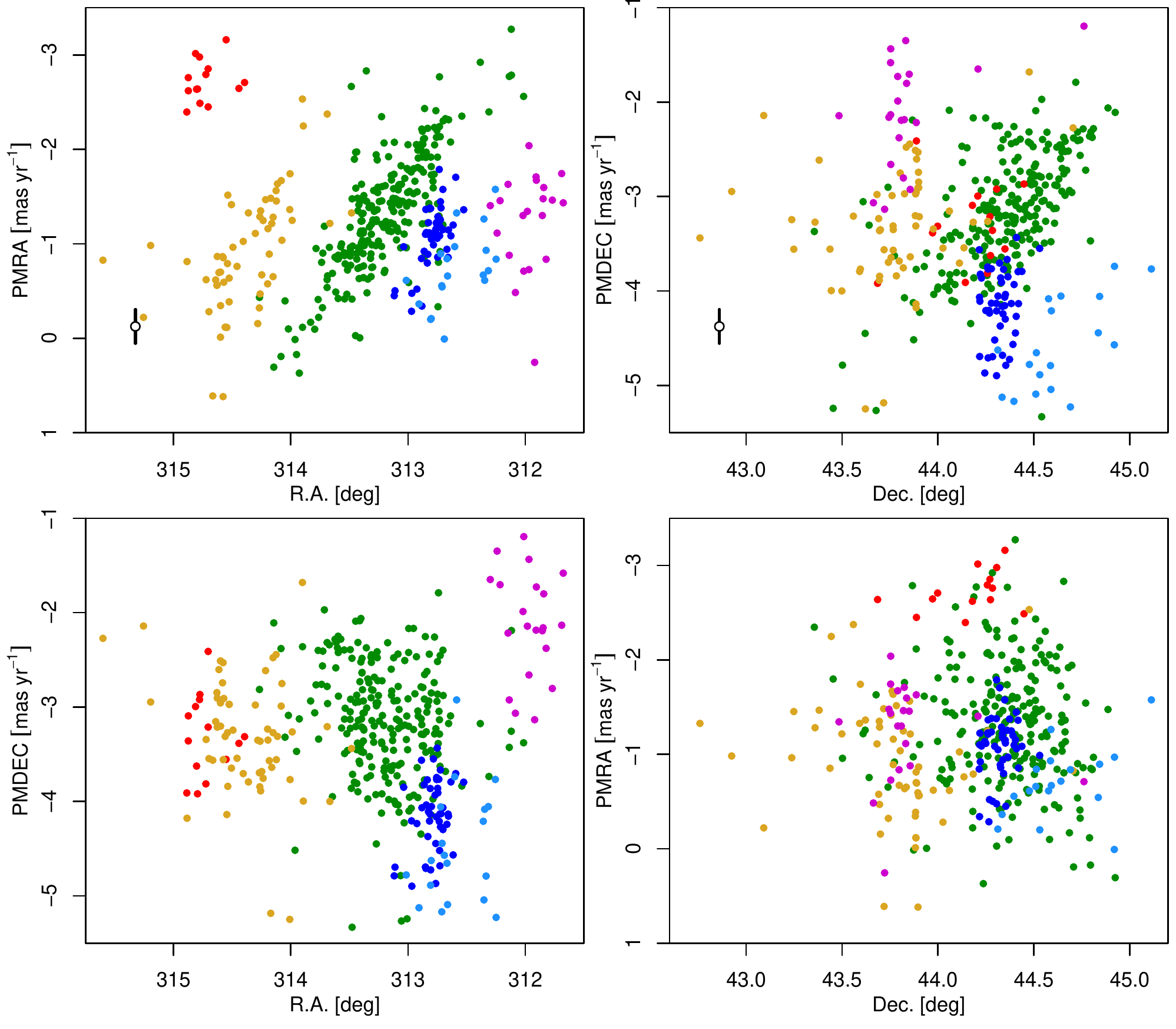} 
\caption{Plots of position coordinates $\alpha$ and $\delta$ vs.\ proper motion coordinates $\mu_{\alpha^\star}$ (PMRA) and $\mu_\delta$ (PMDEC). NAP members are color coded by group using the same hues as Figure~\ref{dss.fig}. The black points in the top panels with error bars indicate the median formal uncertainties in $\mu_{\alpha^\star}$ and $\mu_\delta$. 
 \label{pos_vs_pm.fig}}
\end{figure*}

\begin{figure}
\centering
\includegraphics[width=0.45\textwidth]{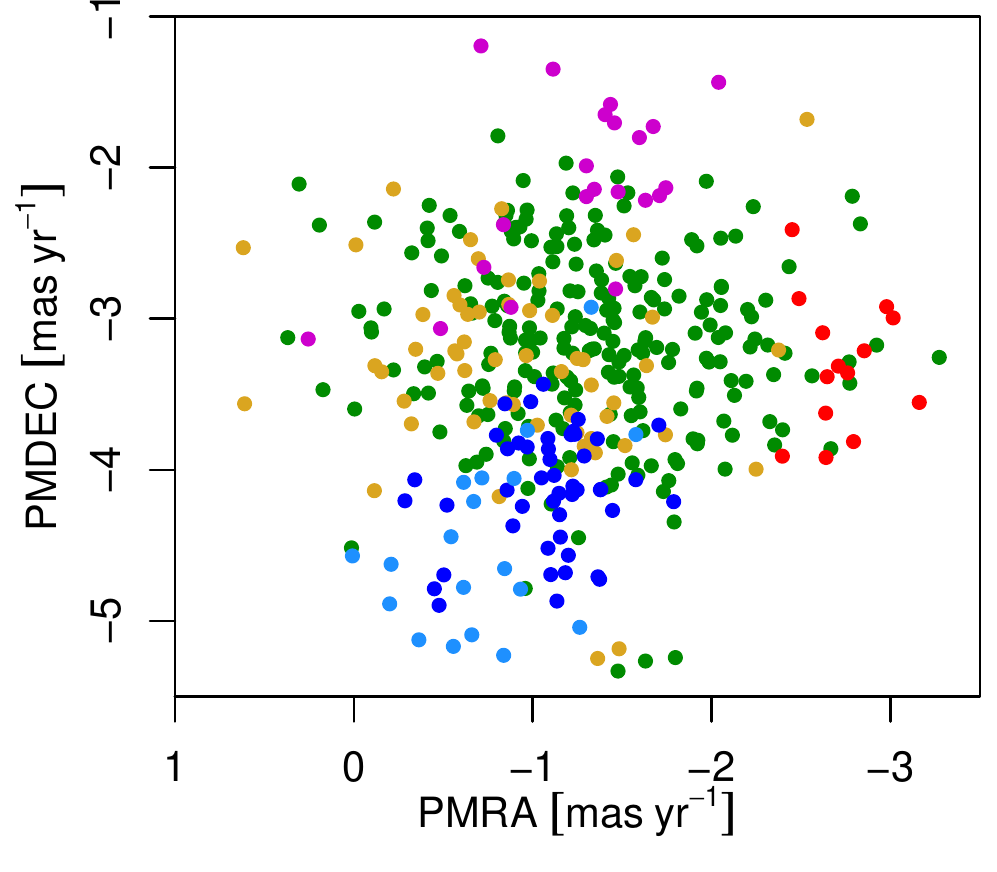} 
\caption{Scatter plot showing $\mu_{\alpha^\star}$ vs.\ $\mu_\delta$. NAP members are color coded by group as in Figures~\ref{dss.fig} and \ref{pos_vs_pm.fig}.
 \label{pm_vs_pm.fig}}
\end{figure}

The distribution of {\it Gaia}-validated NAP members is neither smooth nor centrally concentrated, but has a more complicated, clumpy structure that is apparent both in the stars' spatial positions (Figure~\ref{dss.fig}) and in their distributions in proper-motion space (Figures~\ref{pos_vs_pm.fig} and \ref{pm_vs_pm.fig}). Spatial clustering and sub-clustering of young stars has been appreciated for decades \citep[e.g.,][]{2000AJ....120.3139C,2007prpl.conf..361A}, but only recently have the kinematic data achieved comparable fidelity \citep[e.g.,][]{2017MNRAS.465.1889G,2017ApJ...845..105D}. This analysis is now possible  in a larger number of star-forming regions with {\it Gaia} DR2+ astrometry.

\subsection{Cluster Analysis Algorithm}

When analyzing the stellar population of the NAP region, it is convenient to divide the stars into subclusters. As for all cluster analysis problems, multiple possible strategies can achieve this, with no one approach being ``best'' \citep{everitt2011cluster}. We find that a Gaussian mixture model with hierarchical combination of clusters, outlined below, gives reasonable results that are astronomically useful.

We use a Gaussian mixture model analysis implemented by {\tt mclust} \citep{mclust5} in  R to identify groups of stars in 4-dimensional position--proper motion space. The analysis is performed on the variables $\ell$, $b$, $\mu_{\ell^\star}$, and $\mu_b$, each of which is normalized by subtracting the mean value and dividing by the standard deviation. The program then selects the number of clusters, the free parameters of the Gaussian components, and the values of the parameters via minimization of the BIC. For modeling the clustering of stars, the mixture-model approach has several benefits, which include the ability to identify clusters that differ in both compactness and number of members, the ability to assign stars from the dense cluster core as well as the less-dense cluster wings to the same cluster, and the ability to deal with overlapping clusters. On the other hand, stars clusters are typically not well fit with Gaussian distributions, which can lead to multiple Gaussian components used to fit a single cluster \citep{hmm}. For the NAP stars, {\tt mclust} selects a model with eight ellipsoidal components with equal shape and orientation, where several of the Gaussian components are significantly overlapping or nearly concentric.

The {\tt mclust} package includes the function {\it clustCombi} to deal with such cases by hierarchically merging multiple Gaussian components into single clusters using an entropy criterion \citep{baudry2010combining}. For our case, we find a large change in normalized entropy going from 6 to 7 clusters, so we use this threshold to cut the hierarchical tree at 6 groups. The stars assigned to each of these six groups, labeled A through F, are over-plotted on the optical DSS image of the NAP region in Figure~\ref{dss.fig}.

\subsection{Groups in Position--Proper-Motion Space}

The results of the cluster analysis algorithm can be qualitatively checked for reasonableness by examination of the Group assignments in Figures~\ref{dss.fig}--\ref{pm_vs_pm.fig}. In any one of the projections, the edges of some groups overlap. However, the stars that overlap in one projection tend to be different from the stars that overlap in a different projection, and the groups are mostly well separated. 

In projection on the sky (Figure~\ref{dss.fig}), Group A is fairly isolated in the southwest corner of the region. Groups B, C, and D form a conglomeration in the northwestern portion of the nebula, more-or-less corresponding to the regions known as the ``Pelican's Hat,'' the ``Pelican's Neck,'' and the ``Atlantic,'' respectively.  Groups E and F lie in the southeastern/eastern region -- Group E is composed of several clumps of stars on the dark  ``Gulf of Mexico'' cloud, while Group F is a single clump to the northeast of the ``Gulf of Mexico'' superimposed on an area with bright nebulosity.

The plot of $\alpha$ vs.\ $\mu_{\alpha^\star}$ (Figure~\ref{pos_vs_pm.fig}, upper left) shows the clearest separation of the stars into distinct groups. In the center of this diagram, Group D -- the largest group -- forms a diagonal swath from lower left to upper right. This group is clearly separate from Group E to its left, which follows a parallel diagonal track. On the right, Group C is located adjacent to Group D, but is much more tightly clumped in $\alpha$ and $\mu_{\alpha^\star}$.  On the right side of the diagram, Group B appears partially entangled with Group C, and Group A appears partially entangled with Group B.  However, both Groups A and B are looser than Group C. The spatial separation of A from the rest of the groups make it clear that A is distinct. However, whether B could be considered part of the same group as C is less clear. Group B contains only a few stars from our sample; however, these stars coincide with the deeply embedded ``Pelican's Hat'' cluster of stars discovered by {\it Spitzer} \citep{2009ApJ...697..787G,2011ApJS..193...25R}, most of which are unseen by {\it Gaia}.

In the plot of $\delta$ vs.\ $\mu_{\delta}$  (Figure~\ref{pos_vs_pm.fig}, upper right), the diagonal swath formed by Group~D can still be seen, but the other groups appear distributed in a more vertical direction rather than diagonal. This plot enhances the separation between Groups A, B, and C. Groups C and D are less cleanly separated, but the stars in Group C do not continue the same diagonal tend as Group D. The lower panels in Figure~\ref{pos_vs_pm.fig} show the other combinations of position and proper motion. Finally, on the plot of $\mu_{\alpha^\star}$ vs.\ $\mu_{\delta}$ (Figure~\ref{pm_vs_pm.fig}), Groups D and E are in the center, Group A is at the top extreme, Groups B and C are at the bottom extreme, and Group E is at the right-most extreme.

\begin{figure*}
\centering
\includegraphics[width=1\textwidth]{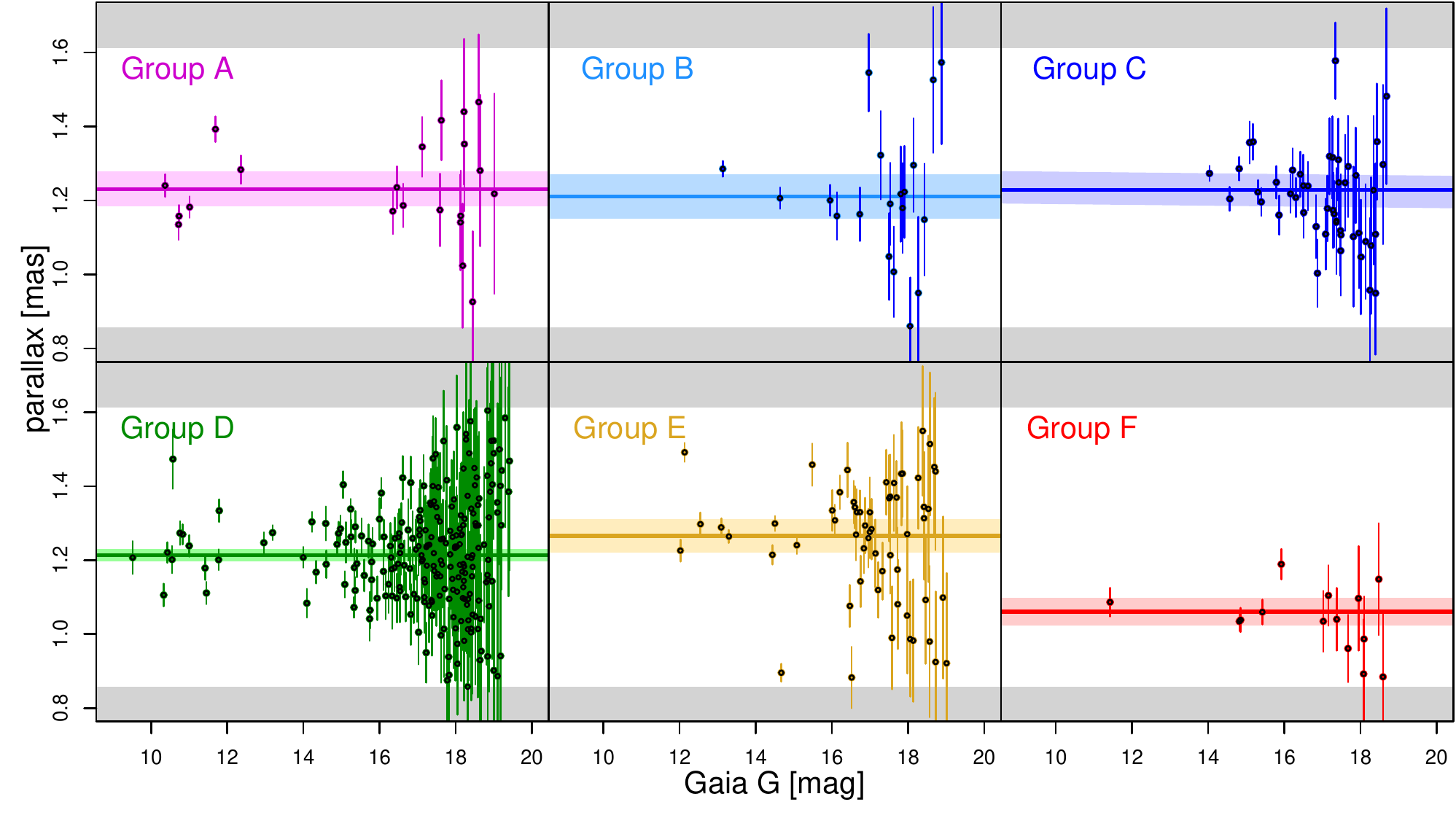} 
\caption{The panels show the mean parallax (colored horizontal line) and scatter plot of each star's measured parallax vs.\ magnitude (points) for stars in each group. Uncertainty on the mean parallax is indicated by the colored band around the line, which shows the 2$\sigma$ confidence interval. The gray regions at the top and bottom of each graph indicate the range of parallaxes that were excluded by the selection cut made when identifying members.
 \label{mean_parallax.fig}}
\end{figure*}

Intriguingly, the Bajamar Star is assigned to Group D even though it is spatially closer to stars from Group E in the ``Gulf of Mexico.'' This is a result of the proper motion of the Bajamar Star, which is rapidly moving toward Group E away from the center of Group D. This proper motion does not match the proper motions of nearby stars in Group E; however, it does fit with the proper-motion gradient seen for stars in Group D (Figure~\ref{pos_vs_pm.fig}).

Our division of stars into kinematic groups is meant to be useful for understanding the structure of the NAP region, but it is not the full description of the cluster structure. For example, nearby groups may have formed from related star-formation events at different locations on the same cloud, and thus, from an astrophysical perspective, it is ambiguous whether these should be considered to be the same group. Furthermore, if the region were nearer \citep[e.g., at the distance of Taurus;][]{2018AJ....156..271L,2019arXiv190406980F} we might be able to detect finer velocity differences with {\it Gaia} that could be used to subdivide the groups further, but if the region were more distant \citep[e.g., at the distance of the Carina Nebula;][]{2019ApJ...870...32K} the measurement uncertainties in the proper motion would be poorer and, thus, it might not have been possible to distinguish the groups.

\section{Three Dimensional Structure of the Stellar Population}\label{distance.sec}

The region is sufficiently close that differences in distance to the groups begin to become detectible by {\it Gaia}. However, these differences are subtle enough that care must be taken in estimating mean parallaxes of each group. We calculate the mean parallaxes using a maximum likelihood method that takes into account both the heteroscedastic measurement errors as well as the truncation of the sample at  $\varpi_\mathrm{min}=0.86$~mas and $\varpi_\mathrm{max}=1.61$~mas (Section~\ref{ysoc_identified.sec}). The log-likelihood as a function of mean parallax, $\bar{\varpi}$, is given by
\begin{equation}
  \mathcal{L}(\bar{\varpi},\sigma^2) = 
  \sum_i \ln \frac {\phi(\varpi_i; \bar{\varpi}, \sigma_{i}^2 + \sigma^2)}
  { \int_{\varpi_\mathrm{min}}^{\varpi_\mathrm{max}}\phi(\varpi_i; \bar{\varpi}, \sigma_{i}^2 + \sigma^2)\mathrm{d}\varpi  }
\end{equation}
where $\phi$ denotes a Gaussian distribution and $\varpi_i$ are the measured parallaxes with uncertainties $\sigma_i$. This model allows stellar groups to have an intrinsic depth, characterized by the parameter $\sigma$. We use the Broyden-Fletcher-Goldfarb-Shanno (BFGS) algorithm to find the value of $\bar{\varpi}$ that maximizes this likelihood, and use the Hessian matrix calculated by the R function {\tt optim} to estimate confidence intervals.

The mean parallaxes and their confidence intervals are reported in Table~\ref{astro.tab} and plotted in Figure~\ref{mean_parallax.fig}. All groups in the northern and western part of the NAP (Groups A--D) have statistically indistinguishable parallax measurements. Of these, the value for Group D is the most precise, with $\varpi_0\approx1.21$~mas (corresponding to $\sim$795~pc). The parallax of Group F differs the most significantly, with a value of $\varpi_0=1.06\pm0.02$~mas, placing it behind the other groups by $\sim$130~pc. Group E has a mean parallax of 1.27$\pm$0.02~mas, which would place it in front of the other groups by $\sim$35~pc, but this difference is only significant at slightly over the 2$\sigma$ level, meaning that the parallax data are not absolutely conclusive. The model also suggests non-zero depths to the groups in parallax ranging from 0.04--0.12~mas. These seem unphysically large when translated to physical sizes (25-80~pc) and may result from outliers with small error bars (Figure~\ref{mean_parallax.fig}). It is possible that some of these outliers could be misclassified stars, either spurious members or members with a misassigned group. 

Uncertainty in our estimate of the absolute distance would be dominated by the systematic $\sim$0.04~mas correlated uncertainty in parallax (Lindegren et al. 2018). This would translate to a shift of $\pm$25~pc.

\begin{figure*}[t]
\centering
\includegraphics[width=0.45\textwidth]{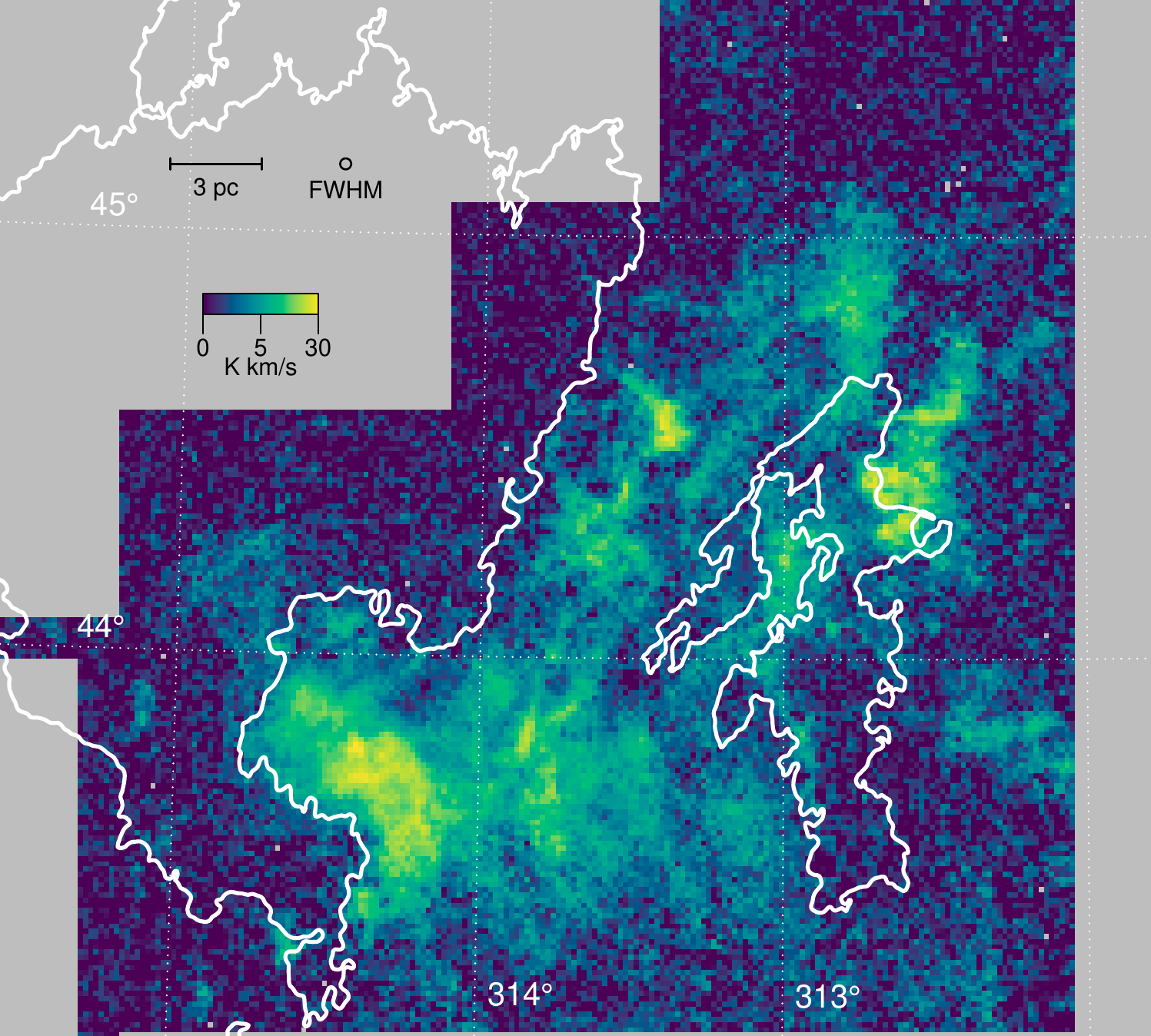} 
\includegraphics[width=0.45\textwidth]{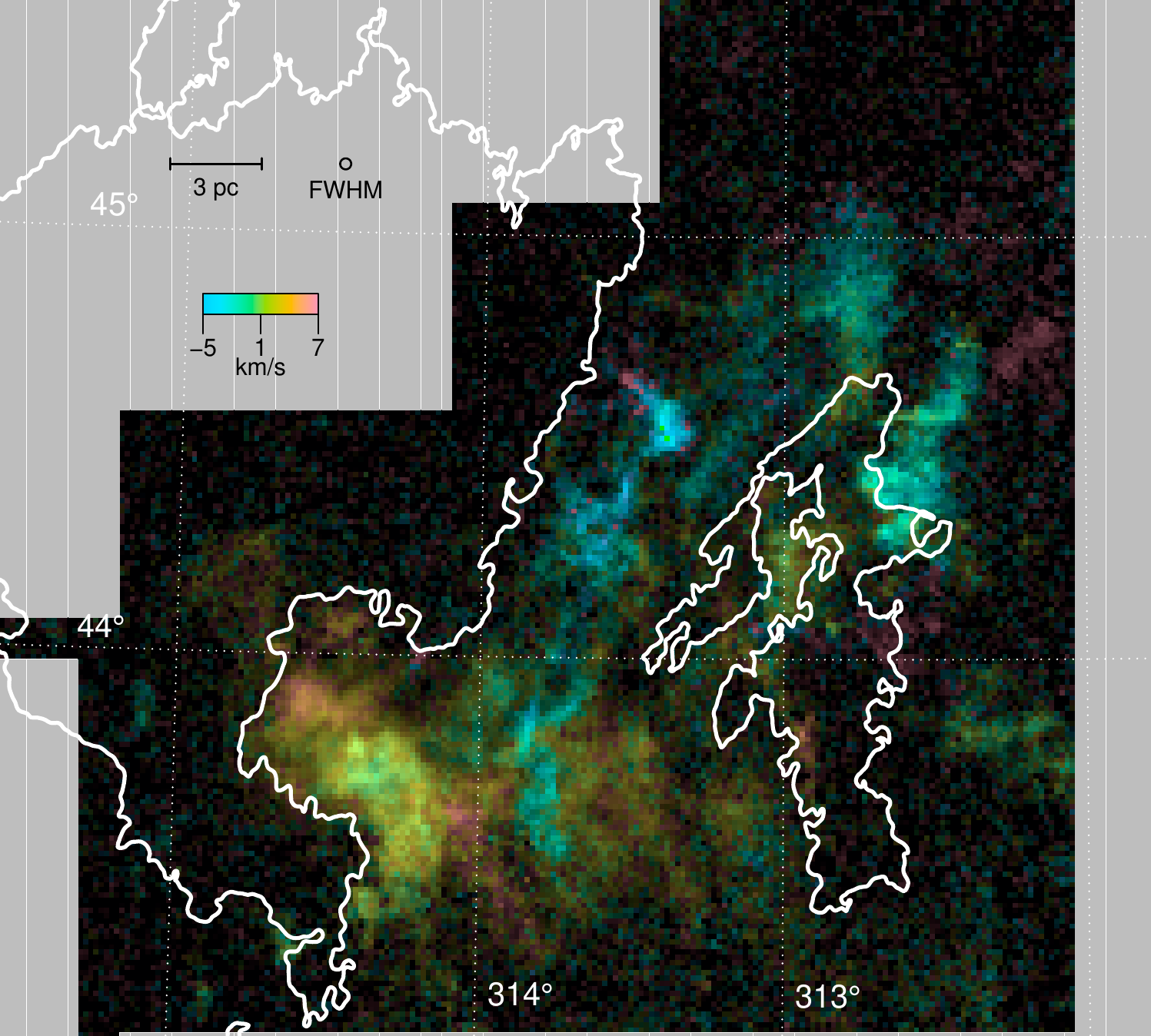} 
\caption{Left: integrated $^{13}$CO $J=1$--0 emission over the velocity range $v_{lsr}=-17$ to 10~km~s$^{-1}$. Right: The color scale for the $^{13}$CO map uses intensity to indicate integrated emission and hue to indicate the first moment. The FWHM of the observations and the physical scale of the map are indicated in the upper left. We also include the outline of the optical emission from the DSS image (white contours).
 \label{int13CO_stars.fig}}
\end{figure*}

\begin{figure*}[t]
\centering
\includegraphics[width=0.8\textwidth]{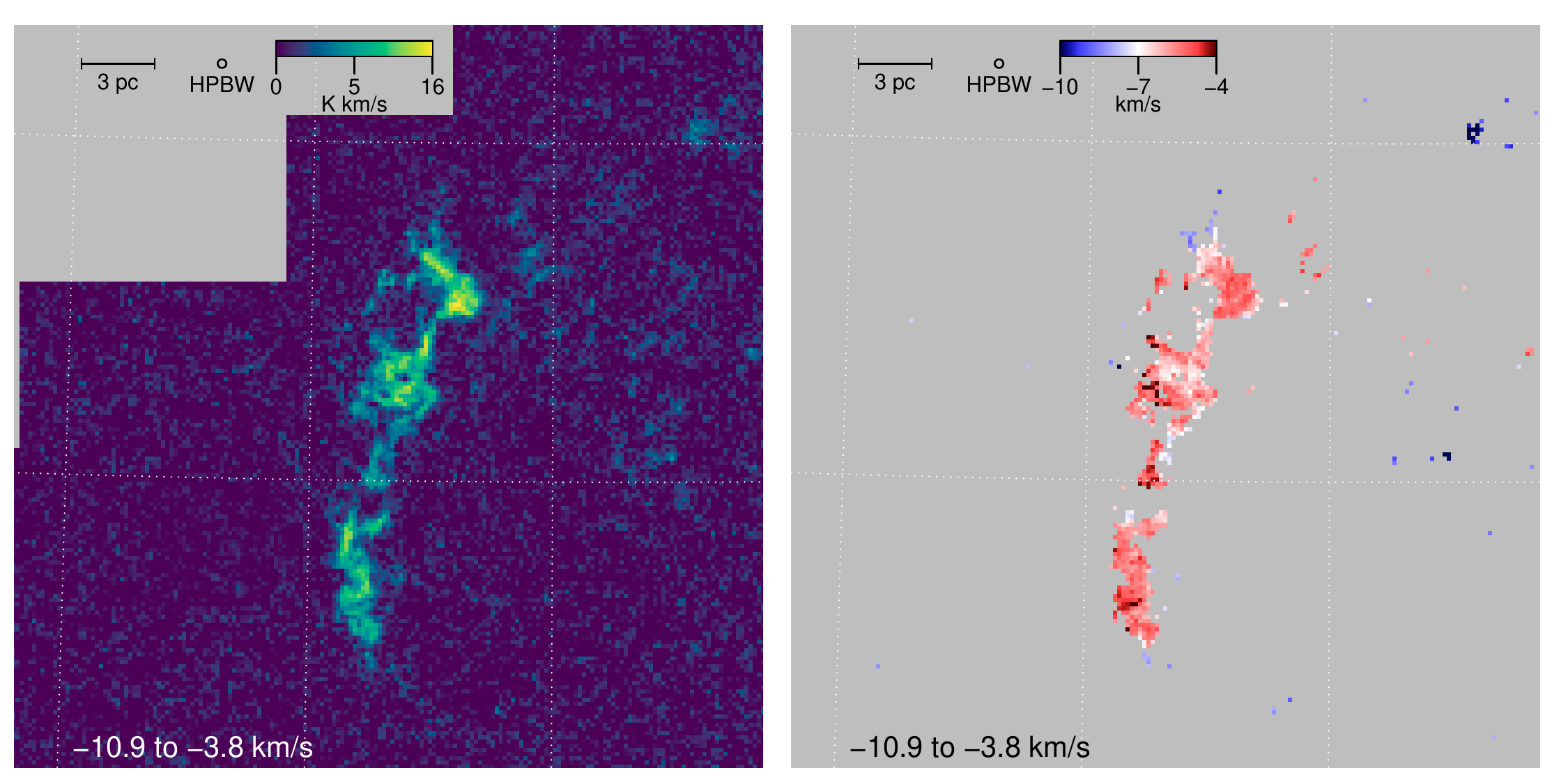} 
\includegraphics[width=0.8\textwidth]{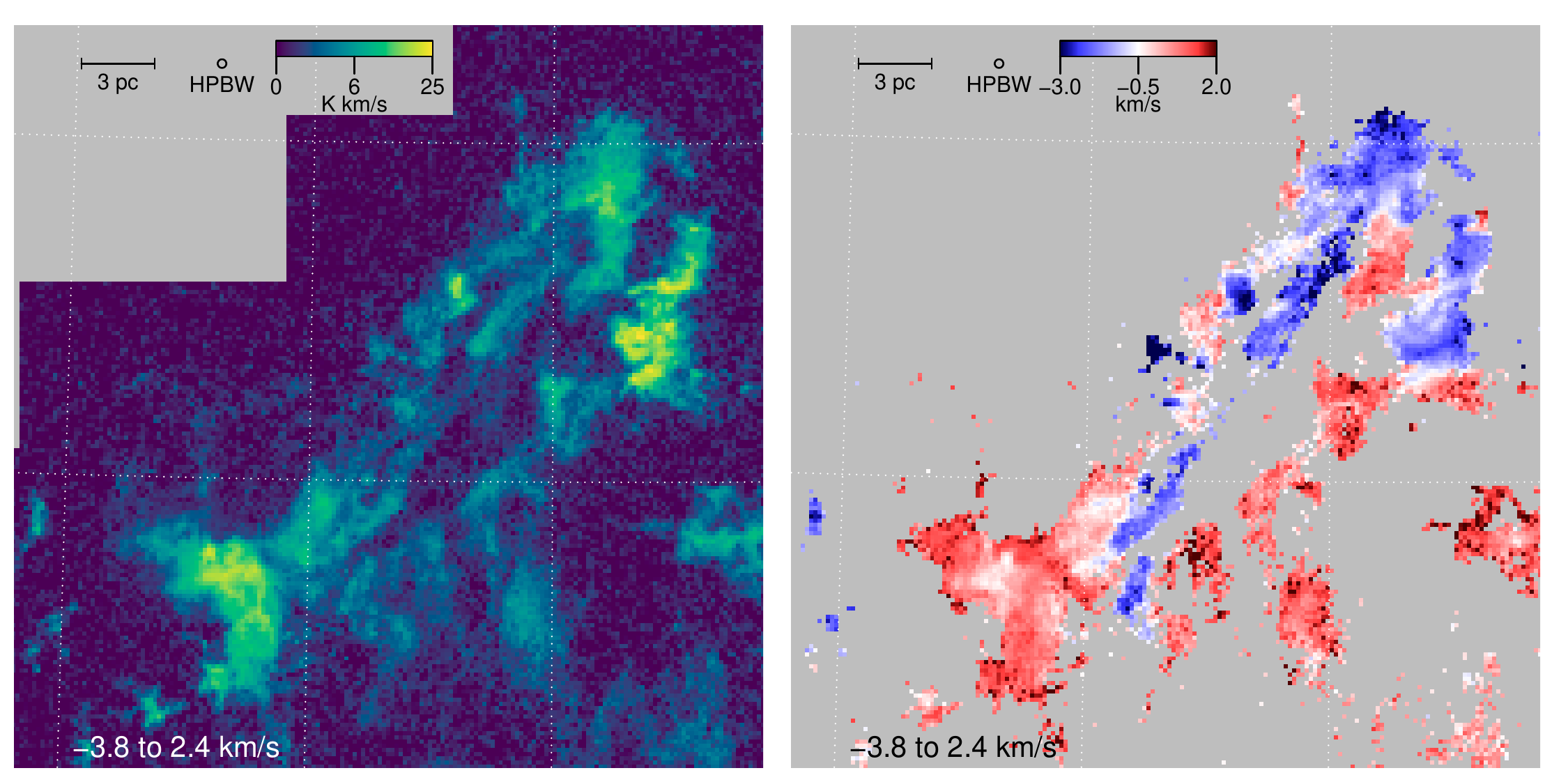} 
\includegraphics[width=0.8\textwidth]{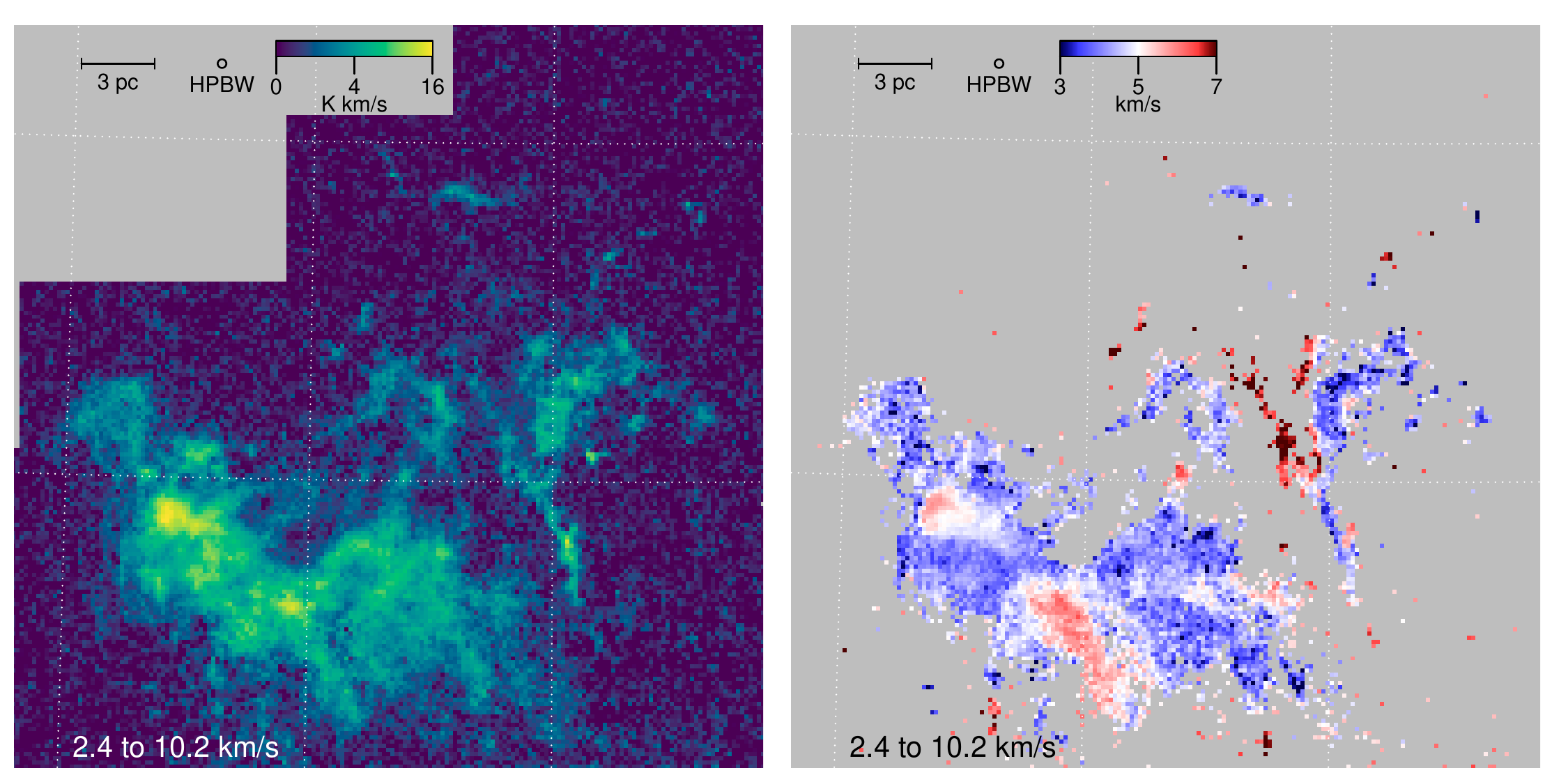} 
\caption{Integrated intensity maps (left panels) and first moment maps (right panels) over three velocity ranges, from top to bottom in order of increasing velocity. These three ranges highlight different aspects of the cloud's structure. The top panel ($-10.9$ to $-3.8$~km~s$^{-1}$) reveals a $\sim$15~pc long filamentary structure, the middle panel ($-3.8$ to 2.4~km~s$^{-1}$) reveals structures spanning the whole region, including a vaguely shell-like structure in the north, and the bottom panel (2.4 to 10.2~km~s$^{-1}$) reveals a large cloud complex concentrated in the ``Gulf or Mexico'' region. 
 \label{moments.fig}}
\end{figure*}

\begin{figure*}[t]
\centering
\includegraphics[width=0.95\textwidth]{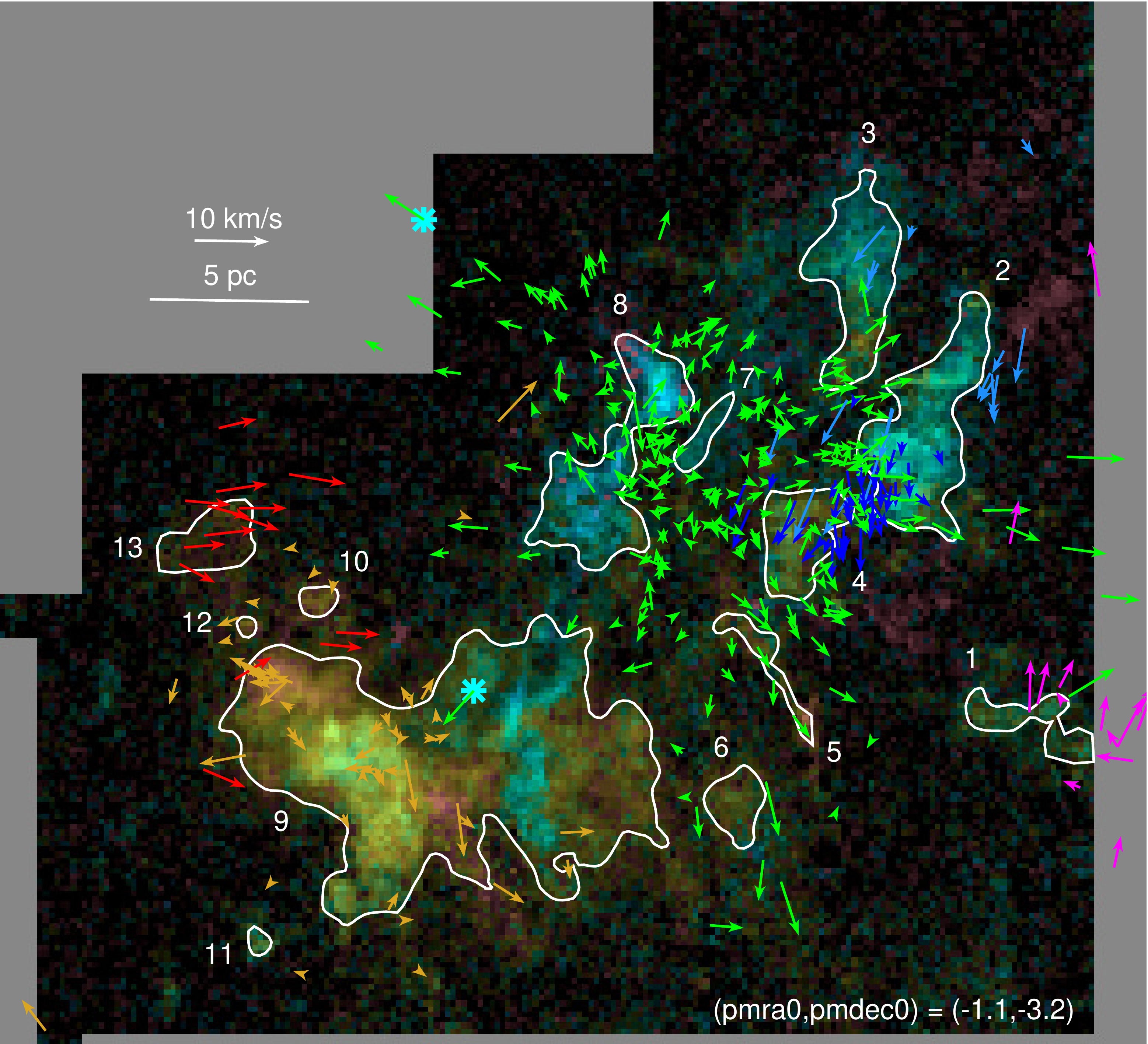} 
\caption{Map of star positions and velocities, indicated by arrows, superimposed on the $^{13}$CO map. Arrows are drawn relative to the median proper motions of the system, which we define to be $(\mu_{\alpha^\star},\mu_\delta)_0 = (-1.1,-3.2)$~mas~yr$^{-1}$. Arrows are color coded by the stellar groups A--F, and the outlines of the molecular clouds 1--13 are also shown. The two O stars are indicated by the cyan asterisks.
 \label{arrowheads_all.fig}}
\end{figure*}

Extinction and nebulosity seen in optical images provide additional evidence to support the order in distance suggested by parallax measurements. Group~F is superimposed on a bright region of the nebula, even though the stars in this group appear reddened on color-magnitude diagrams, suggesting that the H$\alpha$ nebulosity is in front of the group. The stars in Group E are associated with cloud clumps in the ``Gulf of Mexico'' region, which are visible as dark dust lanes in the optical images, suggesting that both the cloud and group are in front of the H\,{\sc ii} region. 

The Bajamar Star is obscured by $\sim$10~mag of cloud near the periphery of the ``Gulf of Mexico.'' Thus, it is implausible that this star could be nearer than the rest of the complex as its measured parallax indicates. More likely, the Bajamar Star is at a similar distance as the complex, and its slightly larger measured parallax is a $\sim$2 standard-deviation statistical error (expected to occur with a probability of 1 in 20) with the binarity of the system possibly playing some systematic role in the offset. Furthermore, although heavily obscured, this star cannot be entirely embedded within the cloud because clear lines of sight are needed for this star to illuminate the H\,{\sc ii} region and the bright rim clouds. The  ``Gulf of Mexico''  is superimposed between the Bajamar Star and the southeast bright rim cloud (``Pacific Coast of Mexico''), requiring the cloud to be in front of the line of sight between this star and this rim. 

The order in distance of the other components is more difficult to ascertain given the data. The cloud in the ``Atlantic'' region (possibly associated with Group~D) is dark in the optical image, suggesting that it is mostly in front of the H\,{\sc ii} region. Group C sits atop the northwest bright rim (``Pelican's Neck''). Thus, a clear line of sight from the Bajamar Star to this bright rim, would put the ``Atlantic Cloud'' and Group D in front of the ``Pelican's Neck Cloud'' and Group C. 
Both the optically dark ``Pelican's Hat'' cloud, associated with Group B, as well as the cloud around Group A can be seen in absorption, suggesting that they are also mostly in front of the H\,{\sc ii} region.

Other group properties listed in Table~\ref{astro.tab} include average proper motions, approximate centers of the groups, and the radii that contain half the groups' stars. To calculate the proper motion, we use the same strategy as \citet{2019ApJ...870...32K} and calculated the weighted median with the conventional $1/\mathrm{error}^2$ weights. To estimate the group centers, we calculate the sigma-clipped mean $\alpha$ and $\delta$ values, with a three standard deviation threshold to avoid strong influences from outlying stars.

\section{Structure of the Molecular Cloud}\label{cloud.sec}

The $^{13}$CO map provides kinematic information about the system that is complementary to the stellar proper motions measured by {\it Gaia}. Figure~\ref{int13CO_stars.fig} shows the integrated $^{13}$CO emission over a velocity range of $v_{lsr} = -17$ to 10~km~s$^{-1}$, which encompasses nearly all the line emission associated with the star-forming region. Dominant clouds are located in the ``Gulf of Mexico,''  ``Pelican,'' ``Pelican hat,''  and ``Atlantic'' regions. These areas tend to have gas at different velocities, with differences ranging from several to tens of km~s$^{-1}$. Figure~\ref{moments.fig} shows integrated intensity maps and first-moment maps for three adjoining velocity ranges. Clouds in the southern ``Gulf of Mexico'' region tend to have more positive velocities, while gas in the centrally located ``Atlantic'' region forms a $\sim$15~pc long filamentary structure with more negative velocities, and the eastern ``Pelican'' and ``Pelican Hat'' regions contains gas with both positive and negative velocities.

 An early radio study of the region by \citet{1980ApJ...239..121B} suggested that the complex is in the process of expansion. \citet{2014AJ....147...46Z} also attribute the cloud morphology in the ``Pelican'' region to an expanding shell around the H\,{\sc ii} region. We find that some parts of the cloud have kinematics consistent with expansion, but that the expansion is not global. The shell-like structure is most distinct at $-1$~km~s$^{-1}$ in the northwest. In contrast, in the south, the ``Gulf of Mexico'' cloud appears to be infalling, not outward-moving. The $\sim$15~pc long filament at $-5$~km~s$^{-1}$, which stretches from the north to the south, appears to be moving away from the H\,{\sc ii} and towards us with a fairly coherent radial velocity across its whole length.

To investigate the molecular cloud structure in more detail, in Figure~\ref{arrowheads_all.fig} we subdivide the system into individual clouds with contours at 6~K~km~s$^{-1}$ from the integrated map. This threshold picks out most of the main cloud components, which are labeled 1 through 12 in both the figure and in Table~\ref{com.tab}. We also include a Cloud 13 defined by contours at 3~K~km~s$^{-1}$ that is associated with stellar Group~F, but does not meet the 6~K~km~s$^{-1}$ threshold. 

From the brightness of the $^{13}$CO $J$=1--0 emission line, it is possible estimate cloud masses if we make certain assumptions. To do so, we must accept  that, as in most CO studies, there are implied systematic effects on the results. For example, the $^{13}$CO emission is less sensitive to diffuse gas traced by $^{12}$CO, but cannot trace dense gas where the $^{13}$CO line becomes too optically thick. To calculate column densities, we assume local thermodynamic equilibrium (LTE), an excitation temperature, and the ratio of $^{13}$CO to H$_2$. \citet{2014AJ....147...46Z} use a $^{12}$CO map from the Purple Mountain Observatory Delingha 13.7~m telescope to estimate excitation temperature throughout the cloud complex. They find temperatures ranging from 5--25~K, with the bulk of the cloud around $\sim$10--15~K. This includes the massive ``Gulf of Mexico'' region, which is mostly $\sim$14~K. However, several areas have higher temperatures (20--25~K), particularly some edges of the clouds in the north of the NAP complex, including the cloud behind the bright rim in the ``Pelican's Neck'' and a cloud filament in the ``Atlantic'' region. Consistent with the higher inferred temperatures, these regions appear geometrically oriented so they would be illuminated by the Bajamar Star. We note that the measured $^{12}$CO excitation temperatures would be representative of the cloud surfaces, but the cloud interiors may not be the same.

We follow the method outlined by \citet{2015PASP..127..266M} using the coefficients from the JPL Molecular Spectroscopy Database\footnote{\url{https://spec.jpl.nasa.gov}}. 
Assuming a $^{13}$CO excitation temperature of $T_{ex}=14$~K (as justified above), we estimate the optical depth of the $^{13}$CO $J$=1--0 transition with the equation
\begin{multline}\label{optical_depth.eqn}
\tau^{13}_{J=1}(v) =  \\-\ln \left\{ 1 - \frac{T_{R}^{13,1}(v)}{5.29\,\mathrm{K}}\left[ \frac{1}{\exp(5.29\,\mathrm{K}/T_{ex})-1}  -0.17 \right]^{-1}  \right\},
\end{multline}
using ${T_R}^*$ as an approximation for the radiation temperature $T_R$ in this equation. This equation is applied to each pixel, channel-by-channel, in the emission-line data cube.
From the distribution of derived optical depths, $\sim$95\% of the $^{13}$CO gas by mass has optical depths $\tau<0.5$, while $\sim$98\% has optical depth $\tau<1$. Although $^{13}$CO is not sensitive to the highest density parts of the cloud, and thus cannot account for mass in these regions, the shape of the distribution suggests that much of the cloud is only moderately optically thick. The $^{13}$CO column density would then be
\begin{multline}\label{ntot.eqn}
N_{tot}(^{13}\mathrm{CO}) \approx \left(6.6\times 10^{14}\,\mathrm{cm}^2\right)
\left(\frac{T_{ex}}{2.64\,\mathrm{K}} + \frac{1}{3} \right) \times\\
\left [
 1 - \exp\left( - \frac{5.29\,\mathrm{K}}{T_{ex}} \right) 
\right ]^{-1}
\frac
{\int \tau^{13}_{J=1}(v) \mathrm{d}v}{\mathrm{km}\,\mathrm{s}^{-1}}.
\end{multline}
Finally, we assume that $N(\mathrm{H}_2) \approx 3.8 \times 10^5 N(^{13}\mathrm{CO})$ \citep{2013ARA&A..51..207B} to calculate $N(\mathrm{H}_2)$ column density. 

Integrating column density over the entire FCRAO $^{13}$CO map yields a total mass of $7.2\times10^4$~$M_\odot$. Masses of individual clouds are given in Table~\ref{com.tab}. Uncertainty in these quantities is likely to be dominated by systematic errors in the assumptions, particularly the $^{13}$CO-to-H$_2$ ratio \citep{2013ARA&A..51..207B}. In addition, clumpiness in the clouds that is not resolved by the FCRAO beam could cause us to systematically underestimate the amount of gas in regions with high optical depth. 

\subsection{Relation of Clouds to Stellar Groups}

There is spatial correspondence between the stellar groups and the clouds (Figure~\ref{arrowheads_all.fig}). Groups~A and F are associated with the minor clouds 1 and 13, respectively. Group~E is embedded in the ``Gulf of Mexico'', which consists of a major cloud component (Cloud 9) and several minor cloud components (Clouds 6, 10, 11, and 13) in the $^{13}$CO map. Group C lies on a bright rim cloud at the southeastern edge of Cloud 2. The stars in Group B are distributed in the northern part of Cloud 2 as well as Cloud 3.

Group D, the most spatially extended stellar group, has the most complicated relationship to the $^{13}$CO clouds. This group is projected over a region that spans several clouds. However, Cloud 8 is the dominant cloud in this part of the NAP region and lies near the center of Group D. The stars selected in our {\it Gaia} analysis do not lie on top of Cloud 8, but instead the densest parts of Group D are next to the cloud. This indicates that Group D is either partially embedded in Cloud 8, which would imply that Cloud 8 is the remainder of the natal cloud that formed this group, 
or that Group D lies behind the cloud. Stars from this group are also superimposed on other nearby minor clouds (Clouds 5, 6, and 7) which may either be smaller remnants of the natal cloud or chance superpositions. The edges of Group D also intersect the edges of Clouds 2, 3, and even (in the case of the Bajamar Star) Cloud 9. 

If the stellar groups and clouds are associated, then it is likely that they have similar mean motions. This enables us to associate mean proper motions with clouds and radial velocities with stars, creating a three dimensional picture of the velocities in this region. 

\subsection{Velocity Structure of Clouds}\label{velocity_structure.sec}

\begin{figure*}[t]
\centering
\includegraphics[width=1.0\textwidth]{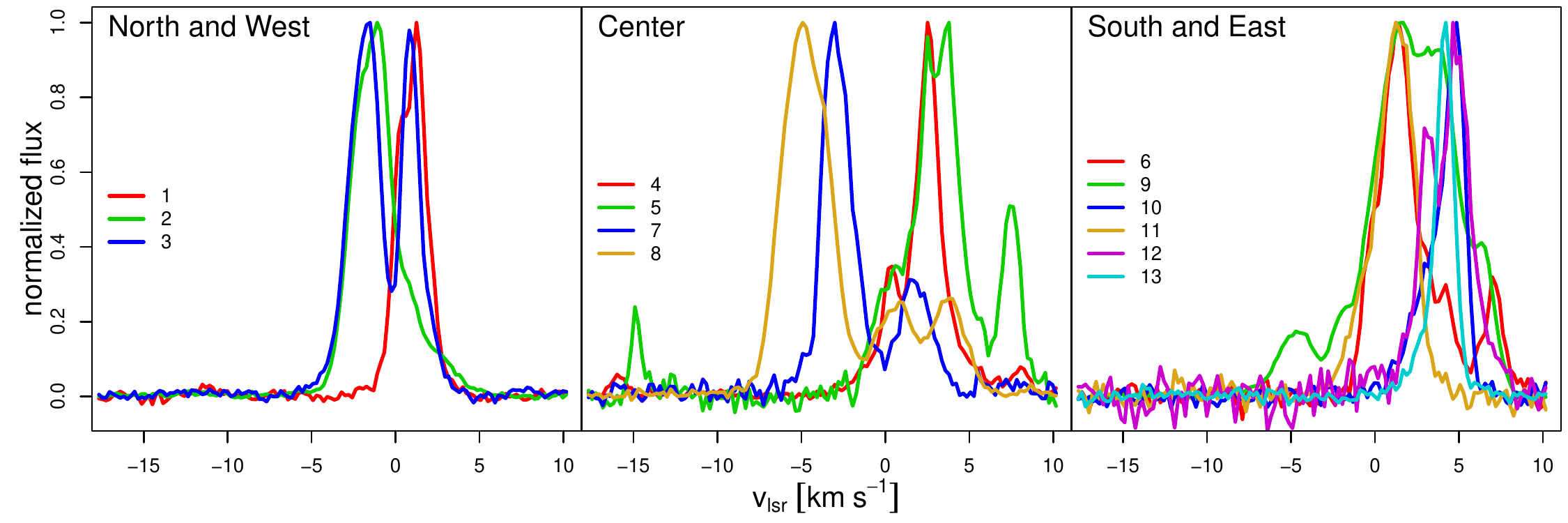} 
\caption{Spatially integrated line flux as a function of $v_{lsr}$ for each of the clouds in Figure~\ref{arrowheads_all.fig}. The three panels show velocity profiles for clouds in three different areas of the NAP region. Molecular gas spans the velocity range $-15$ to 10~km~s$^{-1}$, and several of the clouds have multimodal structure.
 \label{lines.fig}}
\end{figure*}

Line profiles for each of the clouds (Figure~\ref{lines.fig}) were constructed by spatially integrating the $^{13}$CO gas detected within the contours on Figure~\ref{arrowheads_all.fig}. Most of these profiles display several distinct peaks. In Table~\ref{com.tab}, we report basic statistics of the velocity distributions, including the first (mean) and square root of the second (standard deviation) moments, the mode, and the FWHM of the tallest mode. The mode and FWHM ($\Delta v_\mathrm{FWHM}$) may be more characteristic of the kinematics of the clouds because these statistics are less susceptible to merging components with very different velocities that may be from distinct clouds. Thus, we base our discussion of cloud dynamics (Section~\ref{cloud_dynamics.sec}) on these latter quantities. We also estimate the characteristic size of each cloud, defined as the projected radius that contains half the integrated emission.  

\begin{figure*}[t]
\centering
\includegraphics[width=0.48\textwidth]{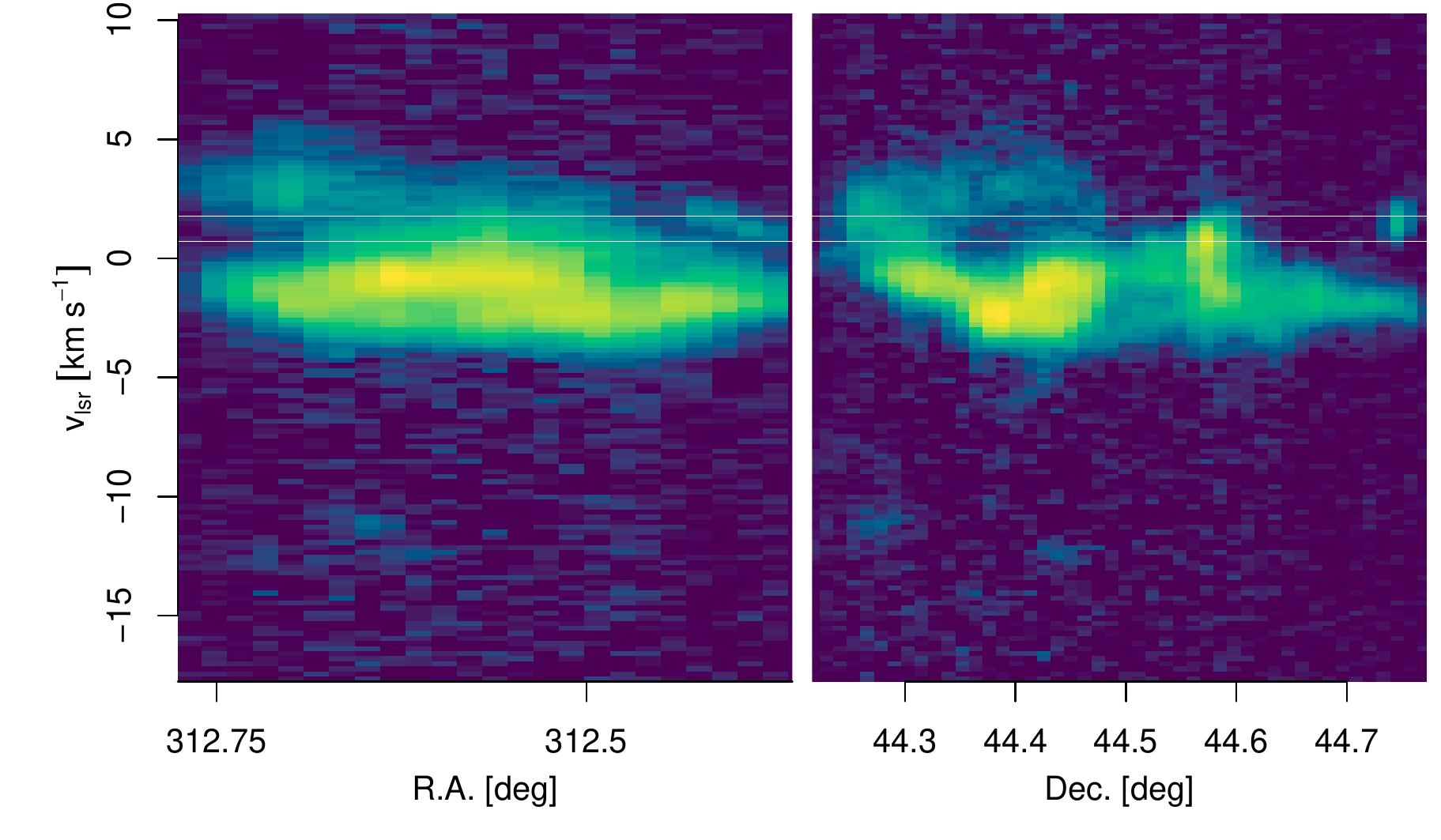} 
\includegraphics[width=0.48\textwidth]{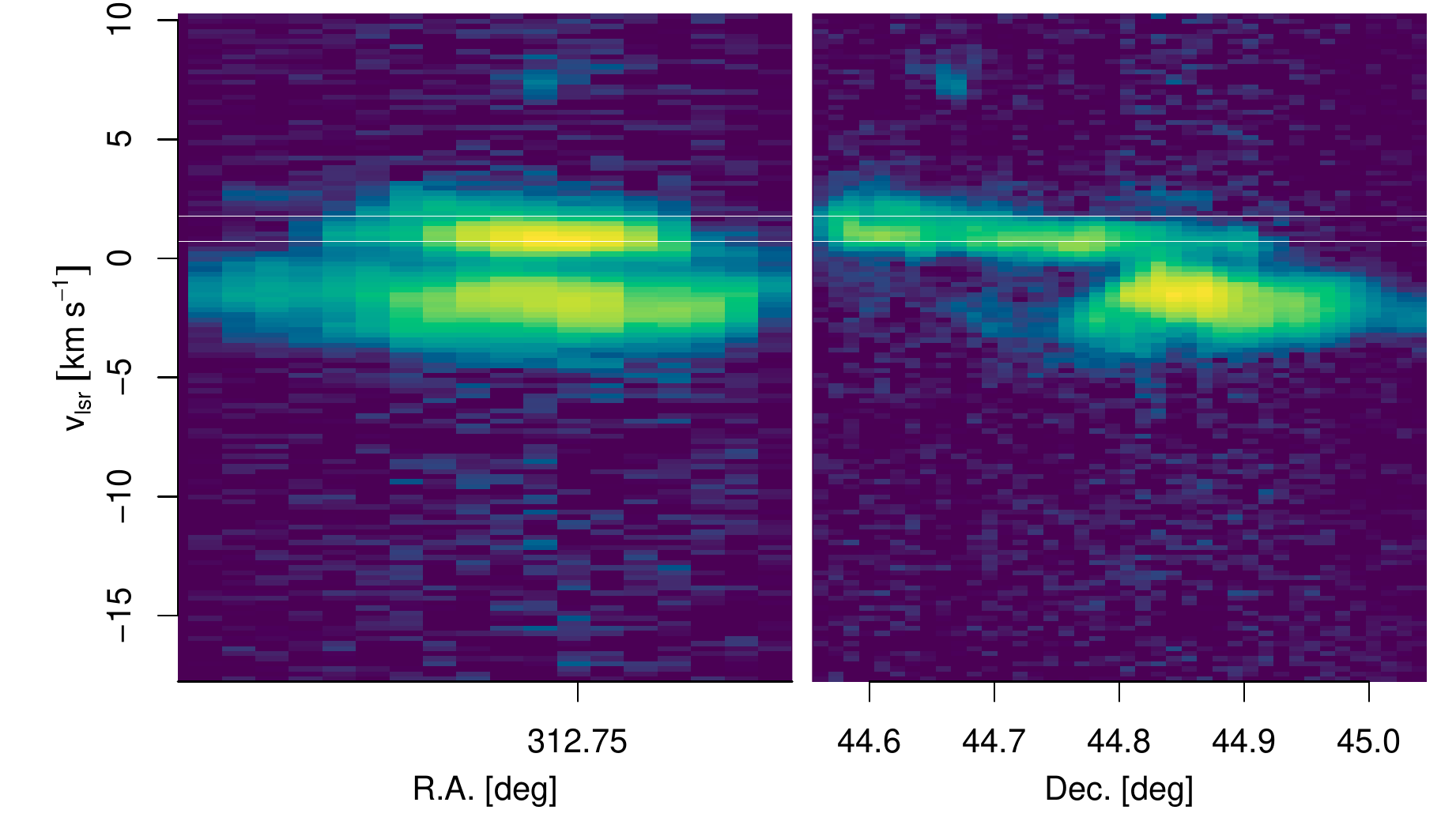} 
\includegraphics[width=0.48\textwidth]{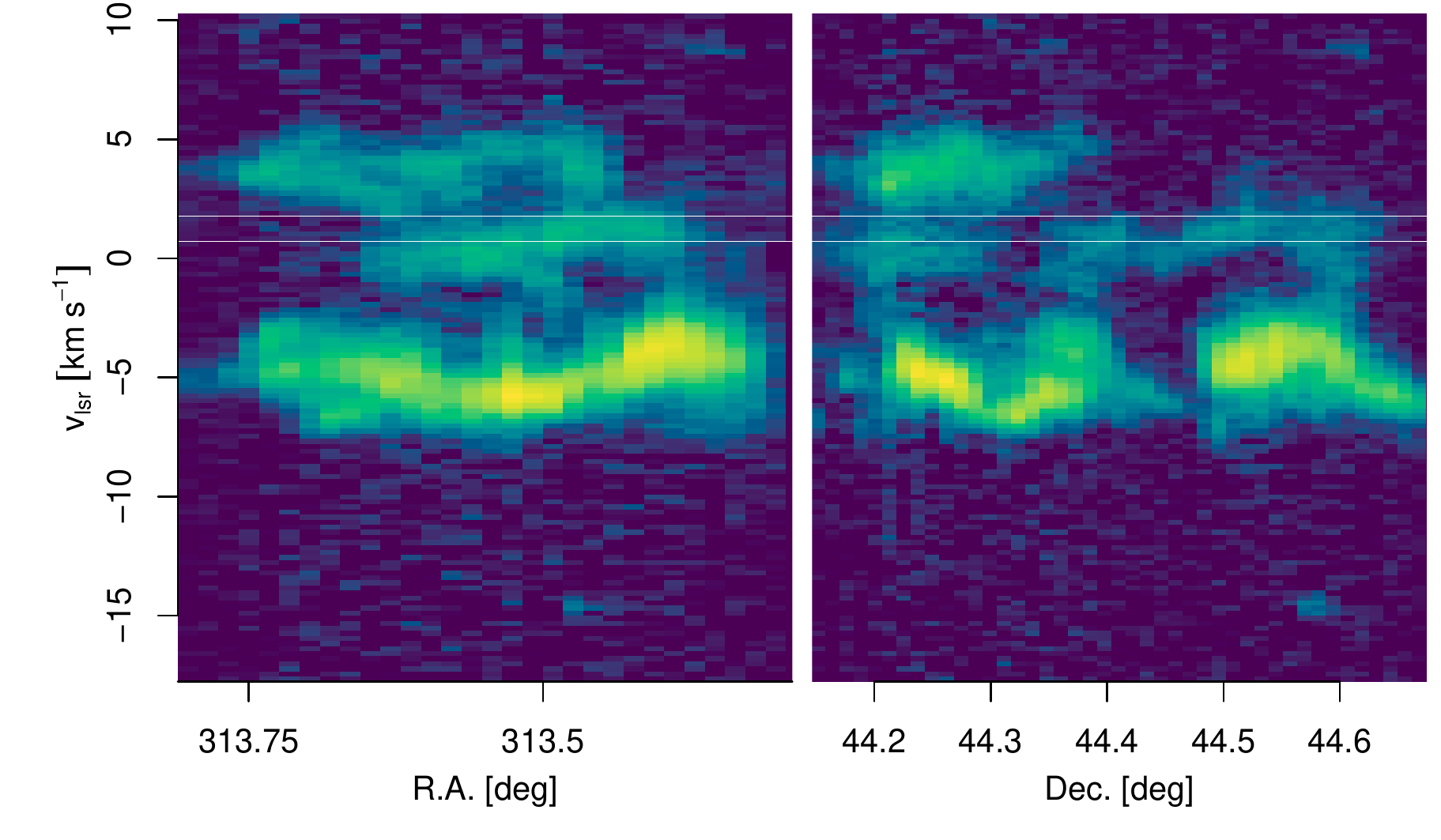} 
\includegraphics[width=0.48\textwidth]{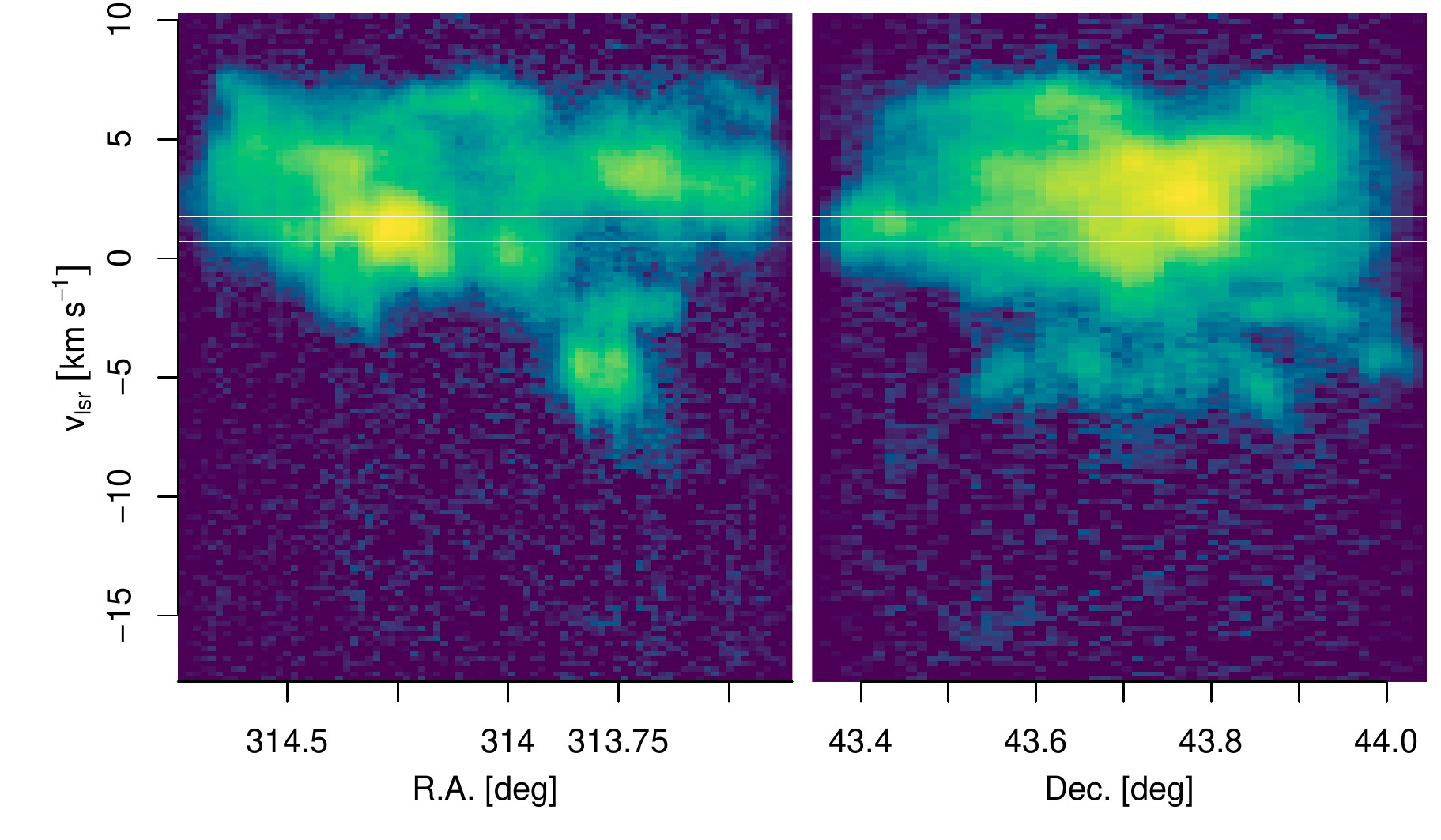} 
\caption{Position--velocity diagrams of $^{13}$CO $J=1$--0 emission are shown for Clouds~2, 3, 8, and 9 (upper left, upper right, lower left, and lower right, respectively).\\
(The complete figure set (13 images) is available.)
 \label{co_vel_vs_pos.fig}}
\end{figure*}

Optical depth can affect line profiles, flattening the peaks and effectively broadening the lines \citep{2016A&A...591A.104H}. To more accurately resolve the velocity structure, we integrated over the optical depths given by Equation~\ref{optical_depth.eqn} rather than  ${T_R}^*$ \citep[see][]{1999ApJ...517..209G}. This results in line profiles that are more sharply peaked with FWHM that are $\sim$20\% narrower. 
 
The interpretation of the multimodal structure becomes clearer with the position--velocity diagrams in Figure~\ref{co_vel_vs_pos.fig}. Four of the major clouds (Clouds 2, 3, 8 and 9) are shown as examples, while the diagrams for the other clouds are accessible via the online figure set. Even some of the minor clouds (e.g., Clouds 5, 6, and 7) have complicated velocity structures.

In Cloud 2, $v_{lsr}$ increases by several km~s$^{-1}$ toward the southern end of the cloud where the optically bright rim and stellar group are located. It is plausible that this could be an effect of the expansion of the H\,{\sc ii} region pushing this end of the cloud away. 

For Clouds 3 and 8, multiple components are visible, corresponding to the multiple peaks in the line profiles. In Cloud 3, there are two velocity components separated by $\sim$5~km~s$^{-1}$, while in Cloud 8 there are 3 components, also separated by $\sim$5~km~s$^{-1}$ from each other. The distinct components in Cloud 3 are also mostly separated along a north--south axis, possibly indicating that this cloud is composed of multiple sub-clouds. However, for Cloud~8 the main component is the dense northern end of the filament at $v_{lsr} \approx -5$~km~s$^{-1}$  (Figure~\ref{moments.fig}), while the other components are related to overlapping filamentary structures at higher $v_{lsr}$. In our analysis, we tentatively associate the stars in Group D with the $-5$~km~s$^{-1}$ gas because this is the dominant component around which the stars are conglomerated, but our data are insufficient to make a definitive conclusion (Appendix~\ref{cloud8.appendix}).

The position-velocity diagram for Cloud~9 in Figure~\ref{co_vel_vs_pos.fig} shows that it has a clumpy structure, with clumps spanning the velocity range from $\sim$0--7~km~s$^{-1}$. A small clump at $-5$~km~s$^{-1}$ is probably not part of this cloud, but is instead a continuation of the long north--south filament seen in Figure~\ref{moments.fig} (top panels). Overall, Cloud~9 has the largest positive $v_{lsr}$ of the complex, which, in combination with our evidence that the ``Gulf of Mexico'' and Group E are in front of the rest of the NAP region, suggests that Cloud 9 is infalling. However, the part of the cloud brightest in $^{13}$CO emission also has a fairly modest velocity of $\sim$1~km~s$^{-1}$. This $^{13}$CO bright patch is fairly near the Bajamar Star and shows up as a having a higher temperature in the $^{12}$CO map from \citet{2014AJ....147...46Z}.

\subsection{Dynamical Properties of the Clouds}\label{cloud_dynamics.sec}

From the kinematic measurements of the cloud in Table~\ref{com.tab}, several dynamical quantities can be calculated that are useful for understanding star formation in this region. For these calculations, we use the line-widths for the most prominent peaks in Figure~\ref{co_vel_vs_pos.fig} to avoid combining unrelated velocity components. The dynamical quantities are given in Table~\ref{cloud_dynamics.tab}. 

If we assume that the kinetic temperature $T_\mathrm{kin}$ of the gas equals the excitation temperature measured for $^{12}$CO ($\sim$14~K) and that the mean molecular mass is $\bar{m}=2.33$~amu, the one-dimensional sound speed in the molecular clouds would be $c_s\approx0.22$~km~s$^{-1}$. For each of the molecular clouds delimited in our analysis, the line width of the tallest mode is larger than can be accounted for by purely thermal broadening. The non-thermal component is 
\begin{equation}
\sigma_{nt} = \sqrt{\left(\frac{\Delta v_\mathrm{FWHM}}{2.35}\right)^2 - \frac{kT_\mathrm{kin}}{m_\mathrm{obs}} },
\end{equation}
where $m_\mathrm{obs}=29$~amu is the mass of the observed molecule, $^{13}$CO. The Mach number is defined to be $\mathcal{M} = \sigma_{nt}/c_s$. For most of the individual clouds, Mach numbers span $2\lesssim\mathcal{M}\lesssim5$. Exceptions are Clouds 8 and 9 -- the most massive clouds -- which have larger velocity dispersions of $\mathcal{M}\approx6$ and 10, respectively. The lowest velocity dispersion is found for Cloud~13 with $\mathcal{M}\approx2$; this also happens to be the cloud with the lowest surface density.  This range of Mach numbers is fairly typical for star-forming clouds, such as the Orion~B cloud \citep[$\mathcal{M}\sim6$;][]{2017A&A...599A..99O} or Taurus \citep[$\mathcal{M}<2$--3;][]{2016A&A...591A.104H}. The position--velocity plots (Figure~\ref{co_vel_vs_pos.fig}) show that some of this non-thermal broadening in the NAP clouds is due to cloud-scale velocity structure. In Cloud 8, the $-5$~km~s$^{-1}$ component appears curved in position-velocity space, while, in Cloud 9, the large velocity dispersion is composed of multiple clumps with different velocities. 

Several dynamically important quantities that can be estimated for each cloud include the mean density $\bar{\rho}$, free-fall timescale $\tau_\mathrm{ff}$, and crossing timescale $\tau_\mathrm{cross}$ defined as
\begin{align}
\bar{\rho} &= M_{^{13}CO}/(4/3\pi)r^2_\mathrm{cloud}\\
\tau_\mathrm{ff} &= \sqrt{3\pi/32G\bar{\rho}}\\
\tau_\mathrm{cross} &= r_\mathrm{cloud}/\sigma_{nt}.
\end{align}
Both the crossing timescales and the free-fall timescales range from 0.1 to 2~Myr. For the main clouds with ongoing star formation, the typical free-fall timescale is $\tau_{\mathrm{ff}}\sim1$~Myr. We note that these timescales are similar to the $\sim$1~Myr ages inferred for the pre--main-sequence stars. 

\begin{figure}
\centering
\includegraphics[width=0.47\textwidth]{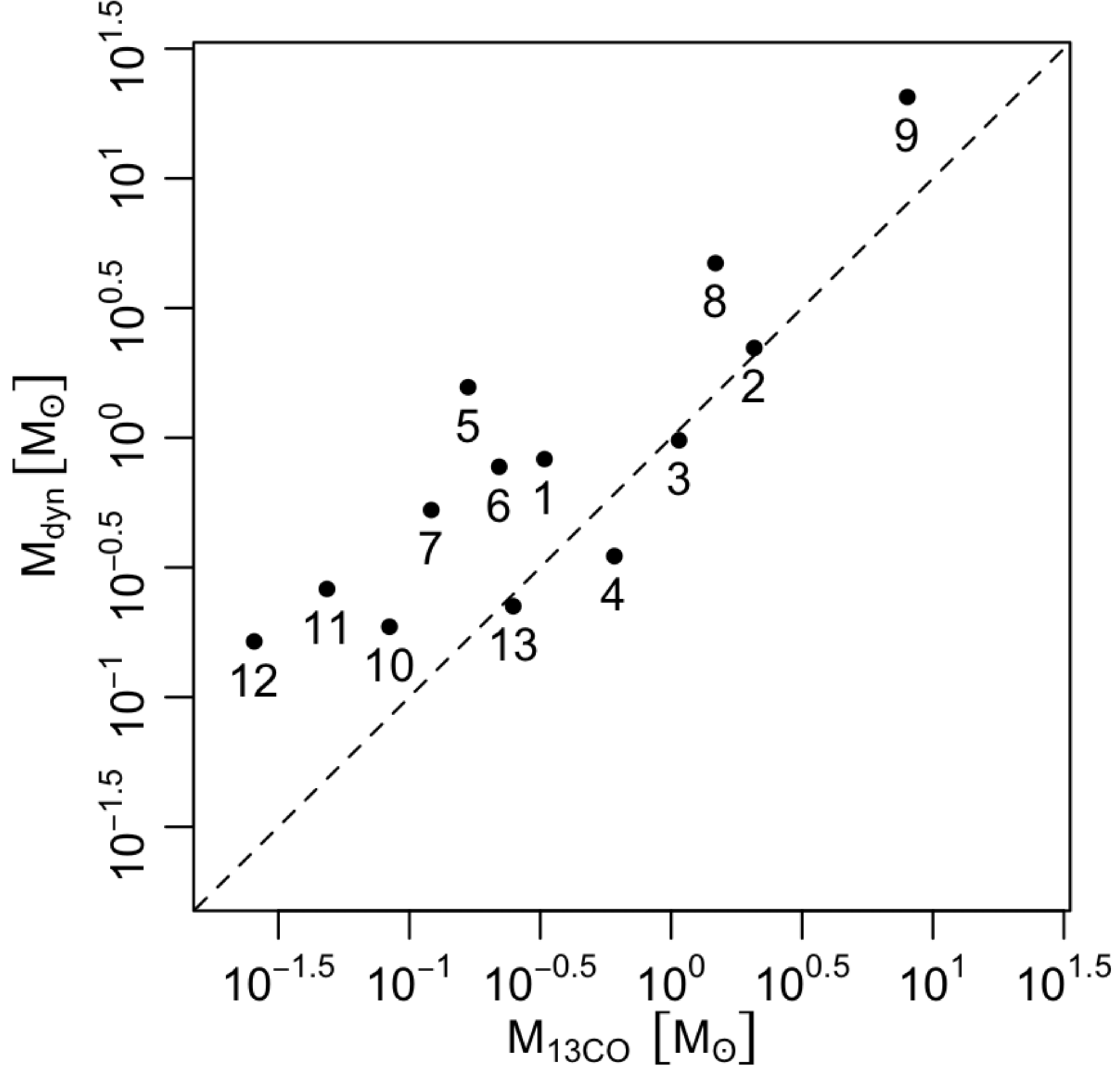} 
\caption{Dynamical cloud masses versus mass estimates from $^{13}$CO column density. The dashed line indicates equivalence between these quantities and the points are labeled by the cloud they represent. Systematic uncertainties on $^{13}$CO based mass estimates may be a factor of several as discussed in the text.
 \label{mass_compare.fig}}
\end{figure}

The velocity dispersions can be used to estimate cloud dynamical masses. However, these calculations depend on the virial ratio of the clouds, $\alpha = 2\mathcal{T}/|\mathcal{W}|$, where $\mathcal{T}$ is kinetic energy and $\mathcal{W}$ is potential energy.  Previous studies typically use the definition from \citet{1992ApJ...395..140B} that assumes a spherical, uniform density cloud, giving
\begin{equation}\label{bm.eqn}
\alpha_\mathrm{BM92} = \frac{5\sigma_{nt}^2 r_\mathrm{cloud}}{GM_\mathrm{cloud}}.
\end{equation}
However, the true value of $\alpha$ depends on the three-dimensional distribution of mass within the cloud \citep{2019ApJ...878...22S,2020MNRAS.492..488G}, with corrections for centrally concentrated mass distributions and filamentary structure typically being a factor of several \citep{1992ApJ...395..140B}. 
If we let $\alpha_\mathrm{BM92}=1$, we obtain a dynamical mass estimate
\begin{equation}\label{dynamical_mass.eqn}
M_\mathrm{dyn} = 5\, r_\mathrm{cloud}\sigma_{nt}^2/G.
\end{equation}
In Figure~\ref{mass_compare.fig}, we plot $\log M_\mathrm{dyn}$ versus $\log M_{^{13}\mathrm{CO}}$ for the 13 clouds we have identified in the NAP region. These quantities are strongly correlated ($p<0.001$ from Kendall's $\tau$ test), with an approximately proportional relationship over a dynamic range of $\sim$200 in cloud mass. However, the dynamical masses are systematically $\sim$0.4~dex higher, with $\sim$0.4~dex scatter in the relation. The similarity in masses derived by the two methods to within a factor of $\sim$3 helps corroborate our $^{13}$CO-based mass estimates in Table~\ref{com.tab}.
\citet{2019MNRAS.490.3061V} point out that, observationally, it is nearly impossible to distinguish between the free-fall and virial velocities, which only differ by a factor of 2 in mass or a factor of $\sqrt{2}$ in velocity.
Table~\ref{cloud_dynamics.tab} provides both $M_\mathrm{dyn}$, calculated from the velocity dispersions using Equation~\ref{dynamical_mass.eqn}, and $\alpha_\mathrm{BM92}$, calculated from both the velocity dispersions and the integrated $^{13}$CO cloud masses using Equation~\ref{bm.eqn}. 

\vspace*{0.25in}
\section{Stellar Kinematics}\label{kinematics.sec}

For the analysis of stellar kinematics, we convert the observed astrometric quantities of parallax and proper motion into physical velocities with units of km~s$^{-1}$. For this we establish a Cartesian coordinate system with positions $(x, y, z)$ and velocities $(v_x, v_y, v_z)$, where we select the origin to be located at the center of the system that we are analyzing and moving with the median velocity of the stars in the system. Given the nature of the data, we are most interested in examining kinematics in the two-dimensional $xy$ plane. We define $x$ to be parallel to $\alpha$ at the origin and $y$ to be parallel to $\delta$ at the origin. The transformations, including orthographic projections, corrections for perspective expansion\footnote{To compute the perspective expansion correction we assume that the stellar group is moving with the radial velocity of the most significant of the cloud components associated with each group (Table~\ref{com.tab}) and transformed to heliocentric coordinates.}, and shifts to the rest frame of the system are described by \citet[][their Equations~1--4]{2019ApJ...870...32K}. 

We are also interested in the component of a star's projected velocity in the direction outward/inward relative to the center of the group ($v_{out}$) as well as the perpendicular azimuthal component ($v_{az}$). These are defined by
\begin{align}
v_\mathrm{out}& = \mathbf{\mathbf{v}}\cdot\mathbf{\hat{r}}\\
v_\mathrm{az}& = \mathbf{\mathbf{v}}\cdot\bm{\hat{\varphi}},
\end{align}
where $\mathbf{\hat{r}}$ and $\bm{\hat{\varphi}}$ are the radial and azimuthal unit vectors relative to the group center in the $xy$ plane. 
Uncertainties in these velocities are calculated by propagation of the {\it Gaia} astrometric errors \citep[see procedure in][]{2019ApJ...870...32K}.

\subsection{Global Kinematics}

In Figure~\ref{arrowheads_all.fig}, we indicate the motions of NAP members using arrows, which, as in the previous figures, are color-coded by stellar group. The figure shows that stars near each other tend to have similar velocities, but the different groups appear to have fairly random motions relative to one another. 

\begin{figure}[t]
\centering
\includegraphics[width=0.45\textwidth]{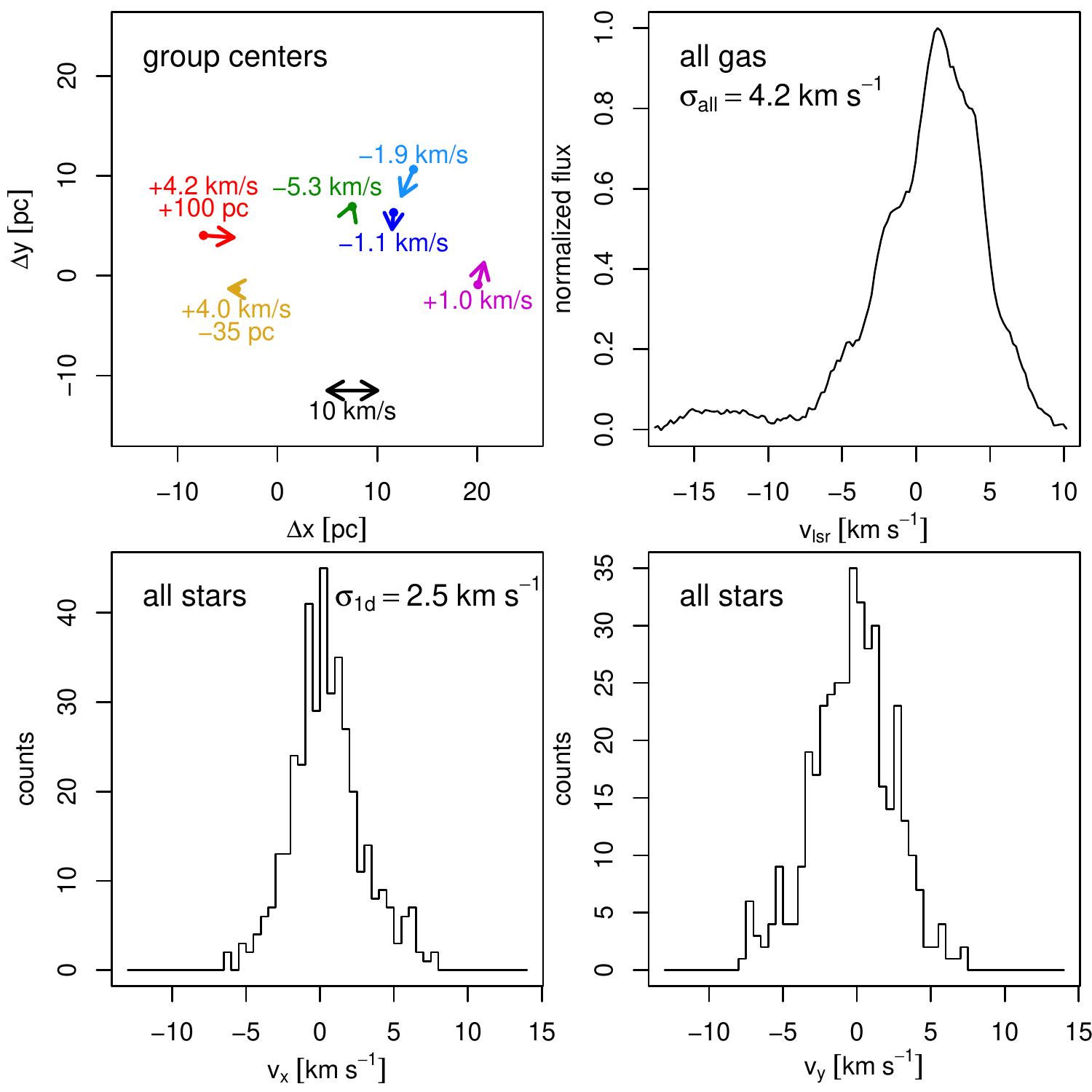} 
\caption{Diagrams illustrating global motions in the NAP region. Upper left: Relative 3D motions of the group centers (reference frame $\mu_{\alpha^\star,0}=-1.1$~mas~yr$^{-1}$, $\mu_{\delta,0}=-3.2$~mas~yr$^{-1}$). Velocities in the plane of the sky from {\it Gaia} are shown by the arrows, while velocities from the molecular gas along the line of sight ($v_{lsr}$) and offsets in distance are indicated by the labels. Upper right: Spatially integrated $^{13}$CO $J=0$--1 line flux for the whole NAP complex. The velocity dispersion is 4.2~km~s$^{-1}$. Lower panels: Velocity distributions ($v_x$ and $v_y$) for the entire NAP stellar population. The total velocity dispersion for stars is $\sigma_{1D} = 2.5\pm0.1$~km~s$^{-1}$. The scale in km~s$^{-1}$ is the same for all three plots of velocity distributions.  
  \label{global_velocities.fig}}
\end{figure}

Figure~\ref{global_velocities.fig} (upper left) shows the relative 3D motions of the group centers inferred from both the mean proper motions of stars and the radial velocities of the associated clouds. This diagram shows no clear pattern of either convergence or divergence.  Stars in Group A tend to be moving mostly north and slightly west -- in a direction tangential to the NAP system center. Stars in both Groups B and C are moving southeast, inward toward the center of the system. Group E is drifting slowly east, away from the center of the NAP complex in the $xy$ plane, but it is likely moving rapidly toward the system center in the third dimension. And, stars in Group F are moving west -- toward the center of the region.  Finally, the center of Group~D appears to be nearly stationary in the $xy$ plane in this reference frame. However, most of the individual stars in this group are directed away from its own center as can be seen in Figure~\ref{arrowheads_all.fig}.

The Bajamar Star is not stationary relative to the center of mass reference frame, but is instead rapidly traveling southeast away from Group D and towards Group E. The other O star, HD~199579, is traveling northeast away from the NAP complex. In this global reference frame\footnote{In Section~\ref{velocity_gradients.sec}, we recalculate the O stars' velocities with respect to Group~D.}, both stars have velocities $>$6~km~s$^{-1}$.  Such speeds would classify both objects as ``walkaways'' \citep{2019A&A...624A..66R,2020arXiv200413730S}, although neither object has traveled outside the star-forming region. 

\begin{figure*}[t]
\centering
\includegraphics[width=0.75\textwidth]{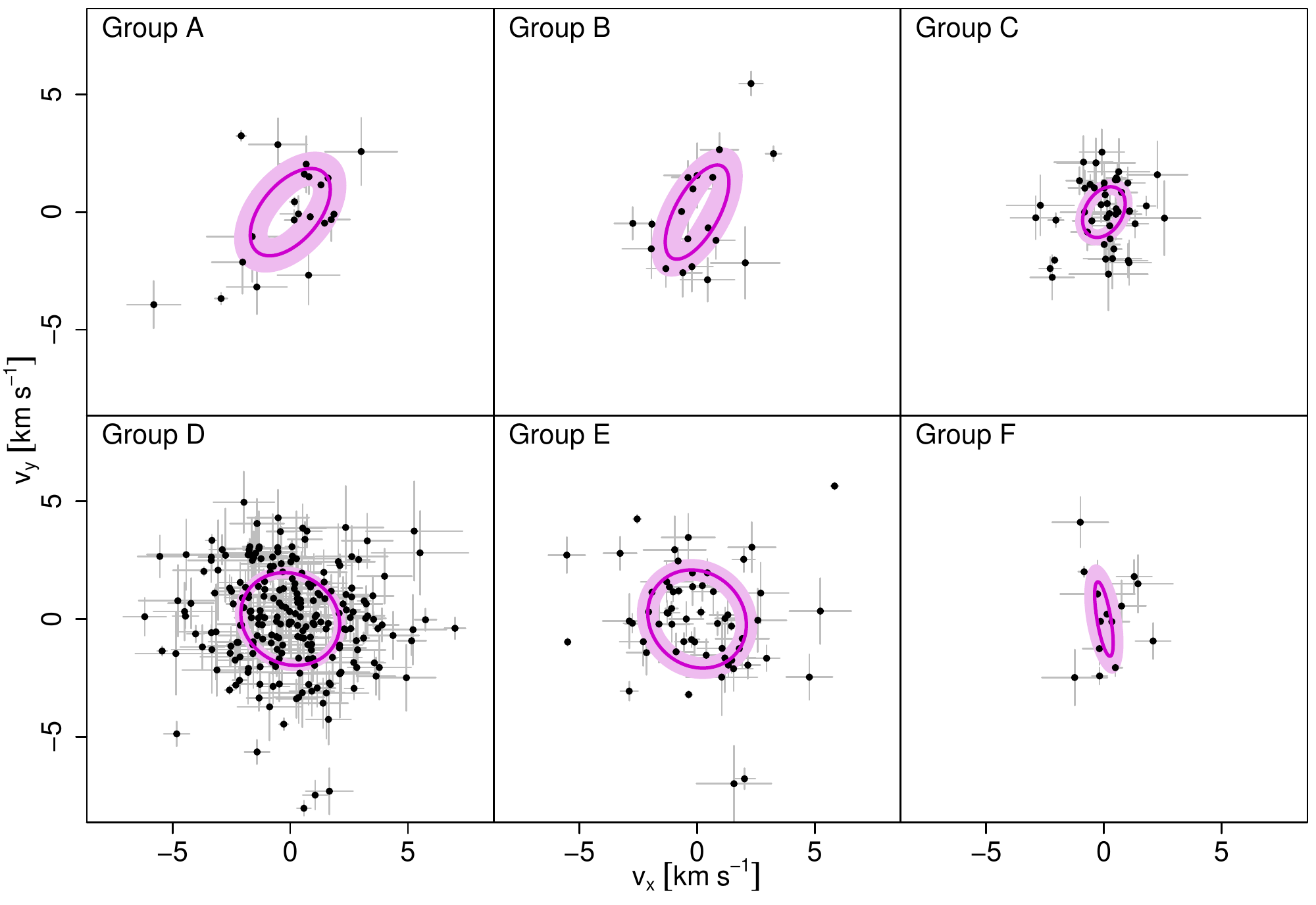} 
\caption{Scatter plots of star velocities. The magenta ellipses indicate the best-fit Gaussians; the ellipses are drawn at Mahalanobis distances of 1 and the shaded region shows the 95\% confidence intervals on the ellipses. 
 \label{vel_disp.fig}}
\end{figure*}

\begin{figure*}[t]
\centering
\includegraphics[width=0.45\textwidth]{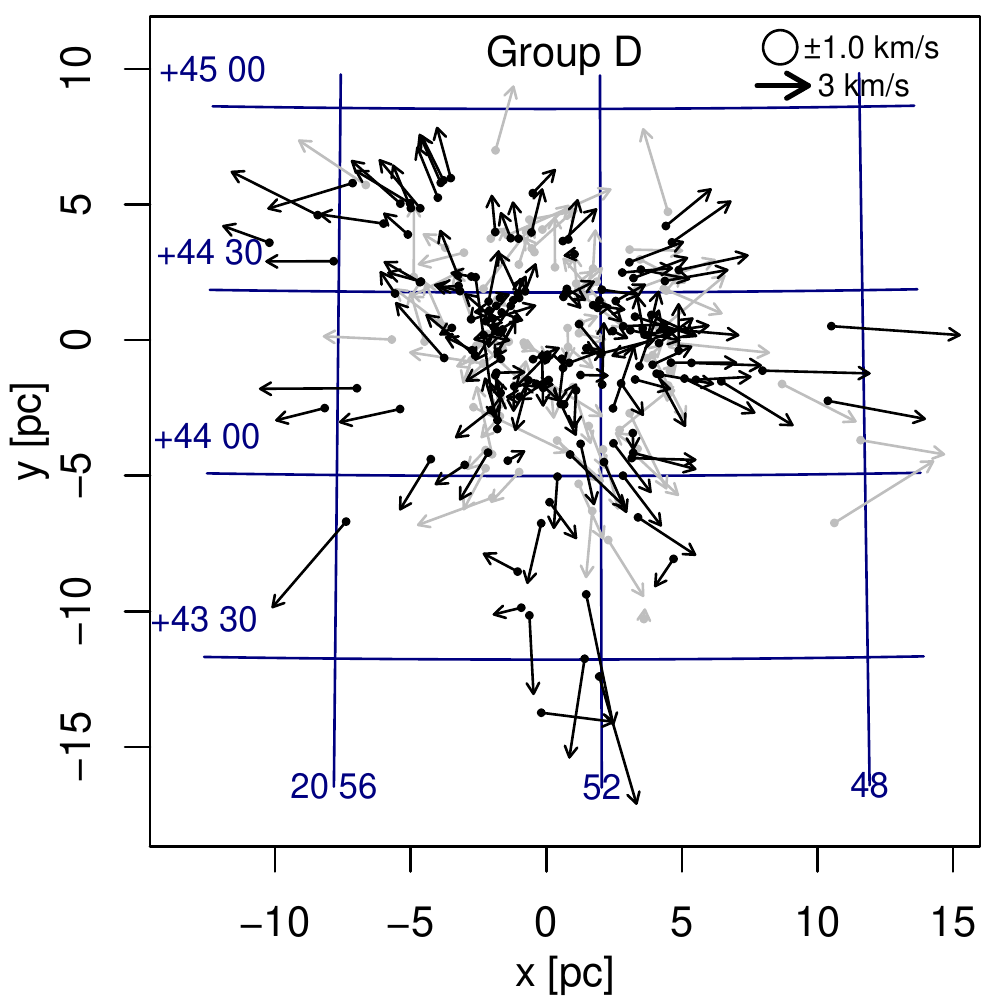} 
\includegraphics[width=0.45\textwidth]{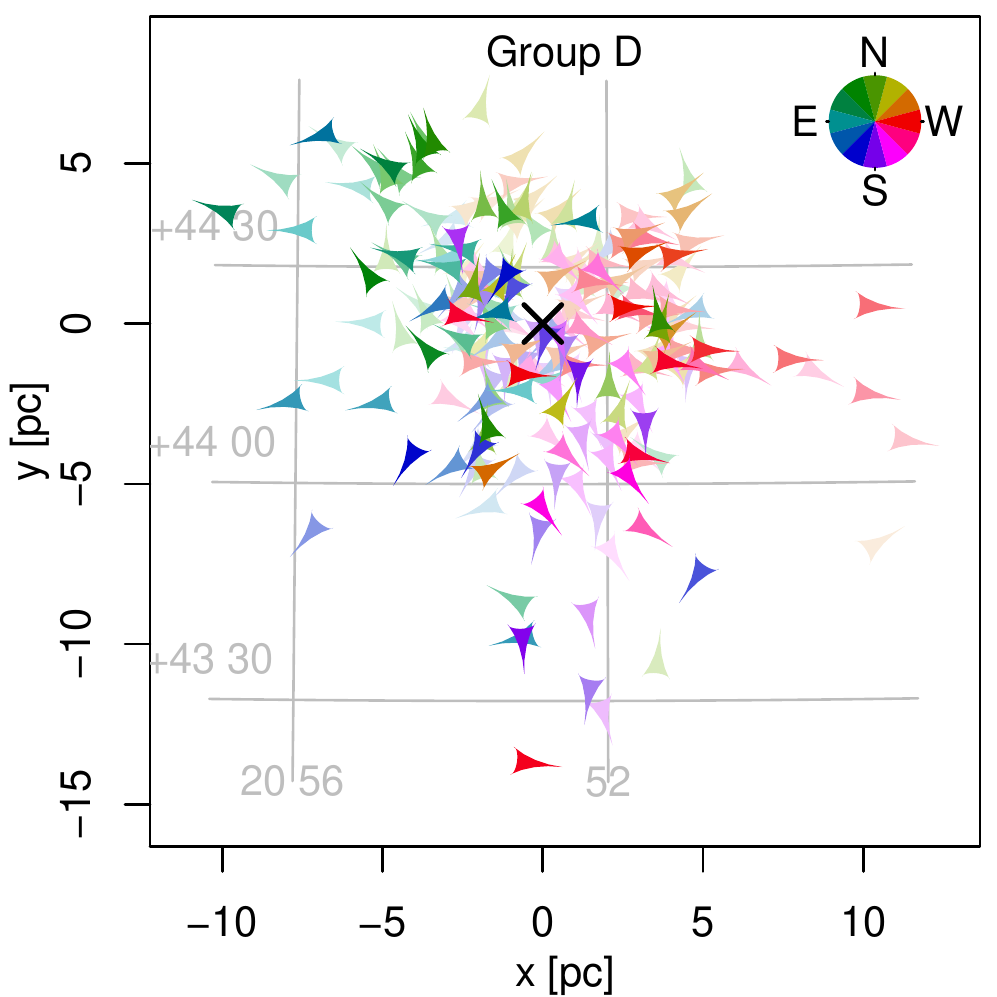}\\
\includegraphics[width=0.45\textwidth]{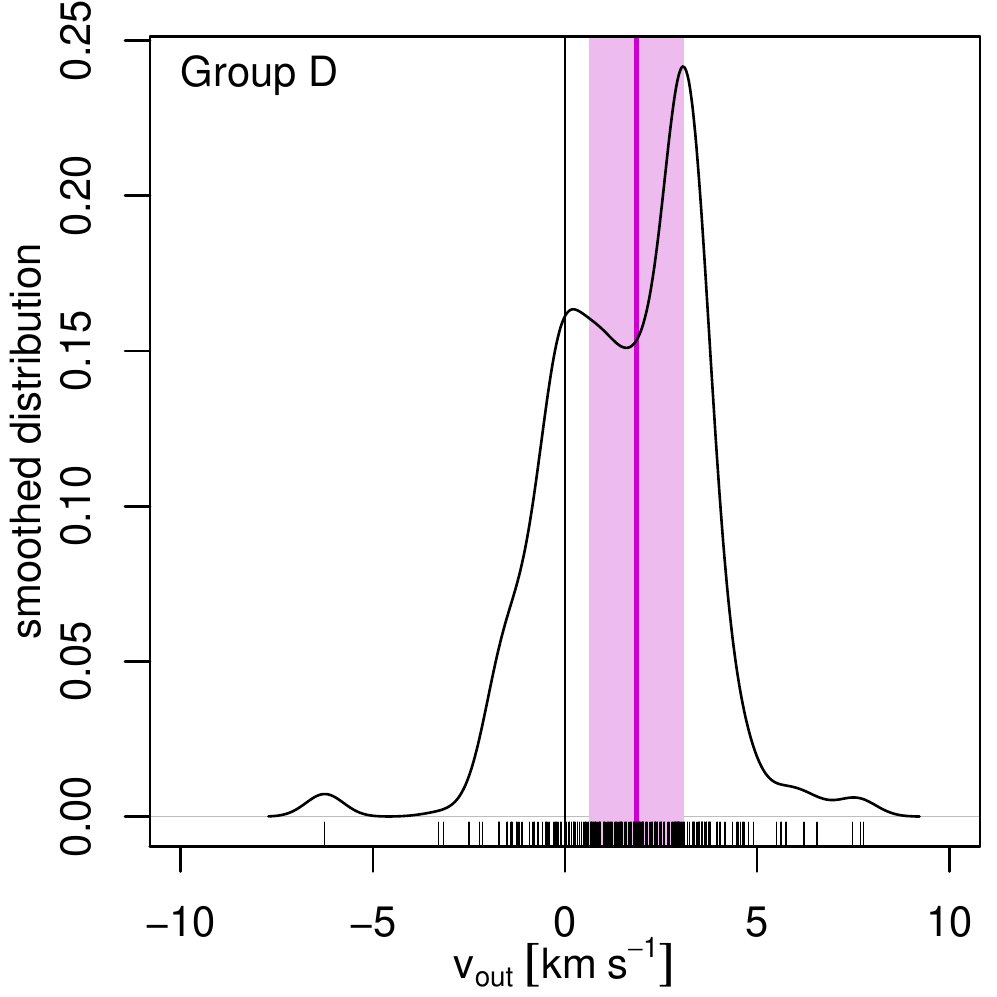} 
\includegraphics[width=0.45\textwidth]{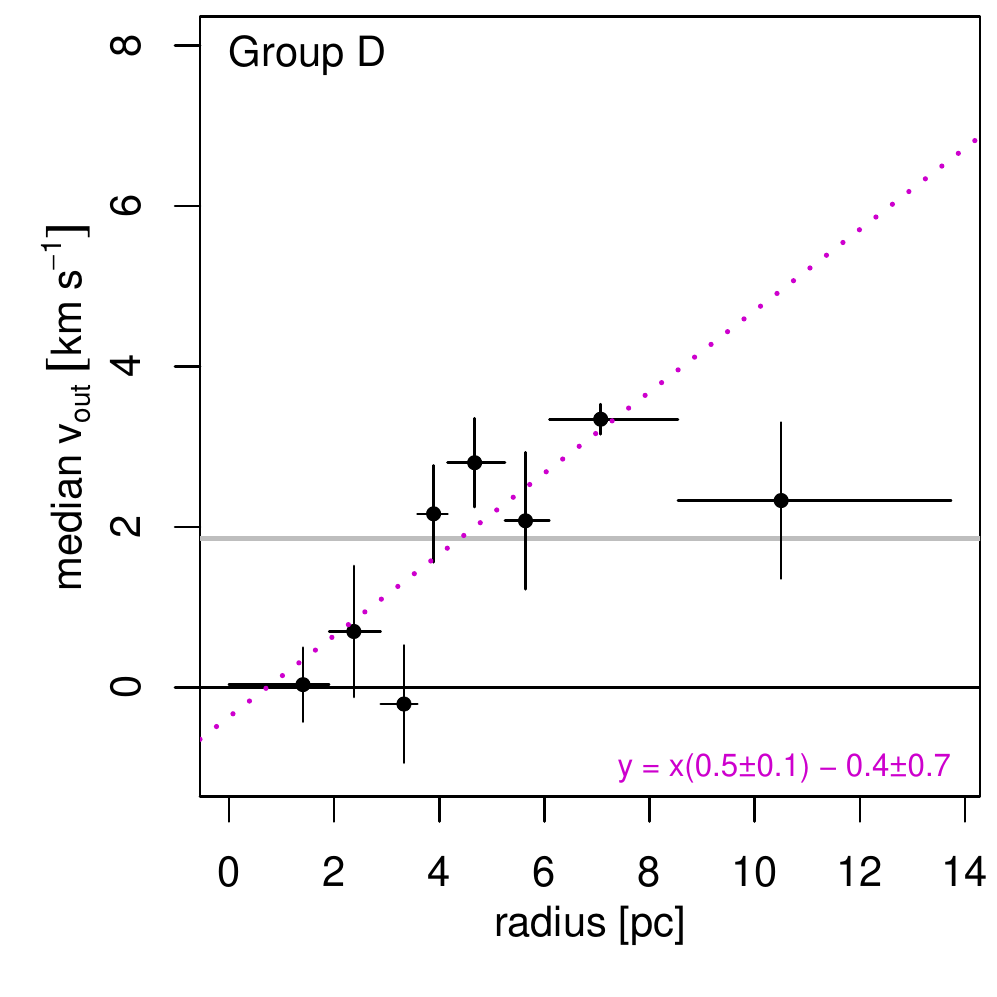} 
\caption{Plots of stellar kinematics for stars in Group D. Top left: Vectors show motions of stars in the $(x,y)$ plane relative to the center of the group. Stars with the most precise velocity estimates ($<$1~km~s$^{-1}$) are shown in black, while stars with less certain velocity estimates are shown in gray. Top right: Arrow heads are color coded by the direction of motion of a star as indicated by the color wheel. Bulk motions such as expansion, contraction, or rotation can be seen as gradients in color across the group. Bottom left: Distribution of the outward velocity component $v_\mathrm{out}$. The median value is indicated by the magenta line and the 3$\sigma$ confidence interval is shown by the shaded area. Bottom right: Median value of $v_\mathrm{out}$ as a function of distance from the center of the region. An error weighted least-squares regression fit is shown.\\
(The complete figure set (14 images) is available.)
 \label{kin1.fig}}
\end{figure*}

\begin{figure*}[t]
\centering
\includegraphics[width=0.75\textwidth]{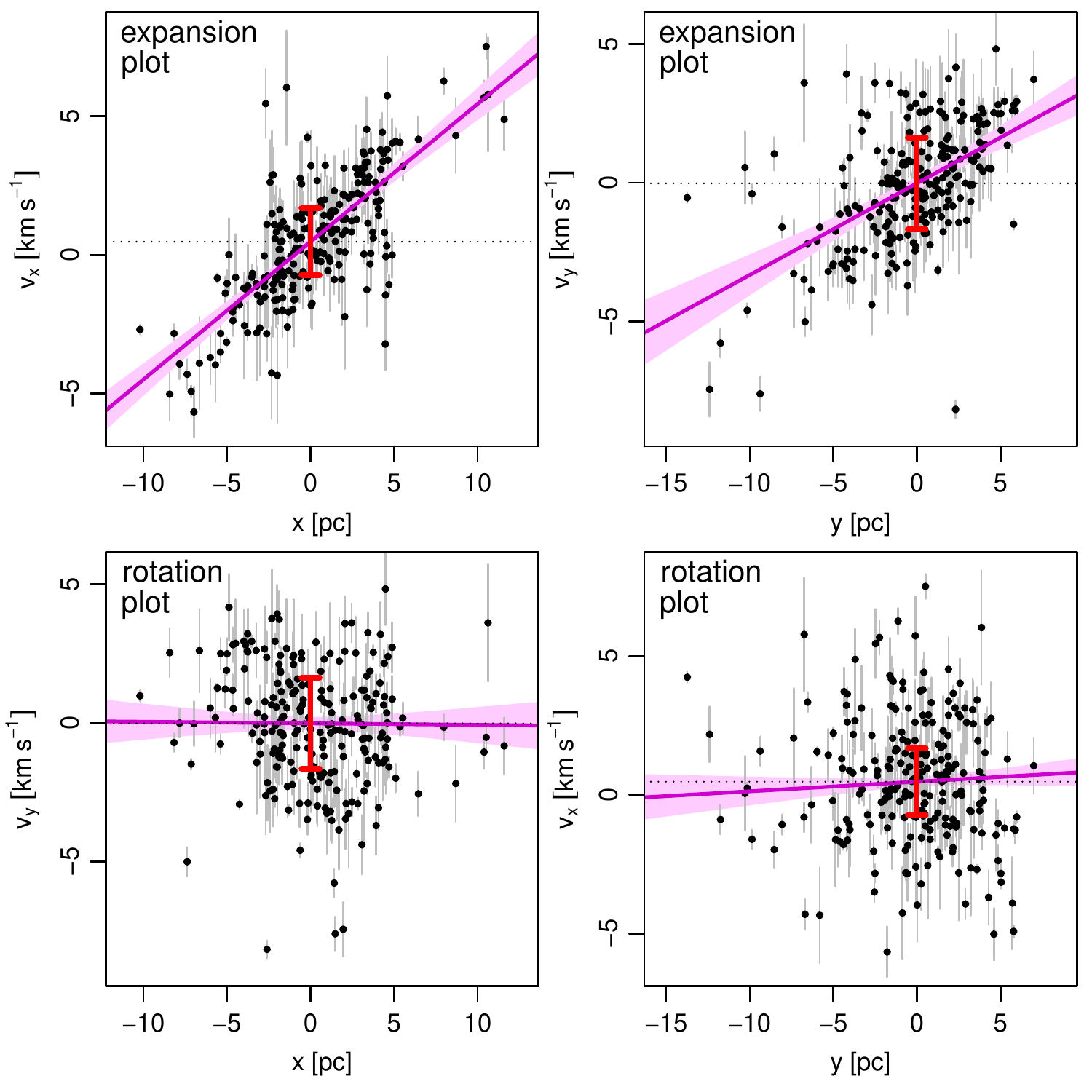} 
\caption{Plots of velocity coordinates $v_x$ and $v_y$ versus position coordinates $x$ and $y$ for stars in Group~D. The top row shows $v_x$ vs.\ $x$ and $v_y$ vs.\ $y$, so expansion/contraction would produce correlation on these diagrams. The bottom row shows the combinations $v_y$ vs.\ $x$ and $v_x$ vs.\ $y$, so correlation here would indicate rotation. We have over-plotted graphical illustrations of our MCMC model fit for comparison to the data. On the top two plots we show lines indicating the relations $v_x = v_{x,0} + xA_{11}$ (left) and $v_y = v_{y,0} + yA_{22}$ (median and 95\% credible region shown in magenta), while the bottom two plots show $v_y = v_{y,0} + xA_{21}$ (left) and $v_x = v_{x,0} + yA_{12}$ (right). The contribution from the intrinsic velocity dispersion, described by the covariance matrix $\Sigma_\mathrm{scatter}$, is indicated by the red bars.\\
(The complete figure set (3 images) is available.)
\label{regression.fig}}
\end{figure*}

\subsection{Kinematics within Groups}\label{within_groups.sec}

In \citet{2019ApJ...870...32K}, we presented several methods for characterizing stellar kinematics and testing for expansion, contraction, or rotation of a stellar system. Here, we apply these methods to each of the stellar groups.

To characterize total velocity dispersion in a group, we fit the $(v_x, v_y)$ distribution with a bivariate Gaussian, taking into account measurement uncertainties, which have the effect of artificially broadening the velocity dispersion.\footnote{The classification of sources using proper motion in Section~\ref{ysoc_identified.sec} restricts the allowable range of velocities, which truncates the tails of the velocity distributions and will slightly decrease the estimated $\sigma_{1D}$. We simulate this effect and find that it will produce a $<$8\% effect on the results for velocity dispersions $<$3~km~s$^{-1}$, which is smaller than the estimated statistical uncertainty on the derived values.} The stellar velocities and the Gaussian fit are shown in Figure~\ref{vel_disp.fig} for each of the six groups. We summarize these distributions using the quantity $\sigma_{1D}$ (Table~\ref{kinematics.tab}), defined as the square root of half the trace of the Gaussian's covariance matrix \citep[][their Equation 17]{2019ApJ...870...32K}. For the six groups, $\sigma_{1D}$ ranges from 1 to 2~km~s$^{-1}$, being smallest for Groups B and F and largest for D and E. Group F is a small group in a small cloud, so it is unsurprising that this group has a small velocity dispersion. In contrast, Groups D and E are large groups, located in a massive cloud (E) or in a partially dispersed cloud (D), so it is unsurprising that they have larger velocity dispersions. However, it is notable that Group C, spatially adjacent to Group D, has a much lower velocity dispersion. In many cases, the velocity dispersions show signs of being anisotropic, similar to the results for other star-forming regions \citep{2019ApJ...870...32K}.

We use several plots, analogous to those in \citet{2019ApJ...870...32K}, to investigate evidence for expansion. Figure~\ref{kin1.fig} shows Group~D, the largest group in our sample where expansion is most evident. 
In this group, 
the fastest moving stars, with velocities $>$5~km~s$^{-1}$ are located near the edges of the region, and they are generally directed away from the center of the group. The upper right plot uses both arrow direction and hue to indicate the direction of motion, with color saturation indicating statistical uncertainty on direction. This use of color allows patterns of motions to be identifiable as a gradient in the color of the marks, in this case showing that stars at different positions in the group are oriented in different directions, all preferentially away from the center. 

The two lower panels show statistical tests for expansion. The plot on the lower left shows the distribution of $v_{out}$ values; if a system is expanding then these will be predominantly positive. For Group~D, the weighted median $v_\mathrm{out}$ value is 1.9$\pm$0.4~km~s$^{-1}$, more than three standard deviations above 0, clearly demonstrating that the group is expanding. This expansion velocity is larger than any of those measured for the 28 systems investigated by \citet{2019ApJ...870...32K}. The plot on the lower right shows the mean expansion velocities for bins at different radii from the group center. Expansion velocity appears to increases with distance from the center, a trend that is seen in many of the expanding young star clusters and associations from  \citet{2019ApJ...870...32K}.

The other groups contain fewer stars than Group~D, and the clear evidence for expansion is not found using these diagrams. For the other groups, arrow diagrams like those in the top two panels of Figure~\ref{kin1.fig} are included in the figure set, but there are too few stars to make the other plots.  For these groups, the weighted median $v_{out}$ values are not significantly different from 0 (Table~\ref{kinematics.tab}) owing to the large statistical scatter. Nevertheless, median~$v_{out}$ may not be the most sensitive test for expansion, given that this statistic combines multiple dimensions and does not take advantage of the tendency, seen in many expanding regions, for velocity to increase with distance from the center.

We apply a further test for expansion, using the non-parametric Kendall's $\tau$ test to test for correlation between $x$ and $v_x$ and between $y$ and $v_y$ (Table~\ref{kinematics.tab}). Statistically significant results would indicate a velocity gradient, which would imply expansion if the gradient is positive and contraction if negative. We find strong statistical evidence for expansion in both dimensions for Group~D, moderate statistical evidence for expansion in both dimensions for Group~B, and strong statistical evidence for expansion in only one dimension ($x$) for Groups C and E.  

\subsection{Velocity Gradients in Expanding Groups}\label{velocity_gradients.sec}

Our results, along with the results from previous {\it Gaia} studies, suggest that stars in expanding young stellar groups often have velocities that are proportional to the distance from the groups' centers \citep[][]{2019ApJ...870...32K,2019ApJ...871L..12R,2019MNRAS.488.3406Z,2019MNRAS.486.2477W,2020MNRAS.493.2339M}. However, \citet{2019MNRAS.486.2477W} point out that the expansion may be anisotropic. In Appendix~\ref{linear_model.sec}, we describe a linear model for velocity as a function of position, using a method that can account for anisotropy and is independent of choice of coordinate system. Table~\ref{kinematics.tab} reports several of the interesting properties from the models, including velocity gradients in the directions of maximum and minimum expansion, the position angle of anisotropy, and the intrinsic scatter that remains after the velocity gradient is accounted for. 

We find statistically significant expansion for Groups~C, D, and E, but no group had statistically significant rotation. Group D has the strongest relation between position and proper motion, with expansion in all directions, but with a larger gradient in the east-west direction (0.5~km~s$^{-1}$~pc$^{-1}$) than in the north-south direction (0.3~km~s$^{-1}$~pc$^{-1}$).  In contrast, Groups~C and E had non-zero expansion gradients only in the east-west direction, but not the north-south direction. We note that the anisotropy position angles of all three groups are similar. The regression analysis and the calculation of the model parameters are described in greater detail in the appendix.

Figure~\ref{regression.fig} shows scatter plots of position versus velocity with the model regression lines overplotted. The graphs of $v_x$ vs.\ $x$ (upper left) and $v_y$ vs.\ $y$ (upper right) exhibit positive correlations if there is expansion, while $v_y$ vs.\ $x$ (lower left) and $v_x$ vs.\ $y$  (lower right) exhibit correlations if there is rotation. 
The regression line and 95\% credible interval from the linear model are shown in magenta. The intrinsic velocity scatter, applied equally to all points, is indicated by the red error bar (1$\sigma$), while the individual 1$\sigma$ measurement uncertainties are indicated by the gray lines. For Group~D, both the $v_x$ vs.\ $x$ and $v_y$ vs.\ $y$ plots show strong correlations, while for Groups C and E correlations are only strong on the $v_x$ vs.\ $x$ plots due to their $\theta\approx90^\circ$ anisotropy position angles.  

The O stars are both among the fastest moving members of Group~D. The Bajamar Star has a velocity of 6.6$\pm$0.5~km~s$^{-1}$ relative to the center of the group, while HD~199579 has a velocity of 6.4$\pm$0.4~km~s$^{-1}$. However, neither of these velocities are discrepant from the linear model. This 
suggests that whatever phenomenon accelerated these stars to their current velocities is also responsible for the expansion of the group as a whole. If we make the simplifying assumption that the velocities have been approximately constant, the Bajamar Star would have been nearest, in projection, to the center of Group~D $\sim$1.5~Myr ago, while HD~199579 would have been nearest $\sim$1.8~Myr ago. 

\subsection{Effects of Spatial Covariances in {\it Gaia} Astrometric Errors}

To ensure our measurements of velocity gradients are robust, we examine how these would be affected by the systematic correlated errors in astrometry \citep[][their Section~14]{2018arXiv180409366L}. They found that the covariance of quasar proper motion as a function of angular separation decreases from $\sim$4000~$\mu$as$^2$~yr$^{-2}$ at 0.07$^\circ$ separation to a local minimum of $\sim$40~$\mu$as$^2$~yr$^{-2}$ at 0.43$^\circ$ separation. This could induce an artificial proper motion gradient of $\sim$0.17~mas~yr$^{-1}$~deg$^{-1}$, which would look like a velocity gradient of 0.045~km~s$^{-1}$~pc$^{-1}$ in a stellar association at a distance of 795~pc. Given that the velocity gradients that we detect are significantly larger than this value, they are most likely astrophysical. 

\section{New Member Candidates from {\it Gaia}}\label{gaia_identified.sec}

In addition to validating previously proposed YSOs in the NAP region, {\it Gaia} astrometry can be used to suggest new member candidates. In most earlier studies, candidates were identified based on criteria that required a star to possess a circumstellar disk (e.g., strong H$\alpha$ emission, infrared excess, large amplitude variability). However, many star-forming regions are dominated by pre--main-sequence stars for which disks are not detected \citep{2013ApJS..209...32B}. Thus, astrometric selection of candidates can reveal whether any populations of NAP members have remained undetected in previous analysis, which, in addition to their possible biases, are also incomplete in their selection of candidate members.

We start with a sample of all {\it Gaia} DR2 sources within 3$^\circ$ of the Bajamar Star that pass the same quality criteria from Section~\ref{gaia_data.sec}, and lie within the same parallax range and the $(\mu_{\ell^\star},\mu_b)$ region identified in Section~\ref{ysoc_identified.sec}. However, we apply further parallax cuts to select only objects that are consistent within 2 standard deviations of the median parallax of the groups, i.e.\ $\varpi + 2\,\sigma_\varpi \geq 1.06$~mas and  $\varpi - 2\,\sigma_\varpi \leq 1.24$~mas. These criteria greatly reduce the number of {\it Gaia} sources in the region, from $\sim$75,000 {\it Gaia} sources in the original parallax range down to $\sim$10,000 sources. 

\begin{figure*}[t]
\centering
\includegraphics[width=1\textwidth]{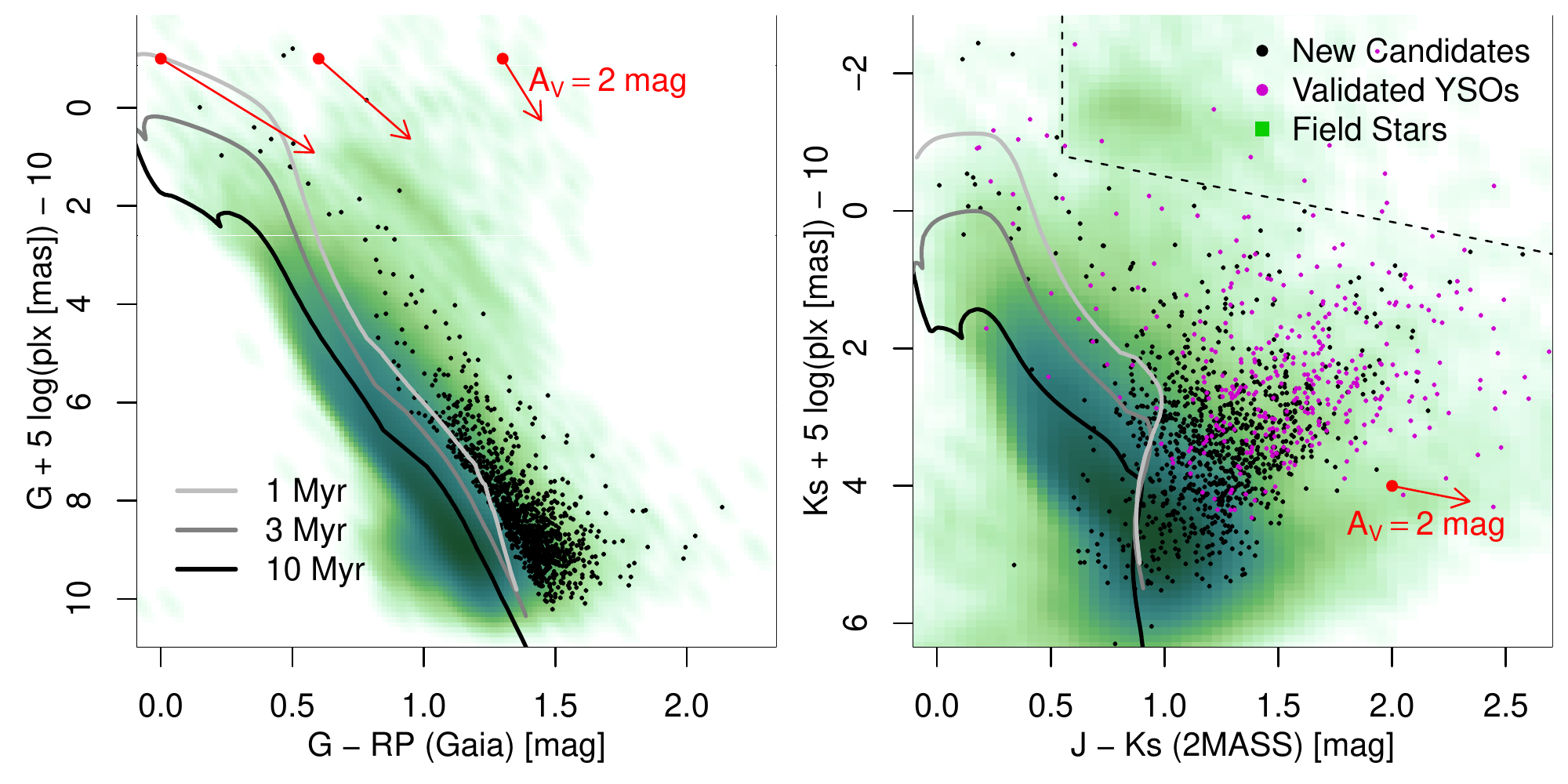} 
\caption{Color magnitude diagrams in the {\it Gaia} bands (left) and near-infrared (right) showing how the distributions of new candidates NAP members (black points) compare to other categories of sources. The green shaded regions show the density of {\it Gaia} sources with the same parallax range, with darker colors indicating higher number densities of sources. On both plots, field stars in the red clump and the giant branch can be seen  near the top of the diagram. We use cuts on the near-infrared diagram (indicated by dotted lines) to remove sources from our candidate sample that have a high probability of being giant stars. On the near-infrared diagram, we also show YSOs from our {\it Gaia} validated literature sample (magenta points). The distribution of these points is shifted to slightly higher $K_s$ band luminosities and redder $J-K_s$ colors. This may be the result of earlier studies selecting stars with disks, and thus more likely to have $K_s$-band excess emission, while the {\it Gaia} selection is independent of whether the object has a disk. 
 \label{new_candidates_cmd.fig}}
\end{figure*}

\begin{figure*}[t]
\centering
\includegraphics[width=1\textwidth]{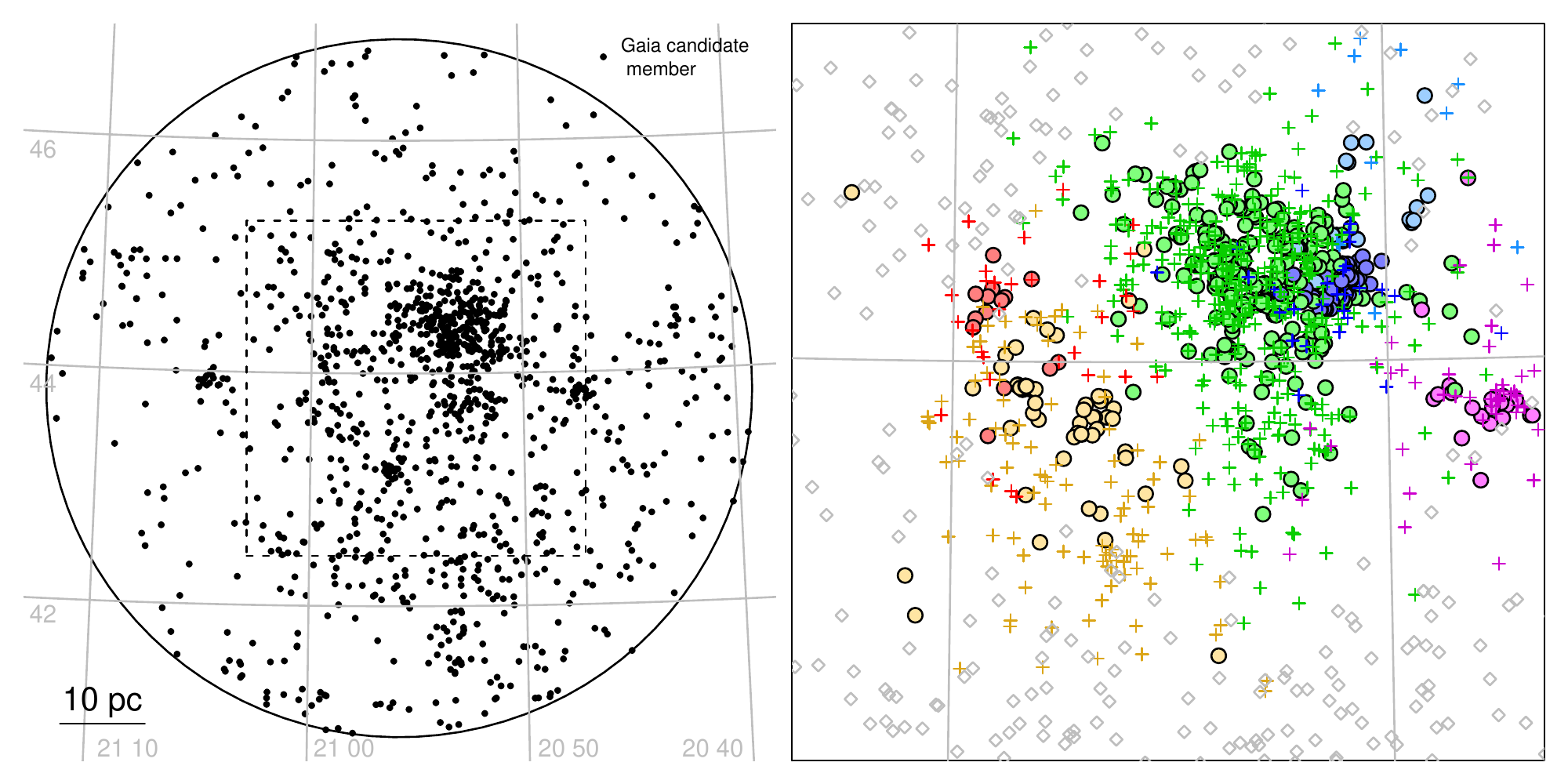} 
\caption{Spatial distributions of new candidate members. Left: All new candidates within 3$^{\circ}$ of the Bajamar Star. The dashed box outlines the blown-up central region shown in the other panel. Right: New candidates are shown as +'s if they are assigned to existing groups or $\diamond$'s otherwise, while the validated YSOs from previous studies are shown as circles. Color-coding of group members is the same as in previous figures. This plot shows that most new candidates in the central region are distributed similarly to the distribution of the previously identified stars. 
 \label{new_candidates_spatial.fig}}
\end{figure*}

We further refine the sample to ensure that the selected objects are likely to be young, using an age limit of 3~Myr that matches the approximate maximum age found for members in Section~\ref{ysoc_identified.sec}. We identify {\it Gaia} sources whose absolute $G$-band magnitude and $G-RP$ color are two standard deviations above the 3~Myr \citet{2012MNRAS.427..127B} isochrone with $A_V=1$~mag (Figure~\ref{new_candidates_cmd.fig}, left). However, remaining contaminants can include main-sequence stars with high extinction as well as post-main-sequence stars. To reduce contamination from reddened main-sequence stars, we use the 2MASS $J-H$ vs.\ $H-K_s$ diagram to estimate extinction using the reddening laws from \citet{1985ApJ...288..618R}. Typical reddening is $\Delta(J-H) \sim 0.3$~mag ($\approx 2.8$~mag in the $V$ band), but ranges from 0 to 0.5~mag, with a tail out to 1.5~mag. Dereddening on the near-infrared color-color diagram can lead to degeneracies between low-mass and intermediate/high-mass stars. In most cases we pick the low-mass solution. However, when a star could be low-mass with $\Delta(J-H) < 0.3$~mag or intermediate mass with  $\Delta(J-H) \geq 0.3$~mag, we pick the higher extinction to avoid classifying reddened intermediate mass background field stars as pre--main-sequence members. Once we have estimated extinction, we require the star to still lie above the 3~Myr {\it Gaia} isochrone with the corresponding reddening. To reduce contamination by giants, we selectively remove objects from the region of 2MASS color magnitude diagram where these objects are likely to dominate the sample: $J-K_s > 0.5$~mag and $M_{K_s} < 0.66 \times (J-K_s) - 0.8$~mag (Figure~\ref{new_candidates_cmd.fig}, right).  This removes objects in the red clump as well as the upper giant branch. Finally, in addition to pre--main-sequence stars, we also include 5 additional stars of spectral type B as compiled by \citet{2009yCat....1.2023S}. The final sample contains 1,187 new candidates (Table~\ref{gaia_candidates.tab}).

The spatial distribution of the new candidates is shown in Figure~\ref{new_candidates_spatial.fig}. Candidates are spatially clustered in the NAP region as expected; however several small clumps were not previously identified, as described below. There is also a distributed population of objects at angular separations of several degrees from the cluster (left panel). In position vs.\ proper-motion space (not shown) the new candidates follow similar structures to those seen in Figures~\ref{pos_vs_pm.fig}--\ref{pm_vs_pm.fig} but with more points scattered between the groups. Many of the distributed objects may be residual contaminants, which would be expected to be distributed approximately uniformly across the field. However, more data (e.g., spectroscopic followup observations) is necessary to determine the rate of contamination. 

To assign candidates to groups A--F as well as a ``distributed'' category, we used the $k$-nearest neighbor ($k$NN) algorithm implemented in the R package {\tt class} \citep{MASS}. For the training examples, we used the 395 previously assigned members of groups A--F in addition to an equal number of randomly selected ``distributed'' stars from {\it Gaia} in the same parallax and proper motion range that were not selected as new candidates.  The $k$NN classifier is run using the normalized $\ell$, $b$, $\mu_{\ell^\star}$, and $\mu_b$ variables, and classifications are based on the 5 nearest neighbors. We assign 52\% of the candidates to groups and 48\% to the ``distributed'' category. 

A previously unidentified group of $\sim$17 new members is centered at 21:04:44 +43:55:30. We call these stars Group~G and list them in Table~\ref{gaia_candidates.tab}. This group is located next to the third-magnitude red-supergiant star $\xi$~Cyg. However, this star has a parallax of $\sim$3.9~mas, meaning that it is much closer, and therefore unrelated to the group. Nevertheless, the angular proximity to such a bright star may explain why the group was not previously identified. 

Another small clump of previously unidentified stars lies just to the south of Group~E. These stars continue the position--proper motion trends seen for the stars in Group~E, so we classify the new stars as a continuation of this group. 

\section{Discussion}\label{discussion.sec}

Observations of the NAP complex reveal that it contains multiple groups of very young stars ($\sim$1~Myr) that are associated with distinct components of the molecular gas. This clumpy distribution of mass covers a region at least 30~pc in diameter, and {\it Gaia} parallaxes suggest that there are several conglomerations along the line of sight spanning up to 150--200~pc. The motions of the groups, as inferred from {\it Gaia} astrometry, appear to be approximately random relative to one another, and several of the groups are rapidly expanding.   

\subsection{Expanding Stellar Groups}

Several mechanisms, including disruption by tidal forces, cloud dispersal, and sequential star formation, may play roles in the expansion of stellar groups in the NAP complex.

A clumpy mass distribution, as observed here, will generate internal tidal forces that affect stellar dynamics and influence the potential assembly of either bound clusters or unbound associations. Tidal fields would naturally produce an anisotropic velocity gradient similar to those observed in the NAP region. Furthermore, tidal forces would preserve clumpy substructure within a group (e.g., D and E) as it is stretched, while processes driven by dynamical relaxation would be more likely to erase these substructures. Tidal forces from giant molecular clouds have long been known to be a major cause of star-cluster disruption \citep{1958ApJ...127...17S,2006MNRAS.371..793G}, and these effects can occur in the star-forming environment itself, inhibiting the initial formation of a bound system \citep{2011MNRAS.414.1339K,2012MNRAS.419..841K}. \citet{2019MNRAS.488.3406Z} have run numerical simulations in which gas clouds, driven away from the newly formed young stellar cluster by feedback, exert tidal forces on the cluster that pull it apart. 

The tidal acceleration $a_{t,\mathrm{axial}}$ per unit of displacement $\Delta r$ along the separation axis due to a mass $M$ at distance $r$ is
\begin{equation}
a_{t,\mathrm{axial}} / \Delta r= 2 G \frac{M}{r^3}.
\end{equation}
For Group~D, the magnitude of the tidal acceleration from Cloud~2 ($2.1\times10^3$~$M_\odot$) located $\sim$5~pc to the west of the center of the group (assuming that the displacement in the third dimension is small) is 0.15~km~s$^{-1}$~pc$^{-1}$~Myr$^{-1}$. This acceleration would be enough to induce the observed $\sim$0.5~km~s$^{-1}$~pc$^{-1}$ east--west velocity gradient in $\sim$3~Myr. Other nearby clouds, notably Clouds~3, 4, and 8 (a potential remnant of Group~D's natal cloud) could also contribute to the tidal forces on this group. These clouds would also exert tidal forces on the other nearby, expanding group, C, located at the edge of Cloud~2. However, precisely calculating the effect of tidal forces is challenging due to uncertainties in the third dimension,  velocities differences between the groups and clouds, and systematic uncertainties in cloud mass. 

Expansion of Group~E is more difficult to explain with tides, since its separation from other parts of the NAP region by $\sim$35~pc along the line of sight means that it would much less strongly affected by tidal forces from other clouds. Thus, any tidal force would need to be generated by clumpy distributions of mass within Cloud~9 itself.

The ongoing dispersal of the molecular clouds due to O star winds and photoevaporation, combined with outflows from low-mass stars \citep{2003AJ....126..893B,2014AJ....148..120B}, will weaken the binding energy of the stellar groups, causing them to expand and/or disperse \citep[e.g.,][and many others]{1978A&A....70...57T,2000ApJ...542..964A,2001MNRAS.321..699K}. In the northwest, the shell-like structure may be the outer rim of a bubble blown by the Bajamar Star, which would have been closer to this part of the cloud $\sim$1.5~Myr ago. Cloud dispersal in this region may have affected the stellar groups that lie near the shell, particularly Groups C and D. In contrast, photoevaporation does not appear to be as significant in the Gulf of Mexico region that contains Group~E. Nevertheless, \citet{2014AJ....148..120B} found dozens of outflows here which are reprocessing a significant volume of the Gulf of Mexico cloud through supersonic shocks. Such outflows may dominate the injection of energy and momentum in the absence of massive stars \citep{2006ApJ...640L.187L,2011ApJ...740...36N}, and may contribute to the cloud's mass loss \citep{2007prpl.conf..245A}.

We examine the scenario in which Group D formed as an embedded cluster in virial equilibrium in a molecular cloud that was subsequently dispersed. Given Group~D's measured velocity dispersion of $\sigma_{1D}=2$~km~s$^{-1}$ and assuming an initial size of $\sim$1~pc, the corresponding virial mass of the system would have been $\sim\!\!9\times10^3$~$M_\odot$. Thus, the natal cloud mass would have been $\sim$6 times greater than the estimated mass of Cloud~8, and $\sim$1.5 times the sum of the masses of all the clouds in the northwestern part of the NAP complex. \citet{2019MNRAS.489.2694W} have demonstrated that expansion in this scenario could be anisotropic if it were preceded by collapse and bounce in an asymmetric gravitational potential, setting up an anisotropic velocity dispersion.

A third explanation for expanding groups is that they are produced by a second generation of star formation in an expanding shell of material. In regions where star formation has been attributed to material swept up in shells around expanding H\,{\sc ii} regions \citep[e.g.,][]{1998ApJ...507..241P} the stars formed in this way would presumably inherit the velocities of the material in the expanding shells. In the NAP region, there is ongoing star formation in several of the clouds making up the shell-structure, so it is plausible that such a scenario could be happening here too. In particular, the Group~C is associated with the bright-rim cloud at the edge of Cloud~2. For several bright rim clouds in other star-forming regions, \citet{2007ApJ...654..316G,2012MNRAS.426.2917G} have shown that the stars in front of the cloud exhibit an age gradient, with the youngest stars nearest the cloud. In such a scenario, as a cloud is driven outward by an expanding shell, the sequential formation of stars could yield a velocity gradient.  

Given the conditions in the NAP region, it seems likely that multiple scenarios operate simultaneously to induce the observed expansion patterns. For example, the lessening of the binding energy due to cloud dispersal would make groups more susceptible to disruption by interactions with neighboring clouds. Although theoretical work has shown that binary-star dynamics can preferentially eject O stars at high velocities \citep[$>$30 km~s$^{-1}$;][]{2016A&A...590A.107O}, this mechanism does not appear to be required here because the O stars and low-mass stars follow the same velocity pattern.

The {\it Gaia} data show that individual groups are expanding, but the relative motions of the groups exhibit no clear pattern of either convergence or divergence (Figure~\ref{global_velocities.fig}, upper left). This is similar to the situation found by \citet{2019ApJ...870...32K} in other large star-forming complexes like NGC~2264, NGC~6357, or the Carina Nebula, and may help explain why expansion has been difficult to detect in young stellar associations analyzed as a whole \citep[e.g.,][]{2018MNRAS.475.5659W,2016MNRAS.460.2593W,2018MNRAS.476..381W}. The velocity dispersion of all stars in the NAP complex is $\sigma_{1D} = 2.5$~km~s$^{-1}$ (Figure~\ref{global_velocities.fig}, lower panels). This, combined with the separations of several to tens of parsecs between groups, means that the mass needed to gravitationally bind the system ($>$10$^5$~$M_\odot$) is orders of magnitude greater than the mass in stars. Thus, it is impossible for the system to coalesce as a bound cluster. Nevertheless, it may be possible in some groups for bound remnants to remain behind even after most stars have escaped \citep{2007MNRAS.380.1589B}.

\subsection{Star-Formation Scenarios}

We can use our inferences about the dynamical state of the stars and gas, as well as the constraints on stellar ages, to test various theoretical scenarios for star formation as applied to the NAP region. These scenarios broadly break down into fast star formation, on the timescale of the free-fall collapse of the molecular clouds \citep[e.g.,][]{2000ApJ...530..277E,2012MNRAS.420.1457H}, and slow star-formation that persists over multiple free-fall timescales \citep[e.g.,][]{2006ApJ...641L.121T,2020MNRAS.tmp..619K}. In the former case, molecular clouds would have little pressure support \citep{2019MNRAS.490.3061V}, while in the latter case turbulent pressure support stabilizes the clouds against collapse and allows for an extended duration of star formation \citep{2011ApJ...730...40P}.

In the NAP region, we demonstrated that the velocity dispersion within individual clouds is consistent with the velocity dispersion that would be expected from either free-fall collapse or virial equilibrium given the systematic uncertainties on the cloud masses (Section~\ref{cloud_dynamics.sec}). When we consider the whole region (Figure~\ref{global_velocities.fig}), the combined velocity dispersion of all gas is larger than that of the individual clouds, with $\sigma_\mathrm{all} = 4.2$~km~s$^{-1}$. In the complex, half the gas is contained within a projected radius of 9.4~pc, so the dynamical mass for the whole system calculated using  Equation~\ref{dynamical_mass.eqn} would be $1.9\times10^5$~$M_\odot$, or about two and a half times the $7.2\times10^4$~$M_\odot$ mass estimated from $^{13}$CO column density. Estimation depends on the virial parameter $\alpha$, which we take to be 1 (virial equilibrium); however, a cloud with $\alpha=2$ (free fall) would yield a dynamical mass estimate half as large, making the differences smaller. As earlier, the systematic uncertainties make distinguishing between these scenarios difficult. Nevertheless, in either case, the velocity differences between the clouds, as well as the velocity dispersions within the clouds, can be accounted for by gravitation. 

The free-fall timescales for the clouds are $\sim$1~Myr (Table~\ref{cloud_dynamics.tab}). This is similar to the $\sim$1~Myr ages of the stars (Figure~\ref{ysoc_cmd.fig}). This implies that, in each cloud, most star formation occurred within the last 1--$2\times\tau_\mathrm{ff}$. There are two considerations that should be taken into account when interpreting free-fall timescales. First, free-fall timescales are typically calculated using the mean density of a cloud as we have done; \ a non uniform density could change the mean $\tau_\mathrm{ff}$, but the correction factor is expected to be of order unity \citep{2020MNRAS.tmp..619K}. Second, we are calculating the present day free-fall timescales; however, in the past, when the stars observed today were forming, the free-fall timescale would likely have been longer \citep{2019MNRAS.490.3061V}. Star cluster formation in a free-fall time is expected for either clouds without turbulent support \citep{1998ApJ...501L.205K} or with decaying turbulence \citep{2000ApJ...535..869K}. Thus, the observation that most of the stars are only one to several $\tau_\mathrm{ff}$ old is consistent with the rapid star-formation scenarios.

The spatial clustering of star formation also favors a scenario of rapid star formation in a turbulent molecular cloud. Overall, the NAP cloud complex has features expected for a giant molecular cloud sculpted by turbulence, including a clumpy, hierarchical density structure and velocity structure that yields a larger velocity dispersion for the whole system than within the individual clouds, broadly consistent with the \citet{1981MNRAS.194..809L} relation. Given that the major stellar groups follow the present-day cloud structure, this supports the view that the stars froze out on a timescale less than the crossing time for the complex \citep{2000ApJ...530..277E}.

The similarity in age ($\sim$1~Myr) of widely separated stellar groups
raises the question of how star formation in the different clouds of the NAP complex became synchronized. 
Although many of the surveys preferentially selected stars stars that are young enough to still possess disks, even the X-ray selected sample in the NAP region appears to follow the same age distribution (Appendix~\ref{ages.appendix}).
The sound crossing time for the complex is $t_s = 9.4\,\mathrm{pc}/c_s = 40$~Myr, meaning that the clouds would not be in thermal contact with each other on the collapse timescale, and we would not expect them to have similarly aged populations. 
A similar problem has been posed by \citet{1999AJ....117.2381P} for the Upper Scorpius OB association and by \citet{2019ApJ...878..111H} for the Serpens molecular clouds.  A possible solution to this is an accelerating star-formation rate, which is a feature of several models, including the conveyor belt model \citep{2014prpl.conf..291L} and global hierarchical collapse \citep{2017MNRAS.467.1313V,2019MNRAS.490.3061V}. With sufficiently high acceleration, most of the stars in star-forming clouds would have formed very recently, even if the first stars to form in those clouds formed at different times.  

Although the NAP clouds appears to be forming stars on free-fall timescales, fast star-formation scenarios face challenges if applied to the entire Galaxy because they would imply a Galactic star-formation rate much higher than observed \citep{1974ApJ...192L.149Z,2006ApJ...653..361K}. To overcome this, theoretical fast star-formation scenarios invoke the quick disruption of the molecular clouds by stellar feedback to keep star-formation inefficient and regulate the Galactic star-formation rate \citep{2020arXiv200406113C}. In the NAP region, where the disruptive influences of the Bajamar Star and outflows from low-mass stars are already apparent, it appears likely that feedback processes could bring star formation to an end before a substantial fraction of the cloud mass is converted into stars.

\section{Summary}\label{conclusion.sec}

In this study we have examined the kinematics of stars and gas in the NAP region using previously published candidate YSOs, {\it Gaia} astrometry, and a newly presented molecular gas map. Our main results are the following. 
\begin{itemize}
\item On the basis of {\it Gaia} parallax and proper motions, we identify a sample of 395 stars as high-probability members of the complex (estimated 3\% residual contamination). 
We also reclassify hundreds of previously cataloged candidates as contaminants. In addition, $\sim$2000 candidates remain ambiguous on the basis of their kinematics alone, though the majority show both spatial clustering with the identified kinematic groups (Figure~\ref{spatial_contaminants.fig}) and traditional signatures of the activity associated with stellar youth.
\item The locations of the confirmed members on the {\it Gaia} color-magnitude diagram suggests that nearly all stars are $<$3~Myr old and that most of them are $\sim$1~Myr old or younger. 
\item Most of the widely dispersed YSO candidates from previous studies are identified as contaminants, while the confirmed members tend to be more tightly clustered. We identify 6 groups using unsupervised cluster analysis in position and proper motion. 
\item The NAP region is $\sim$795~pc from the Sun. However, parallax distributions show slight differences in distances to the individual kinematic groups. Notably, the stars in the ``Gulf of Mexico'' region appear to be $\sim$35~pc closer than the rest of the system, and a small group, containing the famous V1057~Cyg star, is $\sim$130~pc farther away.
\item The locations of each of the stellar groups are spatially correlated with the main components of the molecular cloud. This implies that all parts of the NAP cloud complex, if sufficiently massive, are actively forming stars. 
\item Most of the clouds have complex, multimodal velocity structures. We use the $^{13}$CO map to estimate various cloud properties, including masses and velocity dispersions. We find high correlation between the masses estimated from integrated $^{13}$CO column densities to be in remarkable agreement with dynamical mass estimates, suggesting a strong connection between gravitation and velocity. Mean free-fall times for individual clouds are $\lesssim$1.5~Myr.
\item Relative velocities of the different groups appear randomly oriented, showing no sign of either global expansion or contraction. The radial velocity of the ``Gulf of Mexico'' region implies that it is plunging inward. On the other hand, in the northwest of the NAP complex (``Atlantic'' and ``Pelican'' regions) where the morphology of the CO gas suggests an expanding H\,{\sc ii} region, the relative motions of the stellar groups seem unaffected. 
\item In contrast to the lack of global expansion, several stellar groups are individually rapidly expanding, with velocity gradients of 0.3--0.5~km~s$^{-1}$~pc$^{-1}$. The expansion gradients are anisotropic, and we argue that these could be, in part, attributed to tidal forces from within the clumpy molecular cloud complex.
\item The primary ionizing source in the region, the early-O Bajamar Star, lies between two groups. We suggest that it likely originated as part of a group (which we call Group~D) centered in the ``Atlantic'' part of the NAP region because its trajectory would trace back to this group.  The star's velocity ($\sim$6~km~s$^{-1}$) is consistent with the expansion velocity seen for low mass stars of the same group. Another O-type star, HD~199579, appears to be ejected in a different direction from the same group. 
\item We identify $>$1000 new candidate members of the NAP region from the {\it Gaia} catalog, including a new seventh stellar group (Group~G) located east of the complex. Slightly over half these new candidates are associated with Groups~A--G, while the others are more broadly distributed throughout the 6$^\circ$-diameter selection area of investigation. These objects would require follow-up observations to validate them.  
\end{itemize}

Several lines of evidence suggest a scenario of rapid star formation in a free-fall time \citep[e.g.,][]{2000ApJ...530..277E,2012MNRAS.420.1457H,2019MNRAS.490.3061V} in the NAP complex. The structure of the stellar groups is spatially correlated with the structure of molecular clouds, stellar ages are similar to the free-fall timescales of the star-forming clouds, and the gas velocity dispersions in the clouds are consistent with the velocities expected from gravitational collapse. Furthermore, an accelerating star-forming rate, as predicted by some of the models, would mean that regardless of how long stars have been forming, the majority of stars would have been formed during the last few free-fall timescales. This can explain why several distinct clouds in the NAP region that are sufficiently far apart to be out of direct thermal contact could all have very young stellar populations of nearly the same age. 

Nevertheless, some important properties of the star forming region remain relatively poorly determined. For example, the census of stars in the region is still highly incomplete even after our study, making it difficult to constrain the total stellar mass. While, our sample of 395 {\it Gaia}-validated YSOs is useful for understanding the spatial and kinematic distributions of stars in this region, it is only representative of the optically brightest tail of the full stellar population. 

\appendix

\section{Parameters of the NAP Membership Classifier}\label{classifier.appendix}

Before applying the final step of the NAP membership classifier (i.e.\ classification using proper motions, described by the equations below) we first remove sources that either do not meet our {\it Gaia} quality criteria or have parallaxes outside the range 0.86--1.61~mas (Section~\ref{ysoc_identified.sec}). For YSO candidates that pass these steps, membership probability, $p_\mathrm{mem}(\mu_{\ell^\star},\mu_b)$ is calculated using the Gaussian mixture model (Figure~\ref{selection.fig}, right panel) with the following parameters:
\begin{eqnarray}
d_{\mathrm{mem}}(\boldsymbol{\mu}_i) &=& 0.64\,\phi(\boldsymbol{\mu}_i;\boldsymbol{\mu}_1,\Sigma_1)\\
d_{\mathrm{field}}(\boldsymbol{\mu}_i) &=& 0.36\,\phi(\boldsymbol{\mu}_i;\boldsymbol{\mu}_2,\Sigma_2)\\
p_\mathrm{mem}(\boldsymbol{\mu}_i) &=& d_{\mathrm{mem}}(\boldsymbol{\mu}_i) / \left[d_{\mathrm{mem}}(\boldsymbol{\mu}_i) + d_{\mathrm{field}}(\boldsymbol{\mu}_i)\right]\\
\boldsymbol{\mu}_1 &=& (-3.35,-1.15)\\
\Sigma_1 &=&   \left[ {\begin{array}{cc}
   0.48 ~& 0.06 \\
   0.06 ~& 0.52 \\
  \end{array} } \right]
\\
\boldsymbol{\mu}_2 &=& (-3.52,-1.79)\\
\Sigma_2 &=&  \left[ {\begin{array}{cc}
   78.5 ~& -0.5 \\
   -0.5 ~& 19.1 \\
  \end{array} } \right],
\end{eqnarray}
where the $d$'s represents the density of stars in proper-motion parameter space,  $\phi$ denotes the bivariate normal distribution, $\boldsymbol{\mu} = (\mu_{\ell^\star}, \mu_b)$ are proper motions in Galactic coordinates in units of mas~yr$^{-1}$, and $\Sigma_i$ are the covariance matrices of the Gaussian components.

\section{Selection Effects Related to Stellar Ages}\label{ages.appendix}

Selection of YSOs using features that are connected to disks and accretion means that pre--main-sequence stars that have lost their disks will be missed. This imposes the disk survival function, often modeled as an exponentially decreasing function with an $e$-folding timescale of 2--4~Myr \citep{2009AIPC.1158....3M,2015A&A...576A..52R,2018MNRAS.477.5191R}, as a bias on the observed age distribution.  In contrast, X-rays can identify pre--main-sequence stars both with and without disks \citep{1999ARA&A..37..363F,2018ASSL..424..119F}. Low-mass pre--main-sequence stars maintain high X-ray emission even after 10~Myr \citep[e.g.,][]{2016A&A...589A.113A,2017A&A...605A..85P}, so X-ray selection should detect such a population if it exists.

\begin{figure*}[t]
\centering
\includegraphics[width=0.9\textwidth]{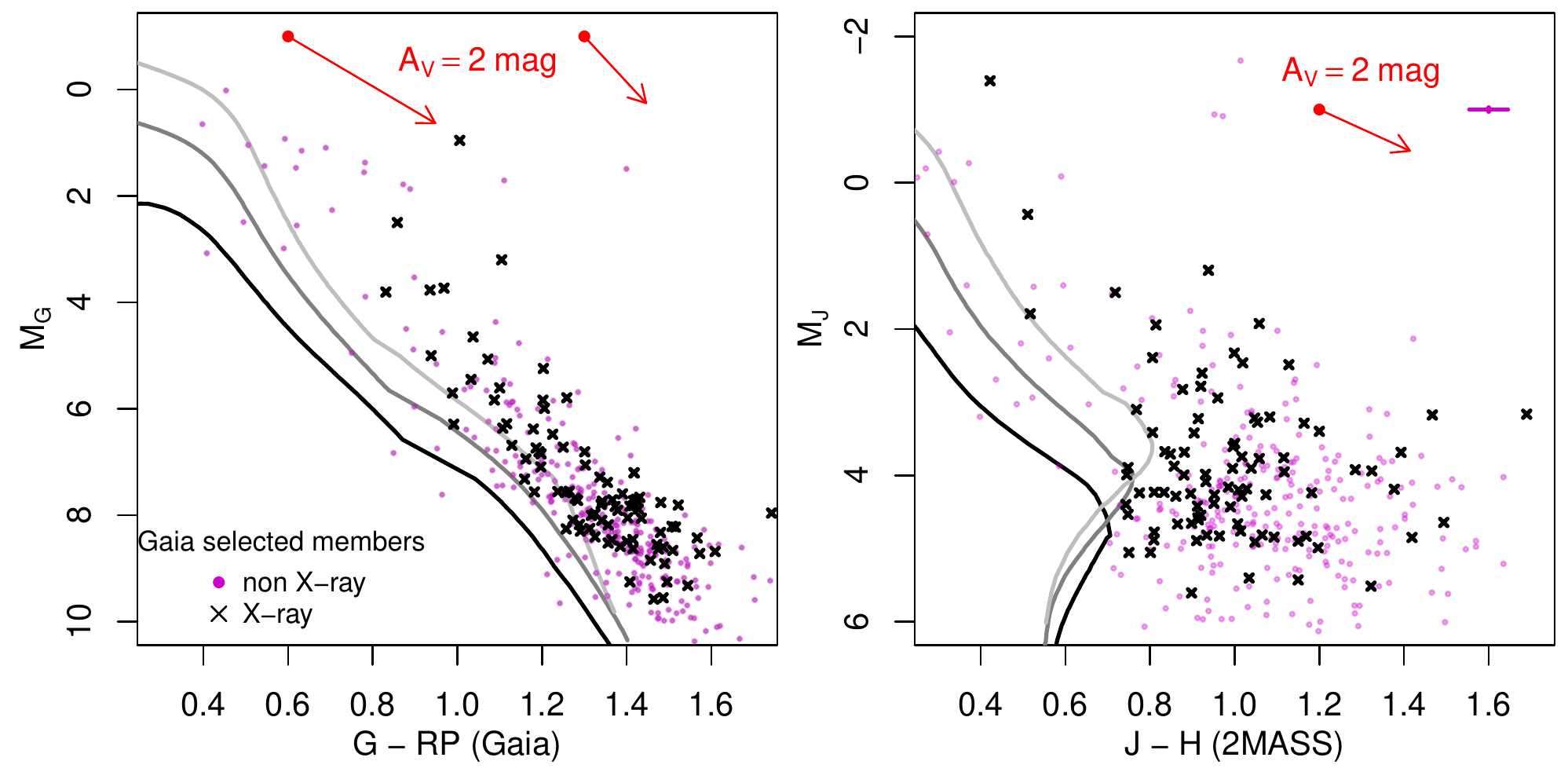} 
\caption{Optical (left) and near-infrared (right) color-magnitude diagrams of the astrometrically validated members, comparing stars selected using X-ray emission (black x's) to stars selected by other criteria (magenta points). The X-ray candidates from \citet{2017AaA...602A.115D} are some of the only stars in our literature-based sample selected with a method that is independent of whether a stars has a disk. Nevertheless, the X-ray sample does not appear older than the other NAP members. We show the same isochrones and reddening laws as in Figure~\ref{ysoc_cmd.fig}. The error bars illustrate typical 2MASS photometric uncertainties. 
\label{xray_cmds.fig}
}
\end{figure*}

The study by \citet{2017AaA...602A.115D} provides an X-ray sample for the NAP region.  In Section~\ref{ysoc_identified.sec}, we used the positions of the NAP members on the {\it Gaia} color-magnitude diagram as evidence that the NAP members are $<$3~Myr old. In Figure~\ref{xray_cmds.fig} we again show the same sources on color-magnitude diagrams, but this time with the X-ray sources marked.  On the $G$ vs.\ $G-RP$ diagram, the X-ray sources lies within the distribution of stars selected by other methods. This suggests that both samples have similar age distributions. On the $J$ vs.\ $J-H$ diagram, relatively few X-ray sources have high $J-H$ values, indicating that the X-ray selected sample is not as highly reddened as the disk-selected sample. Nevertheless, nearly all X-ray sources are above or to the right of the 1~Myr isochrone. Thus, we conclude that there is no evidence for an older pre--main-sequence stellar population within the nebula.

\section{Star Formation within Cloud~8}\label{cloud8.appendix}

Cloud 8 is superimposed near the center, and densest part, of the expanding stellar group D. Given this configuration, it appears that Cloud~8 could be the main remnant of the cloud responsible for producing this stellar group. However, it is alternatively plausible that this superposition is coincidental, and that the cloud is closer to us than the stellar group is. It is particularly difficult to distinguish between these two possibilities because our sample does not include many objects within Cloud 8, presumably due to difficulty detecting obscured YSOs. However, \citet{2002AJ....123.2559C} identified a dense, embedded cluster (number 6 in their catalog; hereafter Cambr\'esy~6) near the center of this cloud. 

\begin{figure}[t]
\centering
\includegraphics[width=0.45\textwidth]{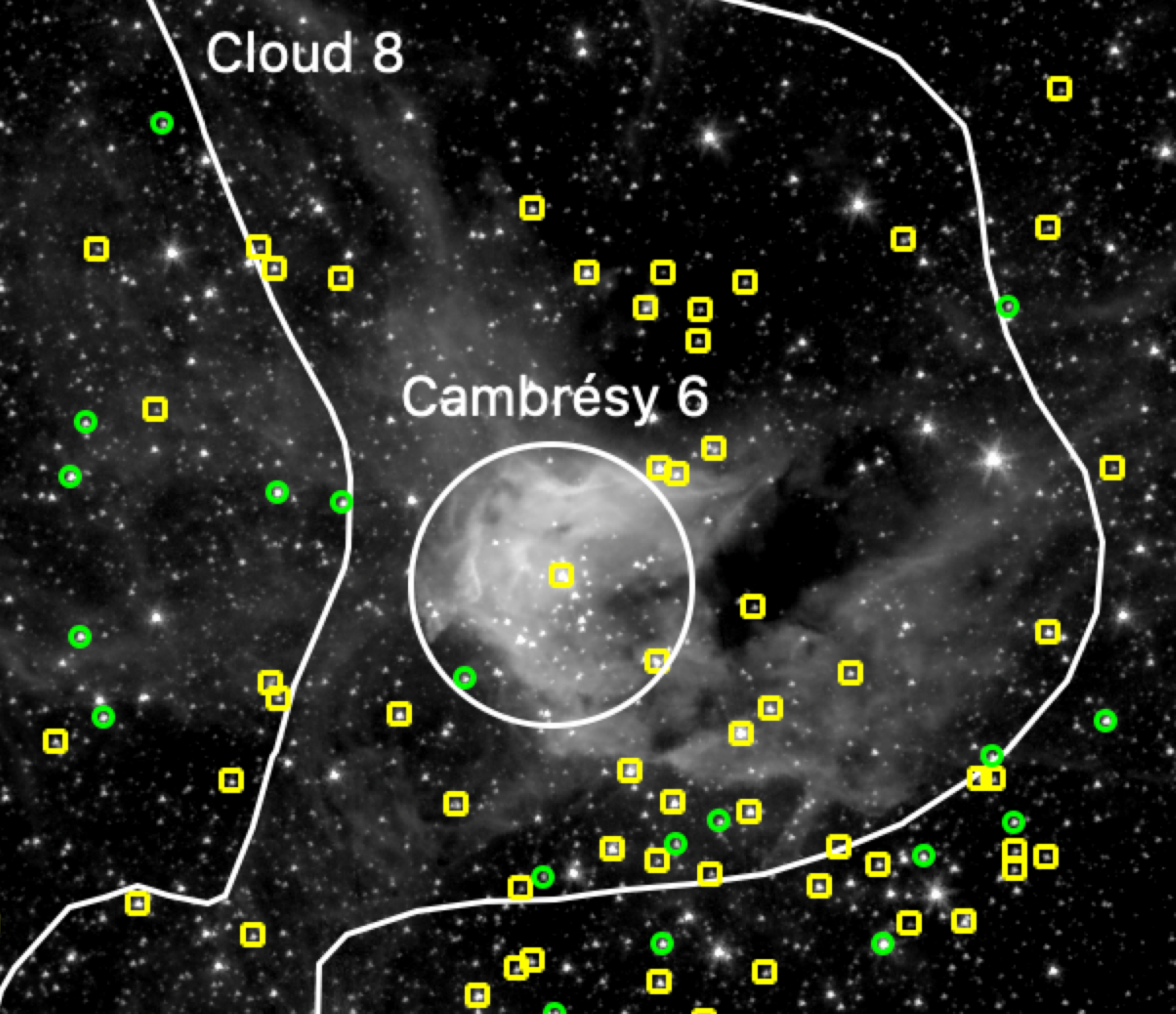} 
\caption{{\it Spitzer}/IRAC 3.6~$\mu$m image of the region around Cloud~8 (white contour). Identified YSO candidates are shown, including sources from Group~D (green circles) as well as {\it Spitzer} candidates from \citet{2011ApJS..193...25R} not included in our study (yellow squares). A 2.5$^\prime$ circle encompasses the cluster Cambr\'esy~6.
 \label{cloud8.fig}}
\end{figure}

Figure~\ref{cloud8.fig} shows the {\it Spitzer}/IRAC 3.6~$\mu$m image along with our {\it Gaia} members (green circles), {\it Spitzer} YSOs from \citet{2009ApJ...697..787G} and \citet{2011ApJS..193...25R} not included in our sample (yellow squares), the contours of Cloud 8 (outer white curves), and the location of Cambr\'esy~6 (white circle). A star cluster can clearly be seen in the {\it Spitzer} image at the location of  Cambr\'esy~6, but almost none the individual cluster members were identified as YSOs by any survey. This suggests that either the stars did not have infrared excess or the mid-infrared nebulosity in the region prevented reliable detection of infrared excess. Nevertheless, we suspect that Cambr\'esy~6 is young because, in {\it XMM Newton} images (not shown), a groups of spatially confused X-ray point sources can be seen at this location. The high absorption of this group and its spatial coincidence with the molecular gas suggests that it lies within Cloud~8 and that the cloud is actively forming stars. Nevertheless, the relation between Cambr\'esy~6 and Group~D remains a matter for future studies. 

\section{Modeling Linear Expansion}\label{linear_model.sec}

Here we describe a linear model to relate the position of a star to its velocity. This model can account for expansion (homologous or anisotropic), contraction, and even some rotational effects. 

We model the velocity of the $i$-th star, $\mathbf{v}_i = (v_{x,i},v_{y,i})$, as being related to its position, $\mathbf{x}_i = (x_i,y_i)$, by the linear regression equation
\begin{gather}
\mathbf{v}_i = \mathbf{v}_0 + M\mathbf{x}_i + \mathbf{v}_{\mathrm{scatter},i},
\end{gather}
where $\mathbf{v}_0$ is a constant velocity shift, $M$ is a $2\times2$ matrix, and the $\mathbf{v}_{\mathrm{scatter},i}$ is a random vector drawn from a bivariate normal distribution with covariance matrix $\Sigma_{\mathrm{scatter}}$. We use $\mathbf{v}_\mathrm{scatter}$ to represent the intrinsic random deviations in velocity. In addition to the intrinsic scatter in velocity, in the observed velocities, $\mathbf{v}_{\mathrm{obs},i}$, we must also account for {\it Gaia}'s measurement uncertainty. 
Thus, $\mathbf{v}_{\mathrm{obs},i}$ is related to its actual velocity by the equation
\begin{equation}
\mathbf{v}_{\mathrm{obs},i} = \mathbf{v}_i + \boldsymbol{\varepsilon}_i, 
\end{equation}
where measurement errors $\boldsymbol{\varepsilon}_i$ are drawn from Gaussian distributions with covariance matrices $\Sigma_{\varepsilon,i}$ obtained from the {\it Gaia} DR2 catalog. Putting this together give us the likelihood equation, 
\begin{equation}
\mathbf{v}_{\mathrm{obs},i} \sim \mathcal{N}(\mathbf{v}_0+M\mathbf{x}_i,\Sigma_\mathrm{scatter}+\Sigma_{\varepsilon,i}),
\end{equation}
from which we estimate the model parameters using a Bayesian approach. 

\begin{figure*}[t]
\centering
\includegraphics[width=0.45\textwidth]{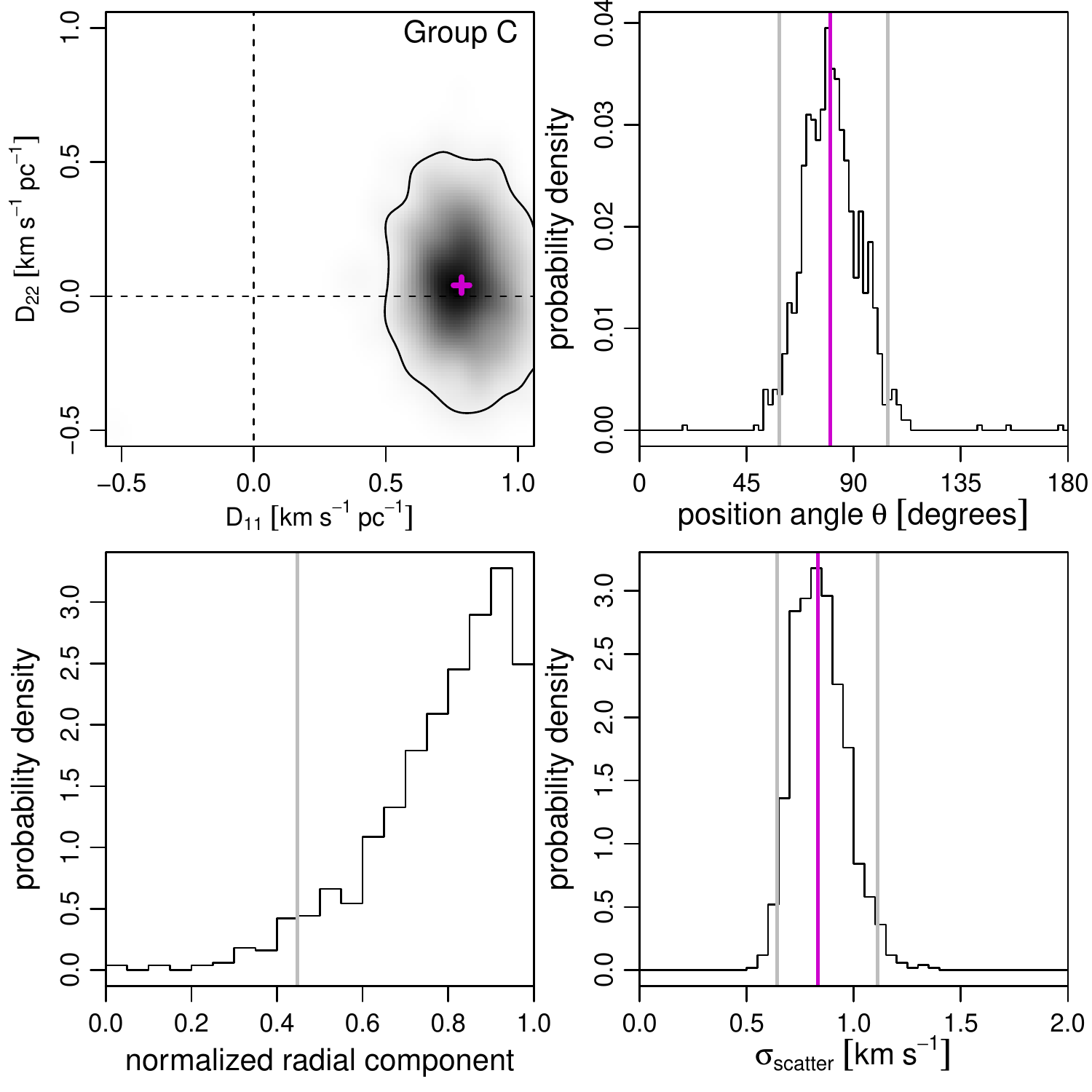} 
\includegraphics[width=0.45\textwidth]{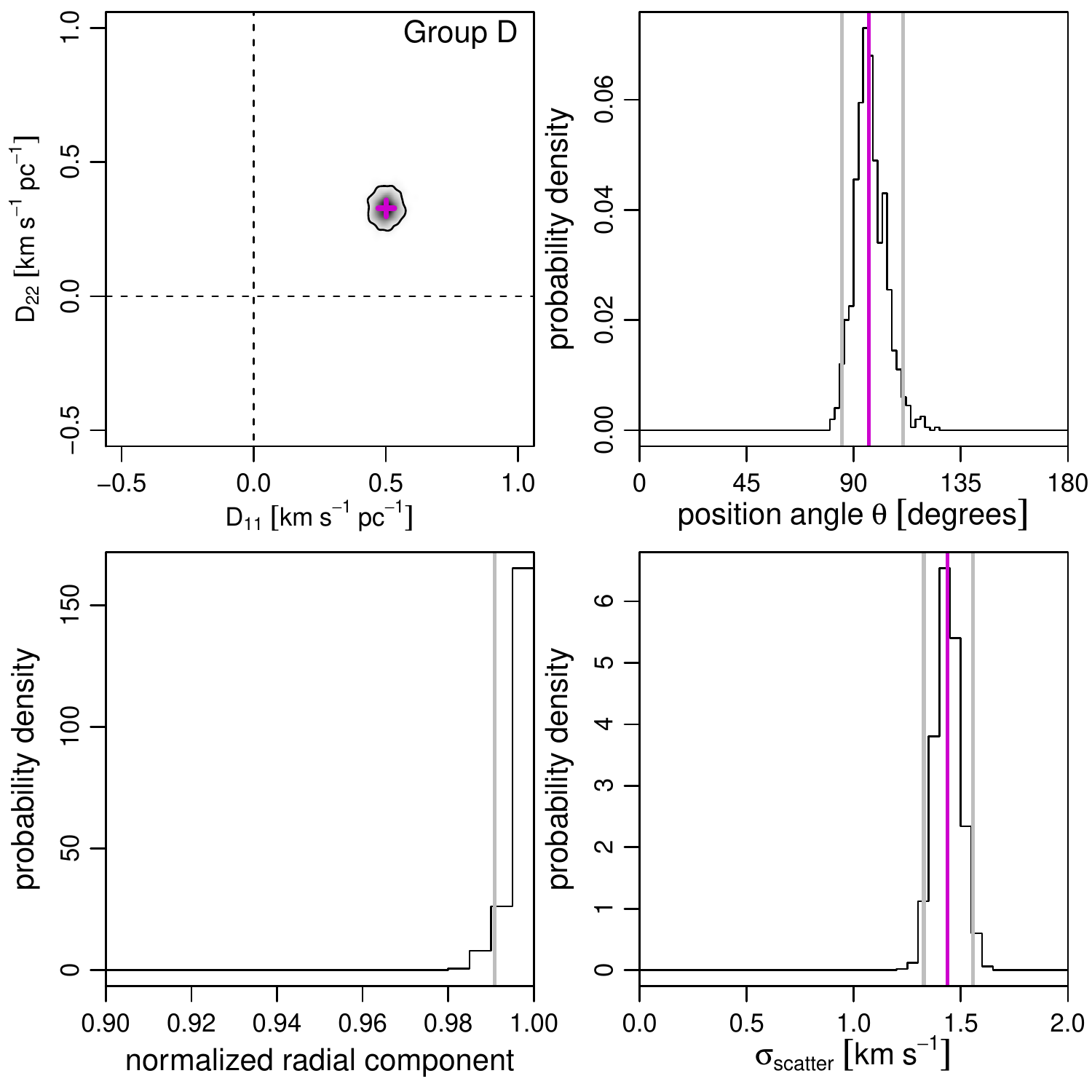} 
\includegraphics[width=0.45\textwidth]{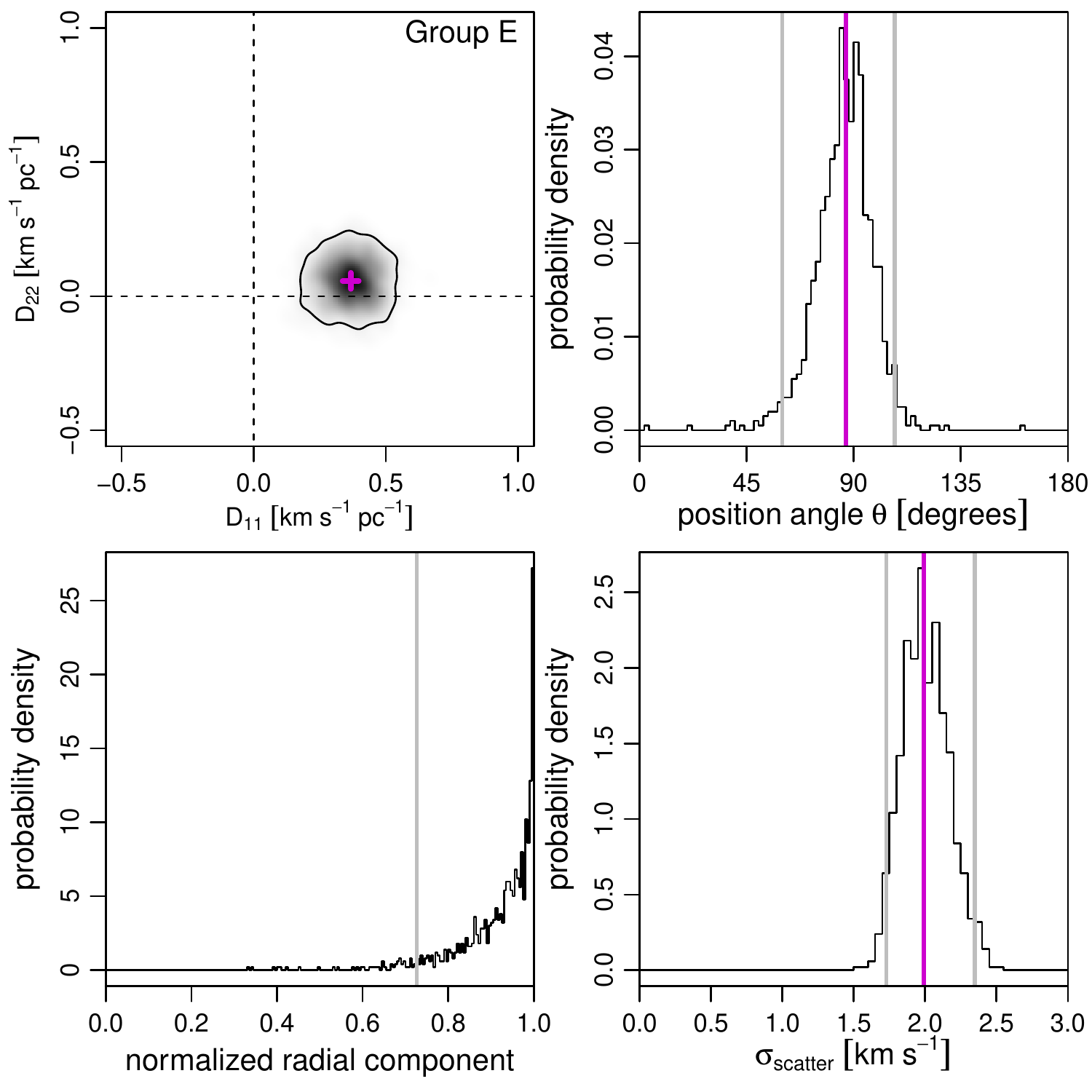} 
\includegraphics[width=0.45\textwidth]{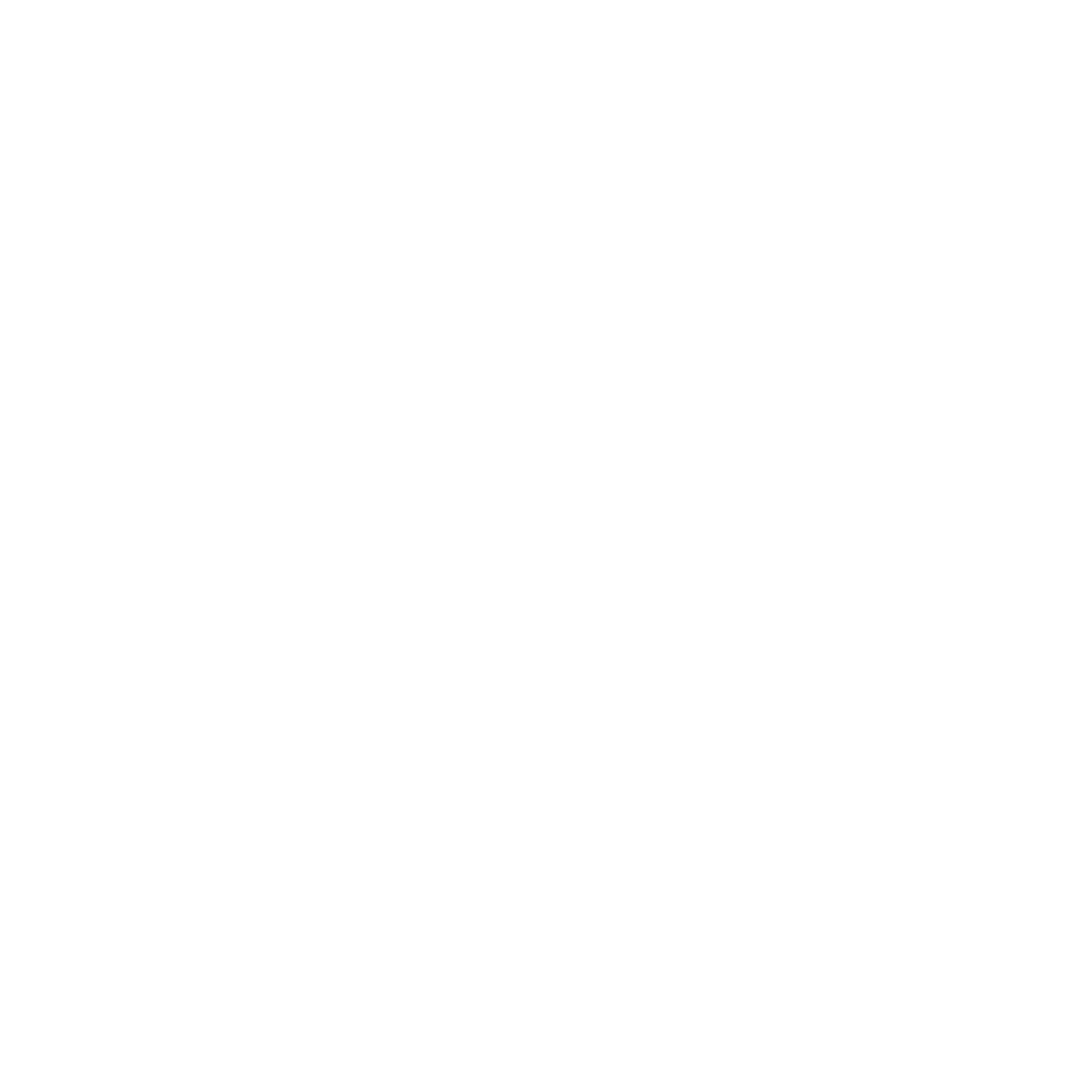} 
\caption{Marginal distributions for parameters of the linear velocity model in Groups C, D and E. These are the three stellar groups with enough sources to provide meaningful constraints on the model. For each fit, distributions are shown for the values of $D$ (upper left), the position angle $\theta$ of the anisotropy (upper right), the radial component of the velocity gradient (lower left), and the random velocity scatter (lower right). The black contour lines (upper left) and vertical gray lines (other graphs) enclose the regions of parameter space with 95\% of the probability. 
 \label{marg.fig}}
\end{figure*}

For the Bayesian model fitting, we use Markov chain Monte Carlo (MCMC) to sample from the posterior distribution, which is implemented by the software package  ``Just another Gibbs sampler'' (JAGS) version 4.3.0 \citep{plummer2017jags}, which is run using {\it rjags} \citep{plummer2019} in  R version 3.6.0. We use ``non-informative'' priors for the parameters, including uniform distributions for $v_{0,x}$ and $v_{0,y}$ from $-$2 to $2$~km~s$^{-1}$, and broad distributions for $M$ and $\Sigma_\mathrm{scatter}$ that cover the reasonable values for these parameters.\footnote{
The prior distribution for the matrix $\Sigma_\mathrm{scatter}$ is made up of chi-squared $\chi^2_1$ distributions, scaled by a factor of 9, for the diagonal entries, and uniform distributions from $-1$ to 1 for the correlation coefficients of the off-diagonal entries. For $A$, we base our priors on the decomposition of this matrix described later on: $D_{1,1}$ and $D_{2,2}$ use Gaussian priors with a mean of 0 and standard deviation of 1.5, and $\theta$ and $\varphi$ use uniform priors from 0 to $2\pi$. 
}
The statistical model was implemented in the BUGS language. For each dataset, three independent chains were run for 5,000 iterations, retaining every fifth sample, after an initial burn-in of 1,000 iterations. Convergence was assessed from inspection of the trace plots as well as the Gelman-Rubin statistic $<1.001$ \citep{1992StaSc...7..457G}. Through experimentation with a variety of priors, we found that changes to the functional form for the priors has little effect on the results, provided that the priors are sufficiently broad. We also find that the results of the Bayesian analysis are approximately consistent with the results from numerical maximum-likelihood estimation. 

The $2\times2$ matrix $M$ describes the dependence of velocity on position, encoding phenomena such as expansion, contraction, or rotation, which may be either radially symmetric or anisotropic. To examine these effects, we decompose $M$ using singular value decomposition (SVD) to write
\begin{equation}
 M = U D V^*,
 \end{equation}
where $U$ and $V^*$ are unitary matrices and $D$ is a diagonal matrix. By multiplying by the identify matrix, written as the product of two reflection matrices, without loss of generality, we can take $U$ and $V^*$ to be rotation matrices that rotate the coordinate systems by angles $\varphi$ and $\theta$, respectively, and $D$ to be a diagonal matrix with entries $D_{1,1}$ and $D_{2,2}$, where the diagonal elements can be positive, negative, or zero, but $|D_{1,1}|\geq |D_{2,2}|$. Thus, $M$ first rotates the coordinate system by angle $\theta$ -- the position angle of the velocity anisotropy, then applies the velocity gradient $D_{1,1}$, with units of km~s$^{-1}$~pc$^{-1}$, along the major axis of the anisotropy and $D_{2,2}$ along the minor axis, and finally rotates the coordinate system again by $\varphi$. The coordinate system rotations mean that the component of the velocity in the outward direction will be $\cos(\varphi+\theta)$ and the component of the velocity in the azimuthal direction will be $\sin(\varphi + \theta)$. 

To interpret these results, we note that statistically significant non-zero values of $D_{1,1}$ or $D_{2,2}$ imply dependence of velocity on position. If the velocity dependence is primarily in the radial direction (i.e.\ $\cos[\theta+\varphi]\approx 1$), then $D_{1,1},D_{2,2} > 0$ implies expansion of the system, while $D_{1,1},D_{2,2} < 0$ implies contraction. If the velocity dependence is primarily in the azimuthal direction (i.e.\ $\sin[\theta+\varphi]\approx \pm1$), non-zero values of $D_{1,1}$ and $D_{2,2}$ imply rotation. 

Figure~\ref{marg.fig} shows the marginal distributions for several of the interesting parameters in the linear models. 
For Group~D, the allowed area of parameter space is very small, so constraints on the velocity gradients are tight. Both velocity gradients are positive; with $D_{1,1}$ being noticeably larger than $D_{2,2}$. For Groups~C and E, the values of $D_{1,1}$ are constrained to be positive, but the uncertainties in $D_{2,2}$ are large enough that this parameter could equal zero. The marginal distributions of the position angles $\theta$ show that these are fairly tightly constrained for all three groups, C, D, and E. We also include marginal distributions of the normalized radial component defined as $\cos(\theta+\varphi)$. These distributions all favor values close to 1, showing that the velocity gradients imply expansion, not rotation. 

 \section{Nearby Clusters}

Examination of the {\it Gaia} $\mu_{\alpha^\star}$--$\mu_\delta$ distribution in the vicinity of the NAP region reveals two salient features that are not directly related to the NAP region. One of these, NGC~6997, is a star cluster superimposed on the ``North America'' region but generally thought to be at a different distance. This group of $\sim$600~{\it Gaia} sources stands out clearly in proper-motion space because it has a smaller velocity dispersion, but it also has significantly different mean proper motions of $\mu_{\alpha^\star}=-4.3$~mas~yr$^{-1}$ and $\mu_\delta = -6.8$~mas~yr$^{-1}$, meaning that confusion between its members and those of the NAP complex is minimal. The group has a parallax of 1.12~mas, meaning that it is $\sim$100~pc farther than the main NAP complex.   

The other feature is a group of $\sim$50 {\it Gaia} sources centered at 21:01:53 +45:12:00 with $\mu_{\alpha^\star}=-1.7$~mas~yr$^{-1}$, $\mu_\delta = -3.8$~mas~yr$^{-1}$, and $\varpi_0=1.0$~mas. 
 
\acknowledgements We thank Min Fang for expert advice about the NAP region, Kevin Fogarty and Michael Grudi\'c for helpful discussions, and the anonymous referee for useful comments. Development of YSO Corral was supported by NASA Award NNX17AF41G.
ARAM was supported by Caltech's Freshman Summer Research Institute (FSRI). 

\facility{FCRAO, {\it Gaia}, Keck:I (HIRES)} 

\software{
          astropy \citep{cite-astropy13, cite-astropy18},
          BUGS \citep{lunn2009bugs},
          class \citep{MASS},
          JAGS \citep{plummer2017jags},
          mclust \citep{mclust5},
          numpy \citep{2011CSE....13b..22V},
          pandas \citep{pandas-proc, pandas-zenodo},
          postgres,
          R \citep{RCoreTeam2018}, 
          rjags \citep{plummer2019},
          SAOImage DS9 \citep{2003ASPC..295..489J},
          scipy \citep{2020NatMe..17..261V},
          scikit-learn \citep{Pedregosa11},
          TOPCAT \citep{2005ASPC..347...29T}
          }

\vspace*{100px}
\bibliography{ms.bbl}

\clearpage\clearpage

\begin{deluxetable*}{llrcrrr}[t!]
\tablecaption{Summary of Candidate NAP Members from the Literature\label{pub.tab}}
\tabletypesize{\small}\tablewidth{0pt}
\tablehead{
  \colhead{Reference} &  \colhead{Method}   &  \colhead{Number of} & \colhead{Ref.} &\colhead{\it Gaia} & \multicolumn{2}{c}{Membership} \\ [-0.2cm]
    \colhead{} &  \colhead{}   &  \colhead{Candidates} & \colhead{Notes} &\colhead{Matches} & \colhead{N.-Mem.} &\colhead{Mem.} \\ [-0.2cm]
    \colhead{(1)} &  \colhead{(2)}   &  \colhead{(3)} &\colhead{(4)} &\colhead{(5)} &\colhead{(6)} &\colhead{(7)}
  }
\startdata
\citet{1949ApJ...110..387M} & 		H$\alpha$	 &		5 &		$a$&			60\% &		3 &				0\\
\citet{1958ApJ...128..259H} &		H$\alpha$  &		68 &			&				40\% &		 6 &				21\\
\citet{1973AaAS....9..183W} &		H$\alpha$ &		142 &	$b$&		89\% &		 113 &				13\\
\citet{1979ApJS...41..743C} &		H$\alpha$ &	21 &			$c$&			57\%  &		1 &				11\\
\citet{1980ApJ...239..121B} &		IR &				11 &			&			36\% &		 3 &				1\\
\citet{1980AJ.....85..230M} &		H$\alpha$ &		7 &			&			57\% &		 3 &				1\\
\citet{2002AJ....123.2597O} &		H$\alpha$ &		32 &			&			47\% &		1 &				14\\
\citet{2005AaA...430..541C} &		Spec. &			8 &		$d$&			88\% &		 5 &				2\\
\citet{2006BaltA..15..483L} &		Phot./H$\alpha$ &	430 &	$e$&		84\% &		 340 &				21\\
\citet{2008MNRAS.384.1277W} & H$\alpha$ &			39 &	$a$&			82\% &		7 & 				25\\
\citet{2008BaltA..17..143S} &		CMD &			5 &			&	 60\% &		 3 &				0\\
\citet{2009BaltA..18..111C} &		Spec.\ &			34 &		$f$&		62\% &		 18 &				3\\
\citet{2009ApJ...697..787G} &		IRE &			1,657 &		& 	24\% &		 125 &				272\\
\citet{2011ApJS..193...25R} &		IRE &			1,329 &	$g$& 	29\% &		 140 &				252\\
\citet{2011AaA...528A.125A} &		H$\alpha$/Clust. &		54 &		& 	22\% &		 1  &				11\\
\citet{2017AaA...602A.115D} &		X-ray &			721 &		&	19\% &		 46 &				93\\
&							H$\alpha$&		123 &		&	66\% &		 22 &				59\\
&							IRE &			179 &		&	51\% &		 58  &				34\\
Combined sample &				All&			         3,473 &		&	35\%&		814 &			  395\\
\enddata
\tablecomments{References for published lists of candidate members of the NAP region. Column~2 is method for selection, including 
H$\alpha$ emission, 
spectral classification,
placement on the color-magnitude diagram (CMD),  
brightness in the infrared (IR) or
IR excess (IRE), spatial clustering,
and X-ray emission. 
Column 3 is the number of candidates in the paper. 
Column 4 gives notes about the reference. 
Column~5 is the percentage of sources with {\it Gaia} DR2 counterparts that pass our quality criteria.  
Column 6 is the number of objects that we reject as members, and Column 7 is the number of objects whose membership we have validated.
The final row gives the statistics of the combined sample; the numbers of sources (Columns~3, 6, and 7) in preceding rows do not sum to the values in this row because many objects have been repeatedly selected by multiple studies. 
}
\vspace*{-0.05in}
\tablenotetext{$a$}{Objects were cataloged as H$\alpha$ emission stars in the referenced papers but later upgraded to potential NAP members by \citet{2011ApJS..193...25R}.}
\vspace*{-0.05in}
\tablenotetext{$b$}{We use the corrected list of 142 objects available from \url{ftp://ftp.lowell.edu/pub/bas/starcats/welin.cyg}, rather than the originally published list of 141 objects.}
\vspace*{-0.05in}
\tablenotetext{$c$}{The H$\alpha$ sample was enlarged by including stars spatially associated with known H$\alpha$ objects.}
\vspace*{-0.05in}
\tablenotetext{$d$}{In addition to the Bajamar Star, we consider all sources from \citet{2005AaA...430..541C} not classified as AGB stars to be candidate members.}
\vspace*{-0.05in}
\tablenotetext{$e$}{We include all 430 stars in the direction of L935. However, a subset of 41 stars are flagged as having H$\alpha$ emission, 20 of which are classified as non-members and 4 as members by our analysis.}
\vspace*{-0.05in}
\tablenotetext{$f$}{We include all objects for which spectra are provided, including 19 stars with emission lines and 15 without. We classify 3 emission-line stars as members and 9 as non-members, as well as 9 non-emission-line stars as non-members.}
\vspace*{-0.05in}
\tablenotetext{$g$}{This combines 1,286 YSO candidates identified in a {\it Spitzer}/MIPS-based search with 43 new IRAC-only candidates.}
\end{deluxetable*}

\begin{deluxetable}{lrrr}[h]
\tablecaption{Astrometrically Validated Members\label{members.tab}}
\tabletypesize{\small}\tablewidth{0pt}
\tablehead{
  \colhead{{\it Gaia} DR2} &  \colhead{$\alpha$}   &  \colhead{$\delta$}   & \colhead{Group} \\
  \colhead{} &  \colhead{(ICRS)}   &  \colhead{(ICRS)} \\
  \colhead{(1)} &  \colhead{(2)}   &  \colhead{(3)} & \colhead{(3)}
}
\startdata
 2163138601938577664&	20 51 19.8&	+44 23 06&	D\\
 2163137742945112960&	20 51 20.6&	+44 20 32&	C\\
 2163138705017719296&	20 51 21.3&	+44 24 05&	D\\
 2163136123738466688&	20 51 12.0&	+44 18 47&	C\\
 2163137742945115136&	20 51 22.6&	+44 21 07&	C\\
 2163148772421081728&	20 51 22.8&	+44 33 42&	D\\
 2163149665774282368&	20 51 23.6&	+44 35 22&	D\\
 2066862546309607552&	20 51 26.4&	+43 53 12&	D\\
 2162947424350344448&	20 51 24.4&	+44 13 04&	C\\
 2163135883219217280&	20 51 24.6&	+44 17 54&	D\\
 \enddata
\tablecomments{{\it Gaia} DR2 counterparts for the 395 previously published YSO candidates that passed our astrometric membership criteria.  
Column~1: Source designation in the DR2 catalog. Columns 2--3: Source positions are truncated, not rounded. We encourage readers to obtain the full-precision {\it Gaia} astrometry for these sources directly from the {\it Gaia} Archive (\url{https://gea.esac.esa.int/archive/}). Column~4: The kinematic group assignment for each star.\\
(This table is available in its entirety in a machine-readable form in the online journal. A portion is shown here for guidance regarding its form and content.)
}
\end{deluxetable}

\begin{deluxetable*}{crrrRRRR}[t]
\tablecaption{Astrometric Properties of Stellar Groups\label{astro.tab}}
\tabletypesize{\small}\tablewidth{0pt}
\tablehead{
  \colhead{Group} &   \colhead{$\alpha_0$}  & \colhead{$\delta_0$}  &   \colhead{$N_\mathrm{samp}$} & \colhead{$\mu_{\alpha^\star,0}$} &  \colhead{$\mu_{\delta,0}$}  & \colhead{$\varpi_0$} & \colhead{$r_{hm}$}\\
\colhead{}  & \colhead{(ICRS)}  & \colhead{(ICRS)}  & \colhead{(stars)} & \colhead{(mas~yr$^{-1}$)} &  \colhead{(mas~yr$^{-1}$)}  &  \colhead{(mas)} & \colhead{(pc)}\\
\colhead{(1)}  & \colhead{(2)}  & \colhead{(3)}  & \colhead{(4)} & \colhead{(5)} &  \colhead{(6)}  &  \colhead{(7)} &  \colhead{(8)} 
}
\startdata
A & 20 47 50& +43 47 27 & 21 & -1.51\pm0.21 & -2.11\pm0.14 & 1.23\pm0.02 & 1.1\\
B & 20 50 20& +44 37 58 & 19 & -0.57\pm0.29 & -4.64\pm0.18 & 1.21\pm0.03 & 1.8\\
C & 20 51 10& +44 19 25 & 47 & -1.11\pm0.19 & -4.13\pm0.17 & 1.23\pm0.01 & 1.1\\
D& 20 52 50& +44 22 13 & 235 & -1.19\pm0.11 & -3.13\pm0.09 & 1.21\pm0.01 & 1.9\\
E & 20 57 30& +43 46 36 & 59 & -0.70\pm0.14 & -3.24\pm0.15 & 1.27\pm0.02 & 1.3\\
F & 20 58 50& +44 09 43 & 14 & -2.68\pm0.06 & -3.35\pm0.25 & 1.06\pm0.02 & 2.0\\
\enddata
\tablecomments{Column 1: Name of the stellar group. Columns 2--3: Approximate coordinates of the group center. Column 4: Number of {\it Gaia}-validated members assigned to each group. Columns 5--6: Mean proper motions of the group. Column 7: Mean parallax of the group.  Column 8: Characteristic radius for the stellar group, defined as the median distance of group members, in projection, from the group center.  All values are in the {\it Gaia} DR2 system, with no correction for zero-point offset.
}
\end{deluxetable*}

\begin{deluxetable*}{crrRRRRRRcc}[t]
\tablecaption{Clouds Properties from the $^{13}$CO Line Map\label{com.tab}}
\tabletypesize{\small}\tablewidth{0pt}
\tablehead{
  \colhead{Desig.} &   \colhead{$\alpha$}  & \colhead{$\delta$}  &   \colhead{$M_{^{13}\mathrm{CO}}$} & \multicolumn{4}{c}{$v_{lsr}$ distribution} &  \colhead{$r_{\mathrm{cloud}}$}  & \colhead{Optical} & \colhead{Group}\\
 &&&&\colhead{mean} &  \colhead{std.\ dev.}  &  \colhead{mode} & \colhead{FWHM}\\
\colhead{}  & \colhead{(ICRS)}  & \colhead{(ICRS)}  & \colhead{($M_\odot$)} & \colhead{(km~s$^{-1}$)} &  \colhead{(km~s$^{-1}$)} &  \colhead{(km~s$^{-1}$)} &  \colhead{(km~s$^{-1}$)}  &  \colhead{(pc)}  & \colhead{} &  \colhead{}\\
\colhead{(1)}  & \colhead{(2)}  & \colhead{(3)}  & \colhead{(4)} & \colhead{(5)} &  \colhead{(6)}  &  \colhead{(7)}  &  \colhead{(8)} &  \colhead{(9)} &  \colhead{(10)}  &  \colhead{(11)} 
}
\startdata
1 & 	20 48 58	& 	+43 48 10	& 	330	& 	0.8	 & 	1.9	 & 	1.2  & 	1.9	&	1.1			&	Dark	&	A	\\
2 &  	20 50 16	& 	+44 29 40	& 	2100	& 	-1.1	 & 	2.2	 & 	-1.1  & 	2.5	&	1.7	 		&	BRC$^\dagger$/Dark	&	B,C	\\
3 &  	20 51 12	& 	+44 48 00	& 	1100	& 	-0.9	 & 	2.4	 & 	-1.5  & 	1.9	&	1.3	 		&	Dark/Neb.	&	B	\\
4&  	20 51 42	& 	+44 12 30	& 	610	& 	1.7	 & 	4.0	 & 	2.5  & 	1.3	&	1.0	 		&	Dark/Neb.	&	D	\\
5 & 	20 52 16	& 	+43 54 20	& 	170	& 	2.8	 & 	4.4	 & 	3.8  & 	2.5	&	1.2	 		&	Neb.	&	D	\\
6 & 	20 52 37	& 	+43 37 10	& 	220	& 	2.2	 & 	2.7	 & 	1.2  & 	2.3	&	0.7	 		&	Dark	&	E	\\
7 & 	20 53 01	& 	+44 27 30	& 	120	& 	-1.7	 & 	3.5	 & 	-3.0  & 	1.9	&	0.7	 		&	Dark	&	D	\\
8 & 	20 54 12	& 	+44 24 40	& 	1500	& 	-2.8	 & 	3.7	 & 	-4.9  & 	3.2	&	2.2	 		&	Dark	&	D	\\
9 & 	20 56 13	& 	+43 41 40	& 	8000	& 	2.0	 & 	3.0	 & 	1.7  & 	5.3	&	3.5			&	Dark	&	E	\\
10 & 	20 57 48	& 	+44 04 00	& 	80	& 	4.4	& 	1.5	& 	 4.9  & 	1.5	&	0.4	 		&	Dark	&	E	\\
11 & 	20 58 26	& 	+43 18 00	& 	50	& 	0.6	 & 	2.0	 & 	1.2  & 	2.5	&	0.2	 		&	Dark	&	E	\\
12 & 	20 58 42	& 	+44 00 20	& 	30	& 	4.2	 & 	1.4	 & 	4.6  & 	2.8	&	0.1	 		&	Dark	&	E	\\
13 & 	20 59 13	& 	+44 12 20	& 	250	& 	3.7	 & 	2.2	 & 	4.2  & 	1.1	&	0.9	 		&	Neb.	&	F	\\
\enddata
\tablecomments{
Column 1: Cloud designation. 
Columns 2--3: Coordinates of the cloud center. 
Column 4: $^{13}$CO mass estimate of the cloud. 
Columns 5: Mean velocity of the cloud. 
Column 6: Standard deviation of the velocity distribution. 
Column 7: Mode of the velocity distribution. 
Column 8: $\Delta v_\mathrm{FWHM}$ for the most-prominent mode. 
Column 9: Characteristic size of the cloud, defined as the radius that contains half the integrated emission. 
Column 10: Appearance of the cloud (or nebular material in the direction of the cloud) in the optical image. 
Column 12: Group of stars associated with the cloud.
}
\tablenotetext{\dagger}{BRC $=$ bright rim cloud.}
\end{deluxetable*}

\begin{deluxetable}{cRRRRRR}[t]
\tablecaption{Cloud Dynamical Properties}\label{cloud_dynamics.tab}
\tabletypesize{\small}\tablewidth{0pt}
\tablehead{
  \colhead{Cloud} &   
  \colhead{$\log \bar{\rho}$} & 
  \colhead{$\tau_\mathrm{cross}$}  &   
  \colhead{$\tau_\mathrm{ff}$}  &   
    \colhead{$\mathcal{M}$} &
        \colhead{$M_\mathrm{dyn}$} &  
          \colhead{$\alpha_\mathrm{BM92}$} \\
  \colhead{}&
 \colhead{(g~cm$^{-3}$)}  &  
 \colhead{(Myr)}&
 \colhead{(Myr)}&
   \colhead{}&
    \colhead{($10^3\,M_\odot$)}&  
   \colhead{}   \\
\colhead{(1)}  & \colhead{(2)}  & \colhead{(3)}  & \colhead{(4)} & \colhead{(5)} &  \colhead{(6)}   &  \colhead{(7)}
}
\startdata
1       &          -20.40   &  1.3  &  1.1&  3.6  &0.8  &      2.5\\
2       &          -20.17   &  1.6  &  0.8&  4.7  &2.2  &      1.1\\
3       &          -20.10   &  1.6  &  0.8&  3.6  &1.0  &      0.9\\
4       &          -20.01   &  1.8  &  0.7&  2.5  &0.3  &      0.6\\
5       &          -20.80   &  1.1  &  1.7&  4.7  &1.6  &      9.3\\
6       &          -19.98   &  0.7  &  0.7&  4.4  &0.8  &      3.5\\
7       &          -20.24   &  0.9  &  0.9&  3.6  &0.5  &      4.3\\
8       &          -20.65   &  1.6  &  1.4&  6.1  &4.7  &      3.2\\
9       &          -20.52   &  1.5  &  1.2& 10.1 &20.6&        2.6\\
10     &           -19.67  &   0.6 &   0.5&  2.8 & 0.2 &       2.2\\
11     &           -19.01  &   0.2 &   0.2&  4.7 & 0.3 &       5.4\\
12     &           -18.38  &   0.1 &   0.1&  5.3 & 0.2 &       6.4\\
13     &           -20.26  &   1.9 &   0.9&  2.1 & 0.2 &       0.9\\
\enddata
\tablecomments{
Column~1: Cloud designation. 
Column~2: Mean density of the cloud, defined as $\bar{\rho} = M_{^{13}\mathrm{CO}}/(4/3)\pi r^3_\mathrm{cloud}$.
Column~3: Crossing timescale, defined as $\tau_\mathrm{cross} = r_\mathrm{cloud}/\Delta v$.
Column~4: Free-fall timescale defined using $\bar{\rho}$ from Column~3.
Column~5: Mach number. 
Column~6: Dynamical mass of the cloud assuming $\alpha=1$ (Equation~\ref{dynamical_mass.eqn}). 
Column~7: The \citet{1992ApJ...395..140B} estimate of the virial parameter, using $M_{^{13}\mathrm{CO}}$ for the mass, $r_\mathrm{cloud}$ for the radius, and $\sigma_{nt}$ for the velocity dispersion.
}
\end{deluxetable}

\begin{deluxetable*}{crrRRRRRRR}[t]
\tablecaption{Internal Kinematic Properties of Stellar Groups\label{kinematics.tab}}
\tabletypesize{\small}\tablewidth{0pt}
\tablehead{
  \colhead{Group} &  \multicolumn{2}{c}{Kendall's $\tau$}  &    \multicolumn{2}{c}{Mean Motions}  &  \multicolumn{4}{c}{Linear Expansion Model Parameters}  & \colhead{Total}\\
  \colhead{} &   \colhead{$p_{x}$} &   \colhead{$p_{y}$}  &   \colhead{mean $v_{out}$}  & \colhead{mean $v_{az}$}  &   \colhead{Gradient 1} & \colhead{Gradient 2} &  \colhead{$\theta$} &  \colhead{$\sigma_{\mathrm{scatter}}$} &  \colhead{$\sigma_{\mathrm{1D}}$} \\
  \colhead{} &   \colhead{} &   \colhead{}  &   \colhead{(km~s$^{-1}$)}  & \colhead{(km~s$^{-1}$)}  &   \colhead{(km~s$^{-1}$~pc$^{-1}$)} & \colhead{(km~s$^{-1}$~pc$^{-1}$)} &  \colhead{($^\circ$)} &   \colhead{(km~s$^{-1}$)} &   \colhead{(km~s$^{-1}$)} \\
\colhead{(1)}  & \colhead{(2)}  & \colhead{(3)}  & \colhead{(4)} & \colhead{(5)} &  \colhead{(6)}  &  \colhead{(7)}  &  \colhead{(8)} &  \colhead{(9)} &  \colhead{(10)}
}
\startdata
A & $>$0.05 & $>$0.05	&	0.7\pm0.4		&	0.0\pm0.5		&\nodata&\nodata&\nodata&\nodata&	1.8\pm0.2\\
B & 0.008 & 0.002		&	0.5\pm0.7		&	-1.4\pm1.0	&\nodata&\nodata&\nodata&\nodata&	1.7\pm0.2\\
C & $<$0.001 & $>$0.05	&	0.7\pm0.6		&	0.6\pm0.4		&0.63\pm0.11	&0.04\pm0.17	&	81\pm12	&	0.8\pm0.1	&	1.0\pm0.1\\
D& $<$0.001 & $<$0.001	&	1.9\pm0.4	 	&	0.3\pm0.2		&0.50\pm0.03	&0.33\pm0.03	&	98\pm8\,~	&	1.4\pm0.1	&	2.0\pm0.1\\
E & $<$0.001 & $>$0.05	&	0.6\pm0.6		&	-0.2\pm0.4	&0.33\pm0.08	&0.05\pm0.07	&	87\pm6\,~&	2.0\pm0.1&	2.1\pm0.2\\
F & $>$0.05 & $>$0.05	&	-0.1\pm0.9 	&	-0.1\pm0.2	&\nodata&\nodata&\nodata&\nodata&	1.2\pm0.3\\
\enddata
\tablecomments{Column 1: Stellar group. Columns 2--3: Non-parametric tests for correlation between position and velocity in right ascension and declination. Columns 4--5: Mean outward velocity and mean azimuthal velocity as calculated in Section~\ref{within_groups.sec}. Columns 6--9: Parameters from the linear model fit to the velocity from Section~\ref{velocity_gradients.sec}, including the velocity gradients along the direction of maximum gradient (Column 6) and in the orthogonal direction (Column 7), the position angle of the anisotropy (Column 8), and the intrinsic velocity scatter left over after the velocity gradient is removed (Column 9). Column 10: The total velocity dispersion.
}
\end{deluxetable*}

\begin{deluxetable}{lrrr}
\tablecaption{New Member Candidates from {\it Gaia}\label{gaia_candidates.tab}}
\tabletypesize{\small}\tablewidth{0pt}
\tablehead{
  \colhead{{\it Gaia} DR2} &  \colhead{$\alpha$}   &  \colhead{$\delta$} &  \colhead{Group} \\
  \colhead{} &  \colhead{(ICRS)}   &  \colhead{(ICRS)} 
  }
\startdata
 2067063447700277504&	20 50 06.05&	+44 17 48.9&	C\\
 2166282204462982528&	20 50 07.33&	+45 49 22.4&	distrib\\
 2067060007428506112&	20 50 07.69&	+44 08 30.1&	D\\
 2163214738816798336&	20 50 10.93&	+44 51 22.0&	D\\
 2065821858551580288&	20 50 12.03&	+41 38 44.1&	distrib\\
 2163266832483920128&	20 50 12.94&	+45 32 43.1&	distrib\\
 2066624635182982144&	20 50 14.35&	+42 15 43.0&	distrib\\
 2067049428926311040&	20 50 15.04&	+43 58 11.8&	D\\
 2166282964674128256&	20 50 15.47&	+45 54 02.3&	distrib\\
 2067049532005529600&	20 50 15.81&	+43 58 59.9&	A\\
 2067060625914494080&	20 50 16.82&	+44 11 53.5&	B\\
\enddata
\tablecomments{{\it Gaia} DR2 sources selected as new member candidates. Columns are the same as in Table~\ref{members.tab}, with the addition of labels for Group~G and distributed stars to the final column.\\
(This table is available in its entirety in a machine-readable form in the online journal. A portion is shown here for guidance regarding its form and content.)
}
\end{deluxetable}

\end{document}